\documentclass[prd, aps, nofootinbib]{revtex4-2}

\usepackage{mathtools}
\usepackage{color}
\definecolor{darkred}{RGB}{196,0,0}

\usepackage{amsmath, amssymb, amsfonts}

\usepackage[hidelinks]{hyperref}
\usepackage{ulem}
    
\newcommand{\be}{\begin{equation}}
\newcommand{\ee}{\end{equation}}
\newcommand{\ba}{\begin{eqnarray}}
\newcommand{\ea}{\end{eqnarray}}
\newcommand{\kh}{\hat k}
\newcommand{\uh}{\hat \nu}
\newcommand{\ut}{\tilde \nu}
\newcommand{\oh}{\hat\omega}
\newcommand{\km}{k_{max}}

\begin{document}

\title {Collective modes of a collisional anisotropic quark-gluon plasma}

\author{Ruizhe Zhao}
\author{Luhua Qiu}
\author{Yun Guo$^*$}
\affiliation{Department of Physics, Guangxi Normal University, Guilin, 541004, China}
\affiliation{Guangxi Key Laboratory of Nuclear Physics and Technology, Guilin, 541004, China}

\author{Michael Strickland$^\dagger$}
\affiliation{Department of Physics, Kent State University, Kent, OH 44242, United States}

\renewcommand{\thefootnote}{\fnsymbol{footnote}}
\footnotetext[1]{yunguo@mailbox.gxnu.edu.cn}
\footnotetext[2]{mstrick6@kent.edu}
\renewcommand{\thefootnote}{\arabic{footnote}}

\begin{abstract}
In this paper we consider the collective modes of a momentum-space anisotropic quark-gluon plasma taking into account the effect of collisions between the plasma constituents.  Our analysis is carried out using a collisional kernel of Bhatnagar–Gross–Krook form and extends prior analyses in the literature by considering all possible angles of propagation of the gluonic modes relative to the momentum-anisotropy axis.  We extract both the stable and unstable modes as a function of the collision rate and confirm prior findings that gluonic unstable modes can be eliminated from the spectrum if the collision rate is sufficiently large. In addition, we discuss the conditions necessary for the existence of unstable modes and present evidence that unstable mode growth rates are maximal for modes with momentum along the anisotropy direction.  Finally, we demonstrate that when there is a finite collisional rate, gluonic unstable modes are absent from the spectrum at both small and large momentum anisotropy. These results pave the way for understanding the impact of collisions on a variety of non-equilibrium quark-gluon plasma observables.
\end{abstract}

\maketitle

\section{Introduction}

At sufficiently high temperatures nuclear matter undergoes a phase transition from hadronic bound states to liberated quark and gluon degrees of freedom.  This phase, called the quark-gluon plasma (QGP), is expected to be weakly-coupled at very high temperatures due to the asymptotic freedom of quantum chromodynamics (QCD).  Despite the fact that perturbative methods can be applied, in order to systematically understand the physics of a high-temperature QGP, one must resum a class of graphs called hard thermal loops (HTLs) \cite{Weldon:1982aq,Braaten:1989mz,Braaten:1989kk,Frenkel:1989br,Braaten:1991gm}.  The HTL dressing of the quark and gluon propagators introduces new mass scales that are proportional to the temperature.  These temperature-dependent masses can be obtained from the static limit of the quark and gluon self-energies, which are non-trivial functions of the temperature and the four-momentum of the excitations \cite{Weldon:1982aq,Braaten:1989mz}.  Computation of these self-energies allows one to determine the collective modes of the system, which in the case of gluons can be determined by finding the poles in the resummed propagators.  In the case of an equilibrium QGP, one finds that there are two gluonic collective modes corresponding to transverse and longitudinal polarizations \cite{Pisarski:1988vb}.  In addition, one finds that a non-trivial cut appears in the self-energies related to the physics of Landau damping of plasma oscillations.

Such resummations are necessary ingredients for the computation of both static and dynamical properties of the QGP. See Ref.~\cite{Ghiglieri:2020dpq} for a recent review of perturbative resummation in QCD. Despite the broad applicability of these results, they are limited in their assumption that collisions among plasma constituents are ignorable. In the high temperature limit this is due to the fact that the time scale between hard collisions (large momentum exchange), given by $t_{\rm hard} T \sim 1/ \alpha_s^2 \log \alpha_s^{-1}$, is much longer than the time-scale for the soft momentum exchanges, given by $t_{\rm soft} T \sim 1/ \alpha_s \log \alpha_s^{-1}$ \cite{Schenke:2006xu}.  As the temperature is decreased and $\alpha_s$ increases in magnitude it becomes necessary to account for both hard and soft collisions. In thermal equilibrium for a variety of observables, e.g., jet energy loss, the systematic inclusion of collisions can be accomplished using the effective kinetic theory of Arnold, Moore, and Yaffe \cite{Arnold:2002zm, AbraaoYork:2014hbk,Kurkela:2015qoa}.

In the case of a non-equilibrium QGP it is still possible to make studies of the collective modes of the system, however, the situation is complicated by the presence of unstable gluonic modes related to the chromo-Weibel instability \cite{Mrowczynski:1993qm,Mrowczynski:1996vh,Romatschke:2003ms,Arnold:2003rq,Romatschke:2004jh}. For a recent review of plasma instabilities in the QGP, see Ref.~\cite{Mrowczynski:2016etf}.  In the collisionless limit such instabilities generically arise if there is any degree of momentum anisotropy in the underlying gluon one-particle distribution function, with the growth rates of the unstable modes depending on the temperature, degree of momentum anisotropy, and the angle at which the mode is propagating relative to the anisotropy axes.  The existence of the chromo-Weibel instability makes it difficult to understand the late time dynamics of momentum-anisotropic plasmas.  For example, in the case of the computation of the imaginary part of the heavy-quark potential, one finds that the chromo-Weibel instability results in a pinch singularity that causes the imaginary part of the integral defining the heavy-quark potential to diverge~\cite{Burnier:2009yu,Nopoush:2017zbu}.  In Ref.~\cite{Burnier:2009yu} the analysis was performed assuming small anisotropy, while in Ref.~\cite{Nopoush:2017zbu} this finding was extended to all anisotropy strengths.  The pinch singularities found were a direct result of the appearance of unstable modes in the spectrum and, in the static limit, there being modes with negative squared masses.  A similar problem will occur in the perturbative calculation of many key quantities used in QGP phenomenology, e.g., jet energy loss and momentum broadening, if the perturbative calculations are carried to sufficiently high order in the gauge coupling.  

In the case of the heavy-quark potential the issue is manifest already at leading order, so this makes it an ideal candidate for understanding the role plasma instabilities play.  As discussed in Refs.~\cite{Dumitru:2007hy,Burnier:2009yu,Dumitru:2009fy,Nopoush:2017zbu} the real and imaginary parts of the heavy-quark potential can be obtained from three-momentum integrations involving the retarded and Feynman propagators, respectively.  As a result, one needs to know the form of the resummed propagators and hence self-energies for arbitrary angle of propagation with respect to the anisotropy axes.
In this paper, we study the gluonic collective modes of a collisional plasma subject to a momentum space anisotropy of spheroidal form, i.e.,
\be\label{aniso}
f({\bf p}) = f_{\rm {iso}}\left(\frac{1}{\lambda}\sqrt{p_T^2 + (1+\xi) p_L^2}\right)=f_{\rm {iso}}\left(\frac{1}{\lambda}\sqrt{{\bf p}^2+\xi ({\bf p}\cdot{\bf n})^2}\right)\, ,
\ee
where $T$ and $L$ indicate momentum perpendicular and parallel to a given anisotropy axis denoted by a unit vector ${\bf n}$, respectively. The degree of momentum-space anisotropy is quantified by an adjustable parameter $\xi$ in the range $-1< \xi < \infty$. In addition, $\lambda$ is a scale which specifies the typical momentum of the particles in the plasma and should be understood as the temperature only in the thermal equilibrium limit~\cite{Romatschke:2003ms,Romatschke:2004jh}. For a spheroidal deformation, there is only one anisotropy axis and the gluon polarization tensor depends on the angle of propagation of the modes relative to ${\bf n}$.  In the collisionless limit, the collective modes of a QGP with a spheroidal momentum anisotropy were determined completely in Refs.~\cite{Romatschke:2003ms,Romatschke:2004jh}.  In this limit, one finds two unstable modes whose growth depends on the strength of the anisotropy, $\xi$, and the angle of propagation $\theta$ of the mode relative to ${\bf n}$.

In order to study the collective modes of the QGP beyond the collisionless limit, in Ref.~\cite{Carrington:2003je} the authors computed the collective modes of an isotropic thermal QGP by making use of a collisional kernel of Bhatnagar–Gross–Krook (BGK) form.  They found that a non-zero collision rate $\nu \neq 0$ pushed the cut associated with Landau damping into the lower half plane, induced damping of the stable collective modes, and caused one of the two stable collective modes to terminate at finite momentum.\footnote{This is due to the mode leaving the first Riemann sheet and going onto the second Riemann sheet through the Landau damping cut.\label{fn1}}  The work of Carrington et al in Ref.~\cite{Carrington:2003je} was extended to the case of a momentum-anisotropic QGP in Ref.~\cite{Schenke:2006xu}, wherein the authors considered a spheroidally deformed QGP, but only considered the special case that the momentum of the gluon was along the momentum anisotropy direction ${\bf n}$.  In this special case, it was possible to obtain analytic results for all four gluon polarization functions.  The authors of \cite{Schenke:2006xu} found that it was possible to eliminate the unstable modes from the spectrum if the collision rate was sufficiently large.

In this paper we extend the analysis of Ref.~\cite{Schenke:2006xu} to include all possible gluon propagation angles and general anisotropy strength $\xi$.\footnote{For recent work which considered the small anisotropy limit using the BGK collisional kernel, see Ref.~\cite{Kumar:2017bja}.  Note that, at leading order in the small anisotropy limit, only three ($\alpha$, $\beta$, and $\gamma$) of the five structure functions considered herein are required.}  For general anisotropy strength and propagation angle it is not possible to obtain purely analytic results, however, the problem can be reduced to numerically evaluating a set of one-dimensional integrals.  To prove this, we decompose the polarization tensor into a set of five independent structure functions. We then determine the full spectrum of stable and unstable modes as a function of both the collision rate $\nu$ and angle of progagation parameterized by $t = \cos \theta$ with $0 \leq t \leq 1$.  Our findings confirm those found in the case $t=1$ \cite{Schenke:2006xu} and we find that for all $\nu > 0 $ the unstable growth rate is maximal at $t=1$ ($\theta = 0$ with respect to the anisotropy axis ${\bf n}$).  Our results set the stage for the computation of the real and imaginary parts of the heavy-quark potential in a collisional and momentum-space anisotropic QGP.  Beyond this, our results can be used in the computation of a large set of inputs necessary for understanding the role of momentum-space anisotropies in the QGP at intermediate couplings, e.g.,  jet energy loss, shear viscosity at higher orders in $\alpha_s$, etc.

The structure of our paper is as follows.  In Sec.~\ref{se} we review how one obtains the gluon polarization tensor from kinetic theory.  In Sec.~\ref{sf} we detail how to obtain the five structure functions necessary to describe a collisional and momentum-space anisotropic plasma.  In this section we also discuss the static limit of the structure functions and their relation to the five non-trivial mass scales appearing in the problem.  In Sec.~\ref{stable} we present our results for the dispersion relations of the (damped) stable modes in the system for various gluon propagation angles and collisional rates. In Sec.~\ref{unstable} we present our results for the dispersion relations of the unstable modes.  Finally, in Sec.~\ref{con} we present our conclusions and an outlook for the future. In App.~\ref{anare}, we present some limiting cases in which it is possible to obtain analytic expressions for the five structure functions and static masses.  Finally, in App.~\ref{nor} we present results obtained when holding the energy density of the system fixed when varying the plasma anisotropy.

\section{The gluon self-energy from kinetic theory}\label{se}

In the high temperature phase of QCD, the gauge coupling constant $g$ becomes small enough and, therefore, there exists a hierarchy of scales and one can construct effective theories at various scales by eliminating the degrees of freedom at higher scales. For example, the collective excitations, which typically develop at a momentum scale $\sim gT$, are well separated from the typical energy of hard partons and can be described by kinetic equations of the Boltzmann-Vlasov type. In the kinetic theory, the distribution of hard particles in the QCD plasma is described by the gauge covariant Wigner functions $n^i({\bf p}, X)$ where the superscript $i$ denotes the species of the plasma constituents, i.e., quarks, antiquarks and gluons, which are treated as massless
particles and assumed to satisfy the mass-shell constraint. Expanding the Wigner function around the color neutral background fields $n^i({\bf p})$, namely, $n^i({\bf p}, X)=n^i({\bf p})+\delta n^i({\bf p},X)$, the linearized kinetic equations read
\be\label{ke}
V\cdot \partial_X \delta f_a^i({\bf p},X)+g \theta_i V_\mu F_a^{\mu\nu}\partial_\nu^{(P)}f^i({\bf p})={\cal C}_a^i({\bf p},X)\, ,
\ee
where $V=(1,{\bf v})$ with ${\bf v}\equiv {\bf p}/p$ and $\theta_q=\theta_g=1$, $\theta_{{\bar q}}=-1$. In the above equation, we have already taken into account the weak-coupling limit and neglected terms of subleading order in $g$. Thus, the theory becomes effectively Abelian. There is no coupling among different color channels in Eq.~(\ref{ke}) and the field strength tensor is simply given by $F^{\mu\nu}=\partial^\mu A^{\nu}-\partial^\nu A^{\mu}$ since the term involving a gauge field commutator can be dropped consistently. In addition, the distribution function $f^i({\bf p})$ and its color fluctuation $\delta f_a^i({\bf p},X)$ are related to the Wigner function via
\be
f^{q/{\bar q}}({\bf p})=\frac{1}{N_c} {\rm Tr}[n^{q/{\bar q}}({\bf p},X)]\, ,\quad\quad f^g({\bf p})=\frac{1}{N_c^2-1} {\rm Tr}[n^g ({\bf p},X)]\,,
\ee
and
\be
\delta f_a^{q/{\bar q}}({\bf p},X)=2 {\rm Tr}[t_a \delta n^{q/{\bar q}}({\bf p},X)]\, ,\quad\quad \delta f_a^g({\bf p},X)=\frac{1}{N_c} {\rm Tr}[T_a \delta n^g ({\bf p},X)]\,,
\ee
where $t_a$ and $T_a$ are the SU($N_c$) group generators in the fundamental and adjoint representations, respectively. On the right hand side of Eq.~(\ref{ke}), we introduce a BGK-type collision term, which is given by
\be\label{ck}
{\cal C}^i_a({\bf p},X)=-\nu \Big[f_a^i({\bf p},X)-\frac{N_a^i(X)}{N^i_{{\rm eq}}}f^i_{{\rm eq}}(p)\Big]\, ,
\ee
with $f_a^i({\bf p},X)=f^i({\bf p})+\delta f_a^i({\bf p},X)$. The above collision kernel ensures an instantaneously conserved number of particles $N_a^i(X)=\int_{{\bf p}} f_a^i({\bf p},X)$. Accordingly, the particle number in equilibrium is given by $N_{{\rm eq}}^i=\int_{{\bf p}} f_{{\rm eq}}^i(p)$. In addition, the collision rate denoted by $\nu$ is inversely proportional to the equilibration time of the hot QCD plasma under collisions between the hard partons. In the remainder of this work, the collision rate $\nu$ is taken to be a free parameter which has no dependence on the momentum and particle species. Determining the magnitude of $\nu$ in the ultrarelativistic heavy-ion experiments is still an open question. Estimates from Ref.~\cite{Schenke:2006xu} place $\nu$ in the range $\nu \sim 0.1-0.2 \, m_D$, however, it is possible that the collision rate is even larger than this.  Herein, we present results for small $\nu$ and postpone the consideration of large $\nu$ to a forthcoming paper.

It is more convenient to solve the linearized kinetic equations for the fluctuation $\delta f_a^{i}({\bf p},X)$ in momentum space which leads to the result for the linearized induced current by each particle species. The total induced current $J_{{\rm ind}\,a}^\mu(K)$ with $K=(\omega, {\bf k})$ is simply a sum of the contributions from quarks, antiquarks, and gluons. Furthermore, the gluon self-energy can be obtained by functional differentiation of $J_{{\rm ind}\,a}^\mu(K)$ with respect to the gauge field $A^b_\nu(K)$ and has already been computed in Ref.~\cite{Schenke:2006xu}. Including the collision kernel as given in Eq.~(\ref{ck}), the gluon self-energy $\Pi^{\mu\nu}_{ab}(K)$ takes the following non-trivial form
\ba\label{defpi}
\Pi^{\mu\nu}_{ab}(K)&=&\frac{\delta J_{{\rm ind}\,a}^\mu(K)}{\delta A^b_\nu(K)}=g^2 \delta_{ab} \int_{{\bf p}} V^\mu \partial^{({\bf p})}_{l} f({\bf{p}})\frac{g^{l\nu}(\oh-\hat{{\bf{k}}}\cdot{\bf{v}})-{\hat K}^l V^\nu}{\oh-\hat{{\bf{k}}}\cdot{\bf{v}}+i \uh}\, \nonumber \\
&+& g^2 \delta_{ab}(i \uh) \int\frac{\mathrm{d} \Omega}{4\pi} \frac{V^\mu}{\oh-\hat{{\bf{k}}}\cdot{\bf{v}}+i \uh}\int_{{\bf p^\prime}} \partial^{({\bf p }^\prime)}_{l} f({\bf{p}}^\prime)\frac{g^{l\nu}(\oh-{\hat {\bf{k}}}\cdot{\bf{v}}^\prime)-{\hat K}^l {V^\prime}^\nu}{\oh-{\hat {\bf{k}}}\cdot{\bf{v}^\prime}+i \uh} \mathcal{W}^{-1}(\hat \omega,\hat\nu )\, .
\ea
In the above equation, $\int_{{\bf p}}\equiv \int\mathrm{d}^3{\bf p}/(2\pi)^3$ and ${\bf{v}^\prime}={{\bf p}}^\prime/p^\prime$. For convenience, we introduce the dimensionless collision rate $\uh=\nu/k$ and define ${\hat K}=(\oh, {\hat {\bf k}})$ with $\oh=\omega/k$ and ${\hat {\bf k}}={\bf k}/k$. In addition, the function $\mathcal{W}(\hat \omega,\hat\nu )$ reads
\be
\mathcal{W}(\hat \omega,\hat\nu )=1-\frac{i \uh}{2}\int_{-1}^{1}\mathrm{d} x\frac{1}{\oh-x+ i\uh}=1+\frac{i\uh}{2} \ln\frac{\oh-1+i \uh}{\oh+1+i \uh}\, .
\ee
Inclusion of the collision kernel doesn't break the transversality of the gluons self-energy and one can show that $K_\mu \Pi^{\mu\nu}=K_\nu \Pi^{\mu\nu}=0$. As the Lorentz components of the gluons self-energy are not all independent, it is sufficient to consider only the spatial components and the temporal components can be simply obtained by using the relations $\Pi^{0i}(K)=k^l\Pi^{li}(K)/\omega$ and $\Pi^{00}(K)=k^ik^j\Pi^{ij}(K)/\omega^2$. Since Eq.~(\ref{defpi}) is diagonal in color, color indices will be suppressed in the following.

In the above equation, $f({\bf p})=2N_c f^g({\bf p})+N_f [f^q({\bf p})+f^{\bar q}({\bf p})]$ and the distribution functions of the hard partons are completely arbitrary up to the requirement that the distribution function used results in convergent integrals for, e.g., the number density, energy density, gluon self-energy, etc.  This sets constraints on the behavior of $f({\bf p})$ at small and large momenta. From here on, an anisotropic parton distribution function as specified in Eq.~(\ref{aniso}) will be adopted. It can be obtained by stretching or squeezing an isotropic distribution $f_{\rm iso}$ along the direction of anisotropy {\bf n} which is conventionally taken to be parallel to the beam-line direction.  Due to early-time free-streaming of the QGP, it is expected that $\xi >0$, corresponding to an oblate momentum-space distribution. As a result, an oblate distribution with $\xi>0$ is the case most relevant to high-energy heavy-ion experiments and hence will be considered in the following. Plugging Eq.~(\ref{aniso}) into Eq.~(\ref{defpi}) and performing the radial part of the integrations, the gluon self-energy using the above distribution reads 
\ba\label{eq:se}
\Pi^{ij}(K)&=&m_D^2 \int\frac{\mathrm{d}\Omega}{4\pi}v^i \frac{v^l+\xi({\bf {v}}\cdot{\bf{n}})n^l}{(1+\xi({\bf {v}}\cdot{\bf{n}})^2)^2}\frac{\delta^{l j}(\oh-{\hat {\bf k}}\cdot{\bf v})+ {\hat k}^l v^j}{\oh-{\hat {\bf k}}\cdot{\bf v}+i \uh}+(i\uh)m_D^2\int\frac{\mathrm{d}\Omega^\prime}{4\pi}\frac{{v^\prime}^i}{\oh-{\hat {\bf k}}\cdot{\bf v}^\prime+i \uh}\,\nonumber \\
&& \hspace{2cm} \times \int\frac{\mathrm{d}\Omega}{4\pi}\frac{v^l+\xi({\bf {v}}\cdot{\bf{n}})n^l}{(1+\xi({\bf {v}}\cdot{\bf{n}})^2)^2}\frac{\delta^{lj}(\oh-{\hat {\bf k}}\cdot{\bf v})+ {\hat k}^l v^j}{\oh-{\hat {\bf k}}\cdot{\bf v}+i \uh}{\cal W}^{-1}(\oh,\uh)\,,
\ea
where
\be\label{md}
m_D^2=-\frac{g^2}{2\pi^2}\int_0^\infty \mathrm{d} p\, p^2\frac{\mathrm{d} f_{\rm iso}(p^2)}{\mathrm{d} p}\,.
\ee

It is important to point out that the above gluon self-energy is not symmetric in Lorentz indices due to the appearance of the BGK-type collision term,\footnote{However, the BGK-type collision term doesn't break this symmetry if the system is isotropic.} namely $\Pi^{ij}(K)\neq \Pi^{ji}(K)$. The same is also found in a magnetized plasma due to the lack of time-reversal symmetry \cite{Wang:2021ebh}. Thus, a symmetric tensor basis as developed for an anisotropic system in the collisionless limit doesn't apply to the decomposition of Eq.~(\ref{eq:se}). Here, we adopt a more general tensor basis and decompose the gluon self-energy into five structure functions as
\be\label{sede}
\Pi^{ij}(K)=\alpha A^{ij}+\beta B^{ij}+\gamma C^{ij}+\delta D^{ij}+\rho E^{ij}\,,
\ee
where the tensor basis reads
\be \label{basis}
A^{ij}= \delta^{ij}-\kh^i \kh^j\, ,\quad\quad\quad
B^{ij} = \kh^i \kh^j\, ,\quad\quad\quad
C^{ij} = {\tilde n}^i {\tilde n}^j/{\tilde n}^2\, ,\quad\quad\quad
D^{ij} = {\tilde n}^i \kh^j\, ,\quad\quad\quad E^{ij} = \kh^i {\tilde n}^j\,.
\ee
In the above equation, ${\tilde n}^i \equiv A^{il}{n}^{l}$, which is orthogonal to $\kh^i$. As compared to the tensor basis used in Refs.~\cite{Romatschke:2003ms,Schenke:2006xu}, the only difference is that the structure functions for ${\tilde n}^i \kh^j$ and $\kh^i {\tilde n}^j$ are not identical any more. 

The determination of the five structure functions can be carried out based on the following contractions
\ba\label{contr}
&&\kh^{i} \Pi^{ij} \kh^{j} = \beta \, ,\quad\quad  {\tilde n}^i\Pi^{ij}  {\tilde n}^j = 
 {\tilde n}^2 (\alpha+\gamma)\, ,\quad\quad  {\rm Tr}\,{\Pi^{ij}} = 2\alpha +\beta +\gamma \, , \nonumber \\
&& {\tilde n}^i \Pi^{ij} \kh^{j}  =  {\tilde n}^2 \delta \, ,\quad\quad \kh^{i} \Pi^{ij}  {\tilde n}^j  =  {\tilde n}^2 \rho\, .
\ea
Clearly, if the gluon self-energy is symmetric, the last two contractions lead to the same result and $\delta=\rho$. The explicit results for the five structure functions in the above decomposition will be presented in the next section.



Given the gluon self-energy, one can further compute the resummed gluon propagator $\Delta^{ij}(K)$ by using the Dyson-Schwinger equation. In temporal axial gauge, the inverse propagator is written as
\be
\left(\Delta^{-1}\right)^{ij}(K)=(k^2-\omega^2+\alpha) A^{ij} + (\beta-\omega^2) B^{ij} +\gamma C^{ij}+\delta D^{ij} +\rho E^{ij}\,.
\ee
Upon inversion, we find that
\begin{equation}
\Delta^{ij}(K) =\Delta_{A} \left[A^{ij} - C^{ij}\right] +
\Delta_{G}\left[(k^2-\omega^2+\alpha+\gamma) B^{ij} +
(\beta-\omega^2)C^{ij} - \delta D^{ij}- \rho E^{ij}\right]~.
\end{equation}

The dispersion relations are obtained by finding the poles of the propagator $\Delta^{ij}(K)$, namely,
\ba\label{defcm}
\Delta^{-1}_{A} &=& \omega^2-k^2-\alpha=0\, ,\\ \nonumber
\Delta^{-1}_{G} &=& (k^2-\omega^2+\alpha+\gamma)(\beta-\omega^2)-{\tilde n}^2\delta \rho= (\omega^2-\Omega_+^2)(\omega^2-\Omega_-^2)=0\, ,
\ea
where
\be\label{defopm}
2\Omega_{\pm}^2={\bar {\Omega}}^2\pm\sqrt{{\bar {\Omega}}^4-4[(\alpha+\gamma+k^2)\beta-{\tilde n}^2\delta \rho]}\,,
\ee
with
\be
{\bar {\Omega}}^2=\alpha+\beta+\gamma+k^2\,.
\ee
Hence, we find three separate equations
\be\label{defmode}
\omega_\alpha^2 = k^2+\alpha(\omega_\alpha)\,,\quad\quad\omega_{+}^2=\Omega_{+}^2(\omega_{+})\,,
\quad\quad
\omega_{-}^2=\Omega_{-}^2(\omega_{-})\,,
\ee
which determine the so-called $\alpha$-mode, $G_+$-mode and $G_-$-mode, respectively.

\section{structure functions and mass scales}\label{sf}

To evaluate the structure functions, we need a master integral of the form
\ba\label{me}
{\cal I}^{[n]}(\xi,\oh,t,\uh)&=&\int_{-1}^{1}\mathrm{d}x  \frac{x^n}{(1+\xi x^2)^2} \int_0^{2\pi}\frac{\mathrm{d} \phi}{2\pi}\frac{1}{\oh-{\hat {\bf k}}\cdot {\bf v}+i \uh}\,\nonumber \\
&=& \int_{-1}^{1}\mathrm{d}x\frac{x^n}{(1+\xi x^2)^2}\frac{{\rm sign}({\rm Re} [\oh]- {\hat k}_z x)}{\sqrt{(\oh-{\hat k}_z x+ i \uh)^2-{\hat k}_y^2(1-x^2)}}\,\nonumber \\
&=&\int_{-1}^{1}\mathrm{d}x\frac{x^n}{(1+\xi x^2)^2}\frac{{\rm sign}({\rm Re} [\oh]- t x)}{\sqrt{(\oh-tx+i\hat\nu)^2-(1-t^2)(1-x^2)}}\, .
\ea
In the above equation, $n$ is an integer and $n=0,1,2,3$ are relevant in the following evaluations. The propagation angle $\theta$ parameterized by $t=\cos \theta$ corresponds to the angle between ${\bf k}$ and ${\bf n}$. In addition, we assume that ${\bf n}$ is parallel to the $z$-axis while ${\bf k}$ lies in the $y$-$z$ plane. More generally, $k_y$ should be replaced by $k_\perp$. The integration variable $x$ is cosine of the polar angle between ${\bf p}$ and ${\bf n}$. In general, $\omega$ is complex-valued.

Then we can express the structure functions as the following \footnote{An overall factor $m_D^2$ is omitted in these expressions. The same is done for the mass scales.}
\ba
\label{al}
\alpha(\xi,\oh,t,\uh)&=&\frac{1}{2( {1 - t^{2}} )}\Big[\hat{\omega}z f_0(\xi)- \xi t^2 f_1(\xi) + \hat{\omega}( 1 - t^{2} - z^{2} )\,{\cal I}^{[0]}(\xi,\oh,t,\uh) \nonumber \\&+& t\big( \xi( 1 - t^{2} ) + 2\hat{\omega}z - \xi z^2 \big)\,{\cal I}^{[1]}(\xi,\oh,t,\uh)
- ( \hat{\omega} - 2\xi z t^{2})\,{\cal I}^{[2]}(\xi,\oh,t,\uh) - \xi t\,{\cal I}^{[3]}(\xi,\oh,t,\uh)\Big]\, ,\\
\label{be}
\beta(\xi,\oh, t,\hat\nu)&=&-\frac{\hat\omega^2}{2\mathcal{W}(\hat \omega,\hat\nu )}\Big[ f_0(\xi)-z\, {\cal I}^{[0]}(\xi,\oh, t,\hat\nu)-\xi t \, {\cal I}^{[1]}(\xi,\oh, t,\hat\nu)\Big]\,,\\
\label{ga}
\gamma(\xi,\oh,t,\uh)&=&\frac{1}{2( 1 - t^{2} )}\Big[\xi( 1 + t^{2} )f_1(\xi)- \hat{\omega}z( 1 + t^{2} )f_0(\xi) - \hat{\omega}\big( ( 1 - t^{2} ) - z^{2}( 1 + t^{2}) \big)\,{\cal I}^{[0]}(\xi,\oh,t,\uh)\nonumber \\
& -& t\big( 4\hat{\omega}z -2 \xi z^2 + \xi( 1 - t^{2} )(1+\hat{\omega}z)\big)\,{\cal I}^{[1]}(\xi,\oh,t,\uh) \nonumber \\
& -& \big( 4 \xi z-2 \hat{\omega}-\xi (1-t^2)(\hat{\omega}+3z) \big)\,{\cal I}^{[2]}(\xi,\oh,t,\uh)  + 2 \xi t\, {\cal I}^{[3]}(\xi,\oh,t,\uh)\Big]\, ,\\
\label{de}
\delta(\xi,\oh,t,\uh)&=&\frac{\hat\omega}{2(1-t^2)}\Big[ t z f_0(\xi)-t z^2\,{\cal I}^{[0]}(\xi,\oh,t,\uh)+z(1-\xi t^2)\,{\cal I}^{[1]}(\xi,\oh,t,\uh)+\xi t\,{\cal I}^{[2]}(\xi,\oh,t,\uh)\Big]\,, \\
\label{rh}
\rho(\xi,\oh,t,\uh)&=&\frac{\hat\omega}{2\mathcal{W}(\hat \omega,\hat\nu )(1-t^2)}\Big[\oh t  f_0(\xi)-\oh t z\,{\cal I}^{[0]}(\xi,\oh,t,\uh)+\big(\oh(1+\xi-\xi t^2)-\xi z\big)\,{\cal I}^{[1]}(\xi,\oh,t,\uh) \nonumber \\
&+&\xi t\,{\cal I}^{[2]}(\xi,\oh,t,\uh)\Big]\, ,
\ea
where 
\ba\label{f0}
f_0(\xi)&=&\int_0^1 \frac{2}{(1+\xi x^2)^2} \mathrm{d} x =\frac{1}{1+\xi}+\frac{\arctan\sqrt\xi}{\sqrt\xi}\, ,\\
f_1(\xi)&=&\int_0^1 \frac{2 x^2}{(1+\xi x^2)^2} \mathrm{d} x=-\frac{1}{\xi+\xi^2}+\frac{\arctan\sqrt\xi}{\xi^\frac32}\, ,
\ea
and $z=\oh+i \uh$.

It is interesting to consider the static limit of the structure functions which defines the corresponding mass scales as the following,
\be
m_\alpha^2=\lim_{\omega\rightarrow 0} \alpha\, ,\quad\quad m_\beta^2=-\lim_{\omega\rightarrow 0} \oh^{-2}\beta\, ,\quad\quad m_\gamma^2=\lim_{\omega\rightarrow 0} \gamma\, ,\quad\quad m_\delta^2=\lim_{\omega\rightarrow 0} {\tilde n}\oh^{-1}{\rm Im}[\delta]\, ,\quad\quad m_\rho^2=\lim_{\omega\rightarrow 0} {\tilde n}\oh^{-1}{\rm Im}[\rho]\, .
\ee
Taking the limit $\oh \rightarrow 0$, the master integral in Eq.~(\ref{me}) reduces to
\ba
{\tilde{\cal I}}^ {[n]}(\xi,t,\uh)&=& \int_{-1}^1  \mathrm{d} x \frac{x^n}{(1+\xi x^2)^2} \frac{{\rm sign}(- {\hat k}_z x)}{\sqrt{(-{\hat k}_z x+ i \uh)^2-{\hat k}_y^2(1-x^2)}}\,\nonumber \\
&=& \int_{-1}^1  \mathrm{d} x \frac{x^n}{(1+\xi x^2)^2} \frac{{\rm sign}(- t x)}{\sqrt{(-t x+ i {\uh})^2-(1-t^2)(1-x^2)}}\, .
\ea

It is important to notice that ${\rm sign}(- {\hat k}_z x)/\sqrt{(-{\hat k}_z x+ i {\hat\nu})^2-{\hat k}_y^2(1-x^2)}$ is complex-valued. Its imaginary part is an even function of $x$ while the real part is odd in $x$. Therefore, the above integral can be written as
\be\label{mi2}
{\tilde{\cal I}}^ {[n]}(\xi,t,\uh)=- \int_{0}^1  \mathrm{d} x \frac{2}{(1+\xi x^2)^2} f^{[n]}(x,t, {\hat \nu})\, ,
\ee
where
\be
f^{[n]}(x,t, {\hat \nu})=
\left\{
\begin{aligned}
& x^n{\rm Re}\Big[\frac{1}{\sqrt{(x -i \uh t )^2-(1+{\hat \nu}^2)(1-t^2)}}\Big]=x^n\frac{\sqrt{s(x,t, {\hat \nu})+x^2-{\hat \nu}^2+ t^2-1}}{\sqrt {2} s(x,t, {\hat \nu})}\,,\quad\quad\,\,\,\,\,\, \text {$n$ is odd} \\
& i  x^n {\rm Im}\Big[\frac{1}{\sqrt{(x -i \uh t )^2-(1+\uh^2)(1-t^2)}}\Big]=i x^{n+1} \frac{\sqrt {2} \uh t/s(x,t,\uh)}{\sqrt{s(x,t,\uh)+ x^2-\uh^2+ t^2-1 }}\,,\quad \text {$n$ is even}\\
\end{aligned}
\right.
\, ,
\ee
and $s(x,t, {\hat \nu})\equiv \sqrt{(x^2-{\hat \nu}^2+t^2-1)^2+4 x^2 {\hat \nu}^2 t^2}$. When $n$ is even, Eq.~(\ref{mi2}) is pure imaginary, while it is real-valued when $n$ is odd.
In the above equation, the sign function 
${\rm sign}({\hat k}_z)$ or ${\rm sign}(t)$ has been dropped. In fact, one can only consider the region $0\le t \le 1$ because as we will see below, the mass scales $m_\alpha^2$, $m_\beta^2$ and $m_\gamma^2$ are symmetric under $t\rightarrow -t$. Although $m_\delta^2$ and $m_\rho^2$ are odd functions of $t$, it is their product $m_\delta^2 m_\rho^2$ that is relevant in our consideration, which is also even in $t$.

In terms of the integral defined in Eq.~(\ref{mi2}), one can obtain the five mass scales.  We find 
\ba
\label{ma}
m_\alpha^2(\xi,t,\hat\nu)&=&-\frac{\xi t}{2(1-t^2)}\big [tf_1(\xi)-(1-t^2+\hat\nu^2)\,{\tilde{\cal I}}^ {[1]}(\xi,t,\uh)-2i \uh t\,{\tilde{\cal I}}^ {[2]}(\xi,t,\uh)+{\tilde{\cal I}}^ {[3]}(\xi,t,\uh)\big]\, ,\\
\label{mb}
m_\beta^2(\xi,t,\uh)&=&\frac{1}{2\mathcal{W}(0,\hat\nu )}\big[ f_0(\xi)-i \uh\,{\tilde{\cal I}}^ {[0]}(\xi,t,\uh)-\xi t\,{\tilde{\cal I}}^ {[1]}(\xi,t,\uh)\big]\, ,\\
\label{mg}
m_\gamma^2(\xi,t,\hat\nu)&=&\frac{\xi}{2(1-t^2)}\big [(1+t^2)f_1(\xi)-(t-t^3+2\hat\nu^2 t)\,{\tilde{\cal I}}^ {[1]}(\xi,t,\uh)-i\hat\nu(1+3t^2)\,{\tilde{\cal I}}^ {[2]}(\xi,t,\uh)\, \nonumber \\
&+&2 t \,{\tilde{\cal I}}^ {[3]}(\xi,t,\uh)\big]\, ,\\
\label{mde}
m_\delta^2(\xi,t,\hat\nu)&=&\frac{1}{2\sqrt{1-t^2}}\big[\hat\nu t f_0(\xi)-i\hat\nu^2t\,{\tilde{\cal I}}^ {[0]}(\xi,t,\uh)+\hat\nu (1-\xi t^2)\,{\tilde{\cal I}}^ {[1]}(\xi,t,\uh)-i\xi t \,{\tilde{\cal I}}^ {[2]}(\xi,t,\uh)\big]\, ,\\
\label{mr}
m_\rho^2(\xi,t,\hat\nu)&=&\frac{-1}{2\mathcal{W}(0,\hat\nu )\sqrt{1-t^2}}\big[\xi\hat\nu\,{\tilde{\cal I}}^ {[1]}(\xi,t,\uh)+i\xi t \,{\tilde{\cal I}}^ {[2]}(\xi,t,\uh)\big]\, .
\ea
The factor $\mathcal{W}(0,\hat\nu )$ in $m_\beta^2(\xi,t,\hat\nu)$ is real-valued and can be expressed as $1- {\hat \nu} \arctan {(1/\hat \nu})$. Therefore, the above five mass scales are all real-valued. For later use, we also define the following two mass scales
\be
2 m_{\pm}^2=M^2\pm\sqrt{M^4-4[m_\beta^2(m_\alpha^2+m_\gamma^2)-m_\delta^2 m_\rho^2]}\, ,
\ee
with
\be
M^2=m_\alpha^2+m_\beta^2+m_\gamma^2\,.
\ee

We point out that when $\hat\omega$ is imaginary-valued, i.e., $\hat\omega=i \hat \Gamma \equiv i \Gamma/k$, the structure functions can be expressed in terms of the integrals defined in Eq.~(\ref{mi2}) which are more efficient than using Eq.~(\ref{me}) when performing the numerical evaluation. It is straightforward to show that ${\cal I}^{[n]}(\xi,\hat \Gamma,t,\uh)={\tilde{\cal I}}^ {[n]}(\xi,t,\hat \Gamma+\uh)$, therefore, the five structure functions are all real-valued for imaginary-valued $\oh$.
In general, the above structure functions and the corresponding mass scales must be evaluated numerically. However, it is possible to obtain analytical results in some limiting cases, which we discuss in App.~\ref{anare}.

\section{dispersion relations for the stable modes}\label{stable}

\begin{figure}[htbp]
\begin{center}
\includegraphics[width=0.32\textwidth]{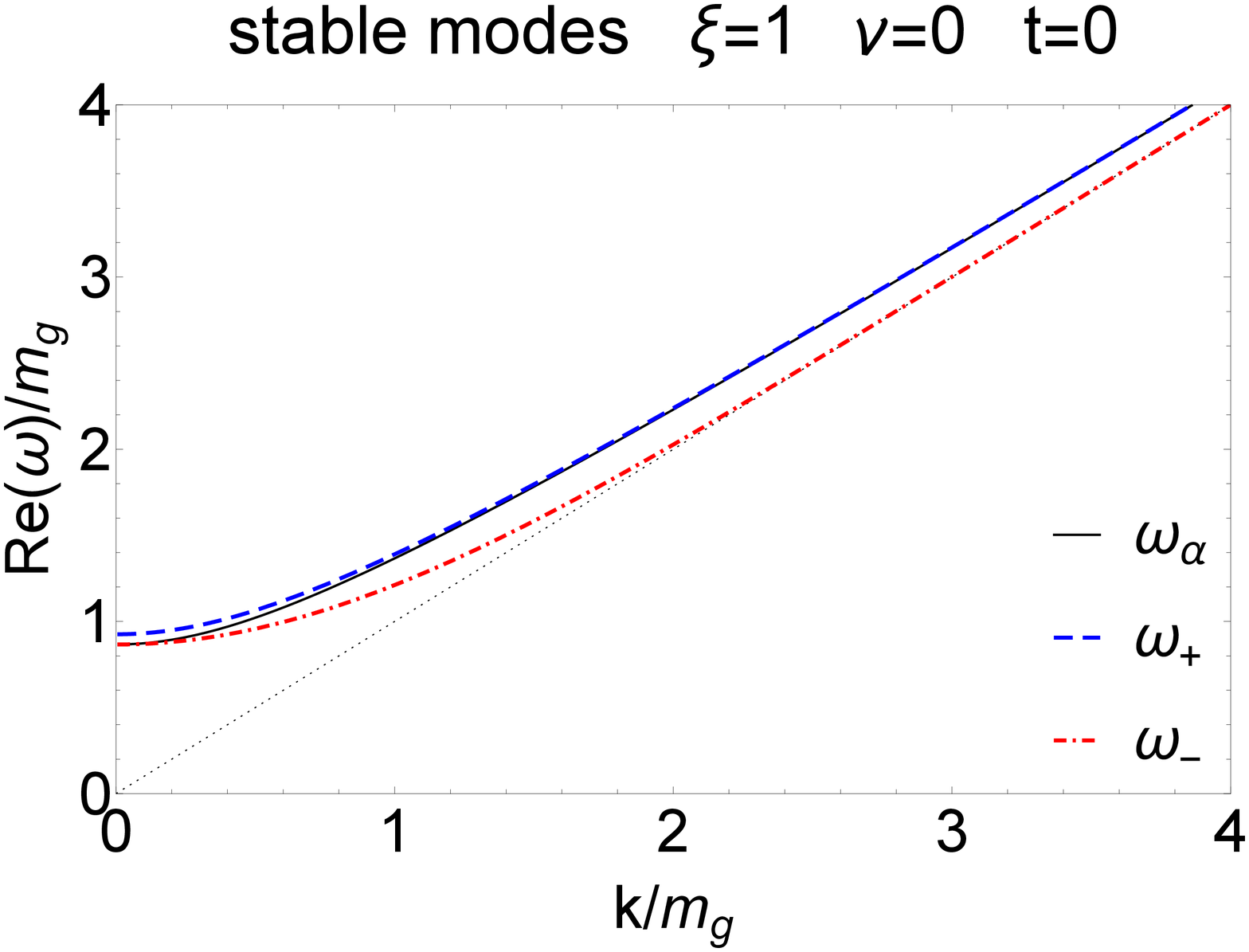}
\includegraphics[width=0.32\textwidth]{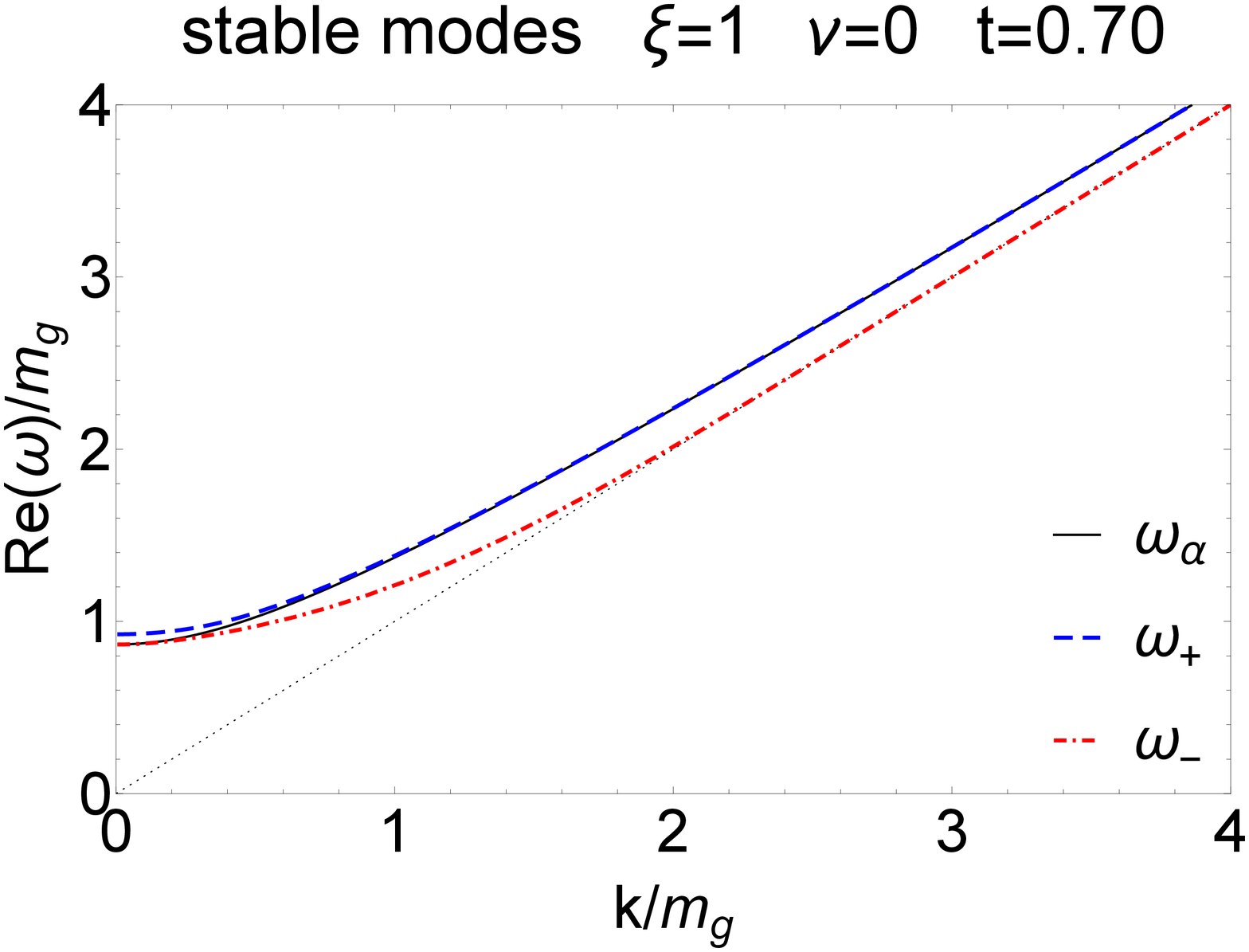}
\includegraphics[width=0.32\textwidth]{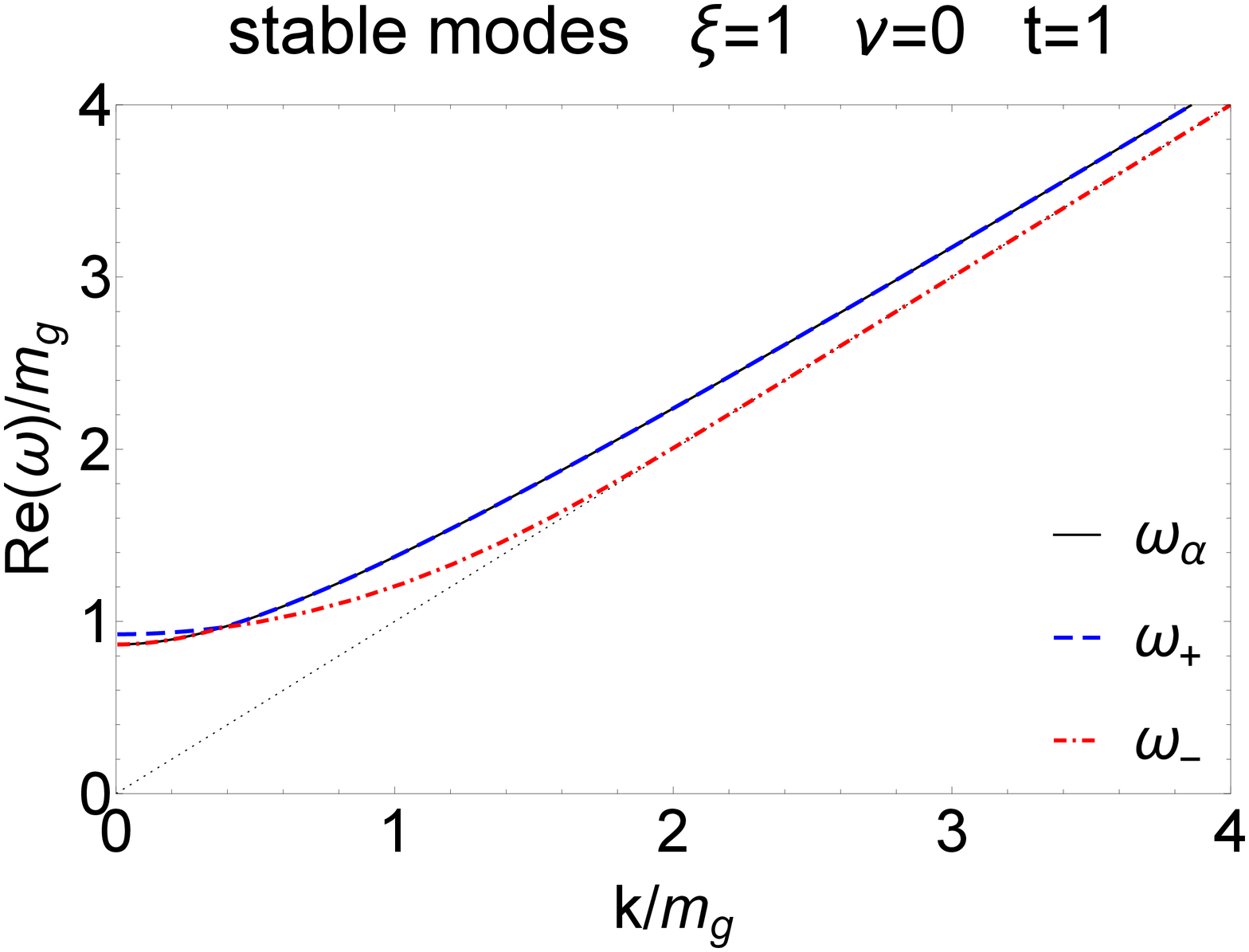}
\caption{The dispersion relation of the stable modes in the collisionless limit for different propagation angles. }
\label{sxi1v0}
\end{center}
\end{figure}

\begin{figure}[htbp]
\begin{center}
\includegraphics[width=0.32\textwidth]{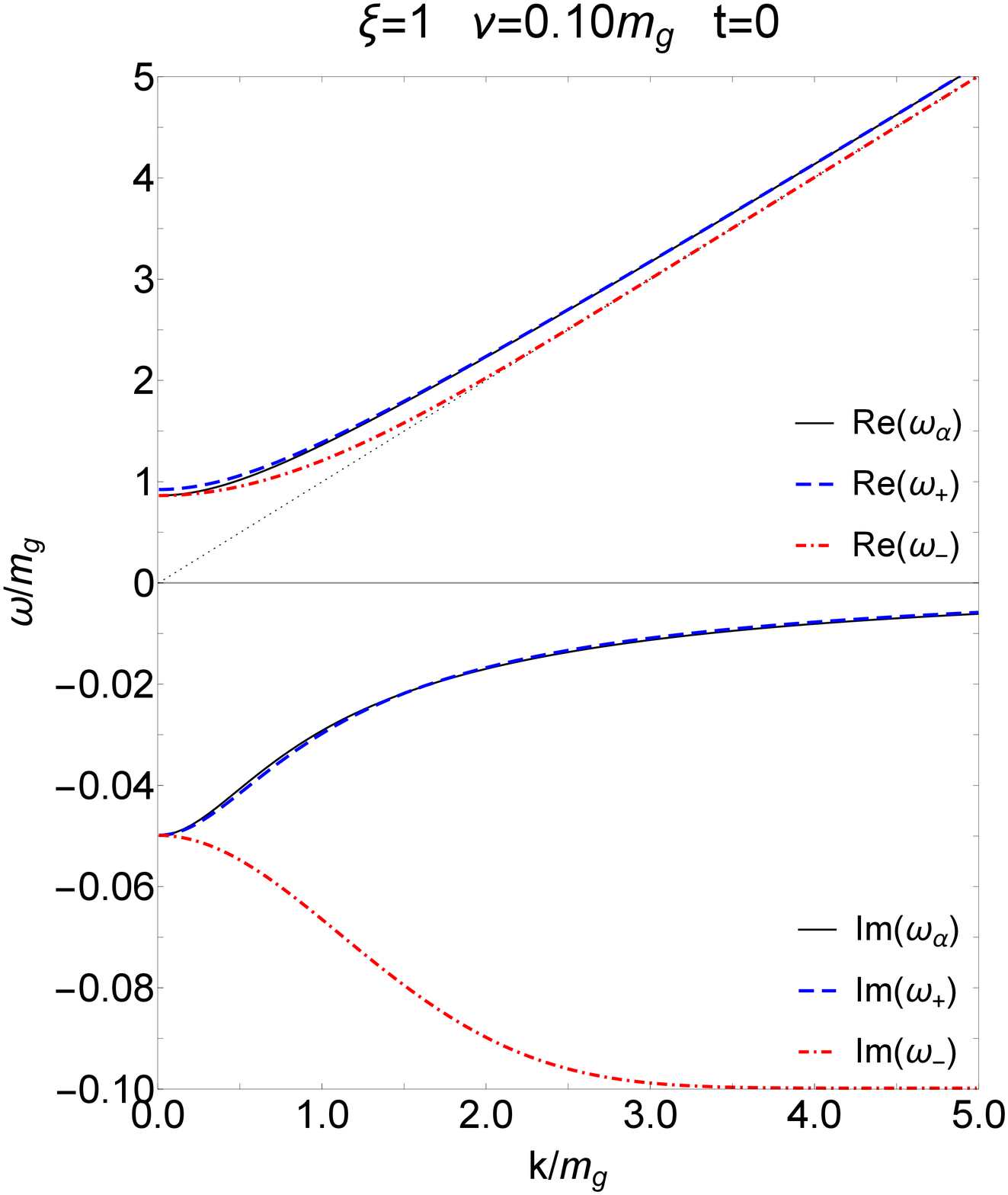}
\includegraphics[width=0.32\textwidth]{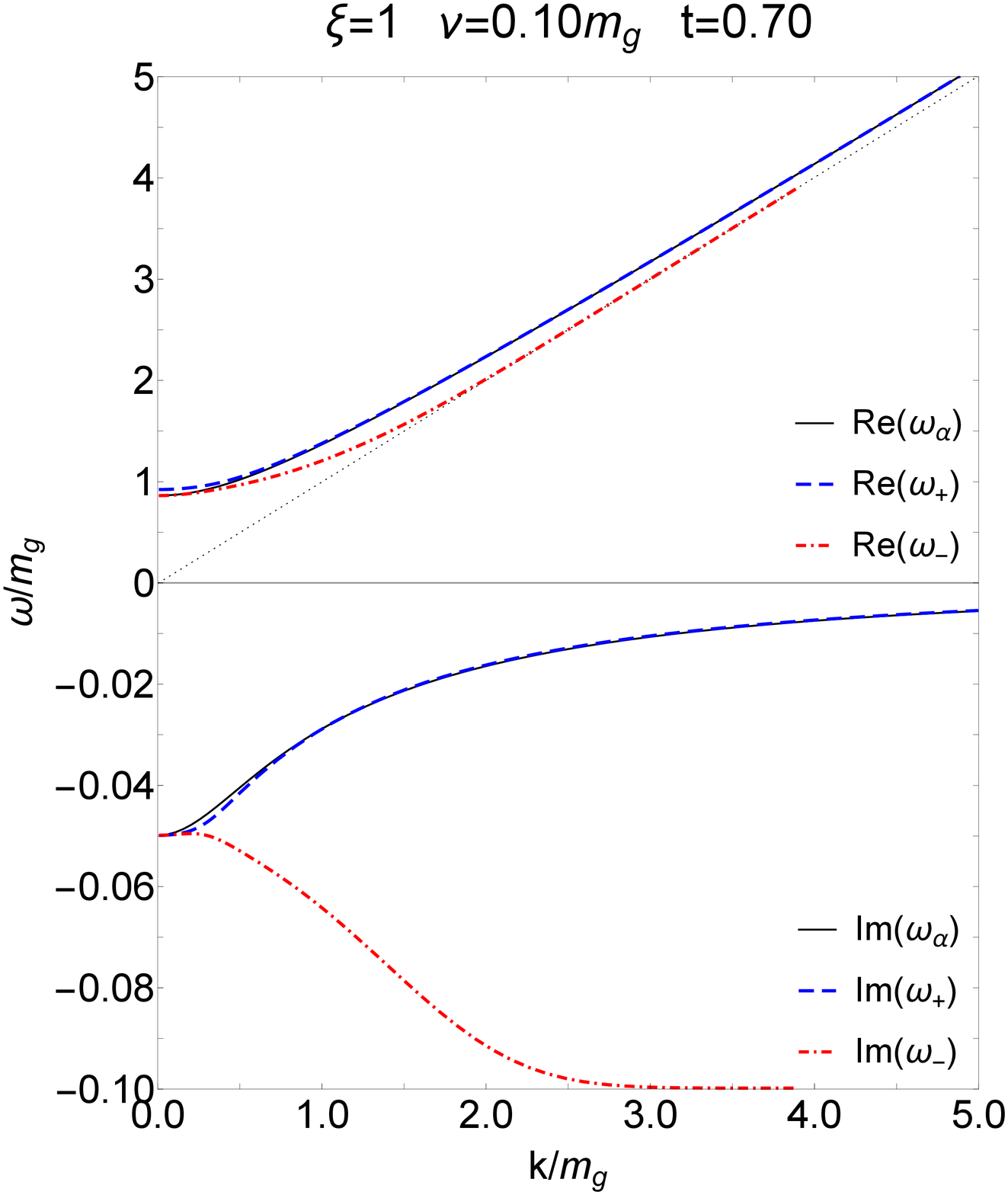}
\includegraphics[width=0.32\textwidth]{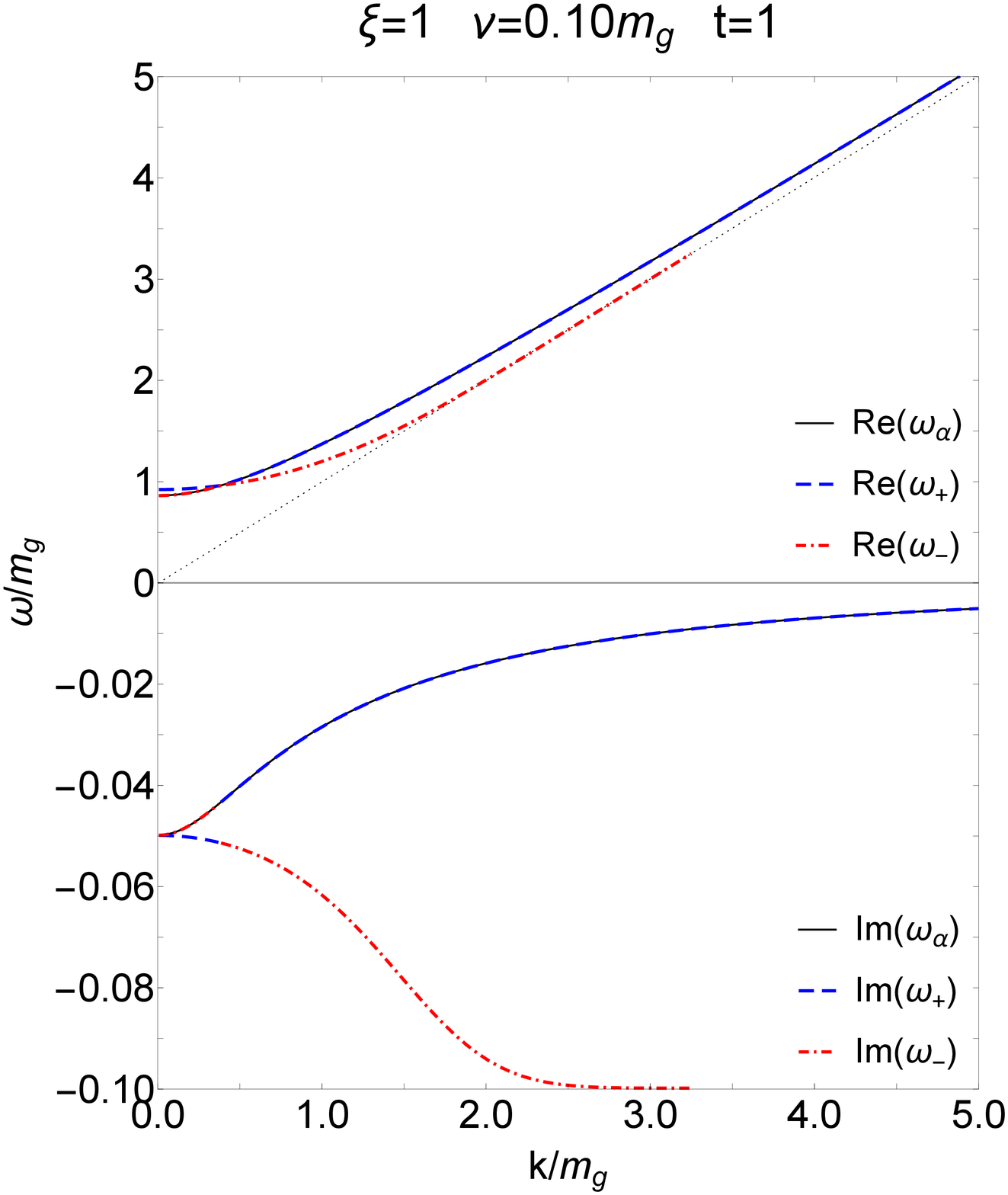}
\caption{Real and imaginary part of the dispersion relation of the stable modes with $\nu=0.1 m_g$ for different propagation angles.}
\label{sxi1v01}
\end{center}
\end{figure}

\begin{figure}[htbp]
\begin{center}
\includegraphics[width=0.32\textwidth]{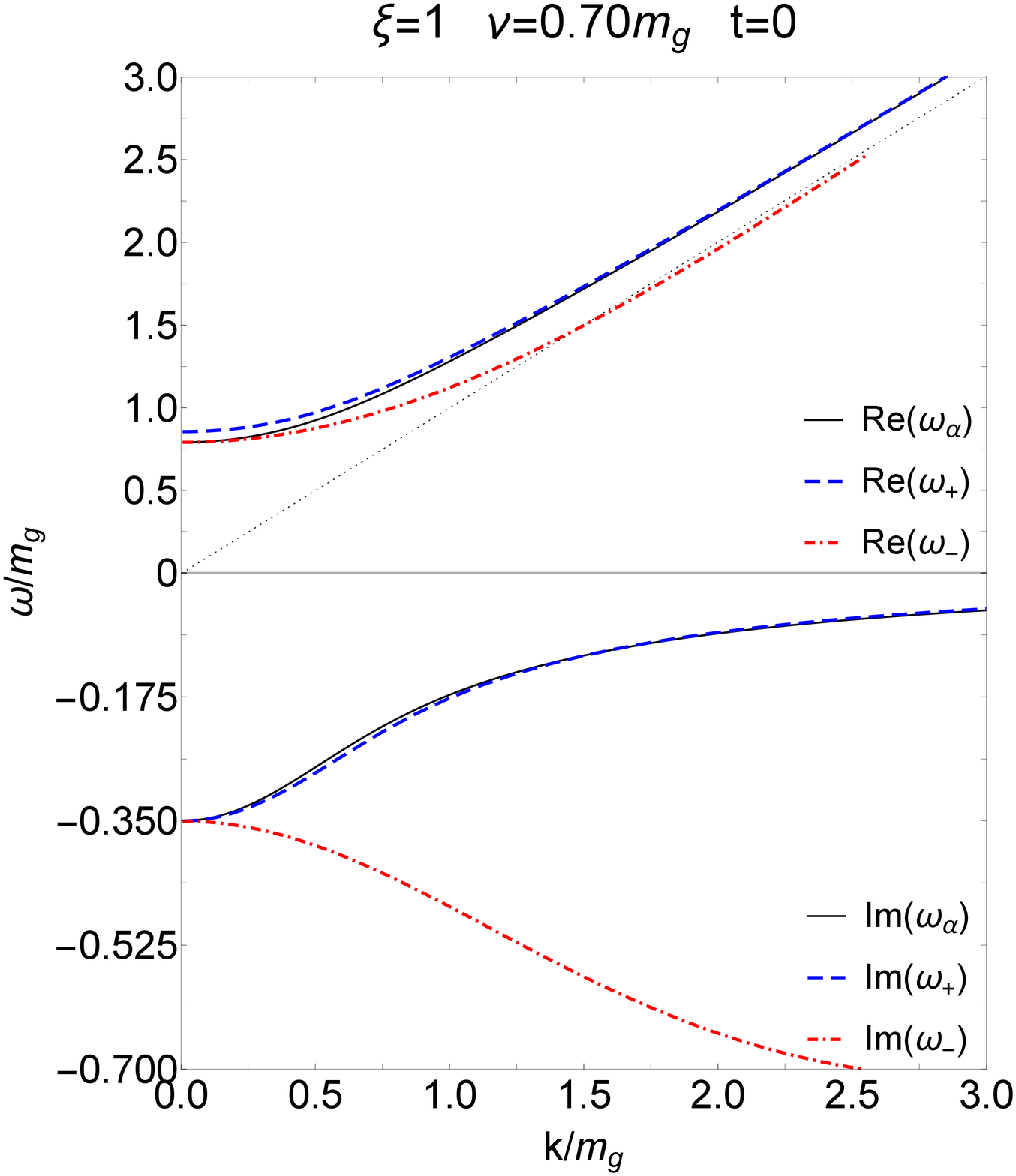}
\includegraphics[width=0.32\textwidth]{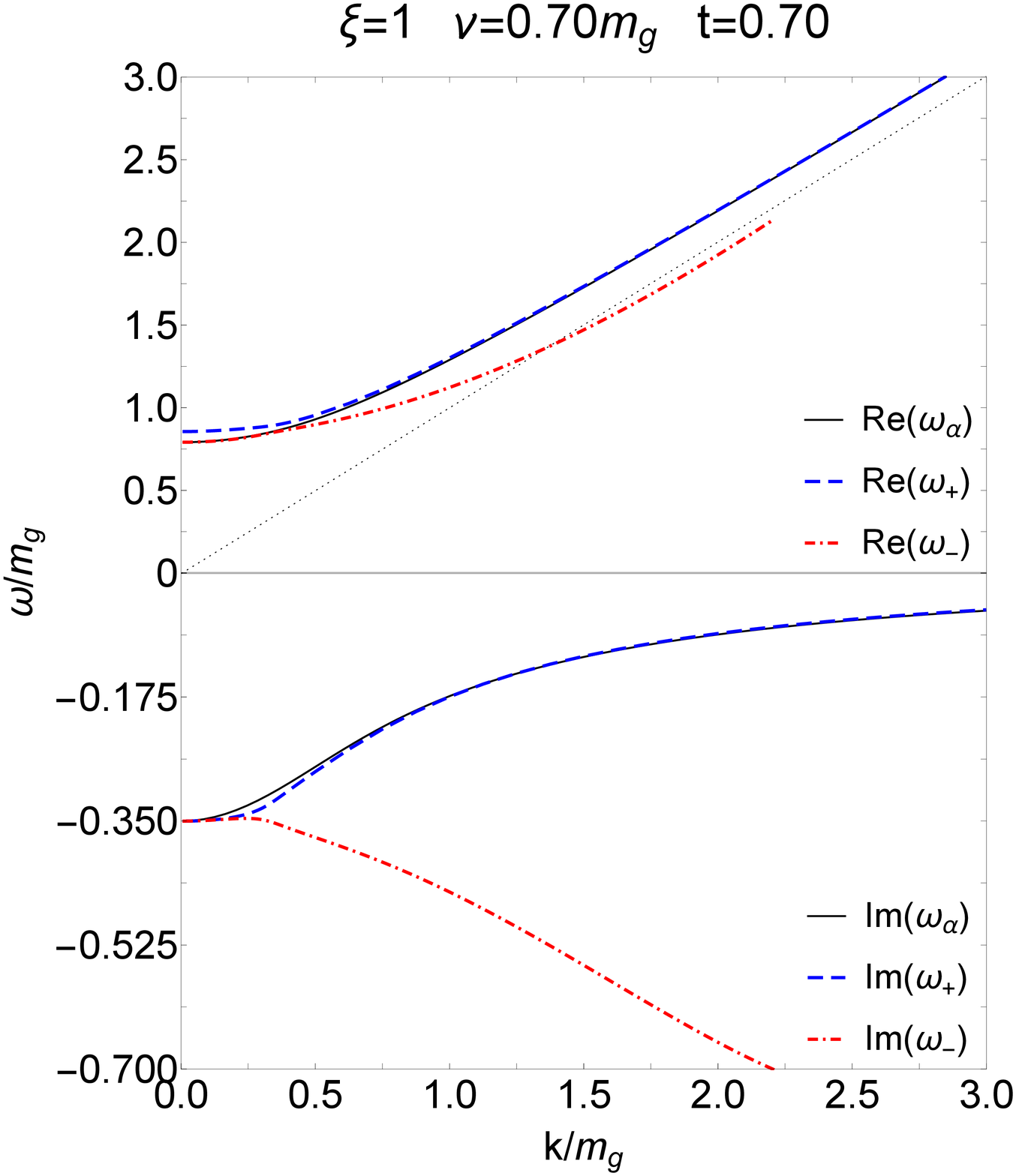}
\includegraphics[width=0.32\textwidth]{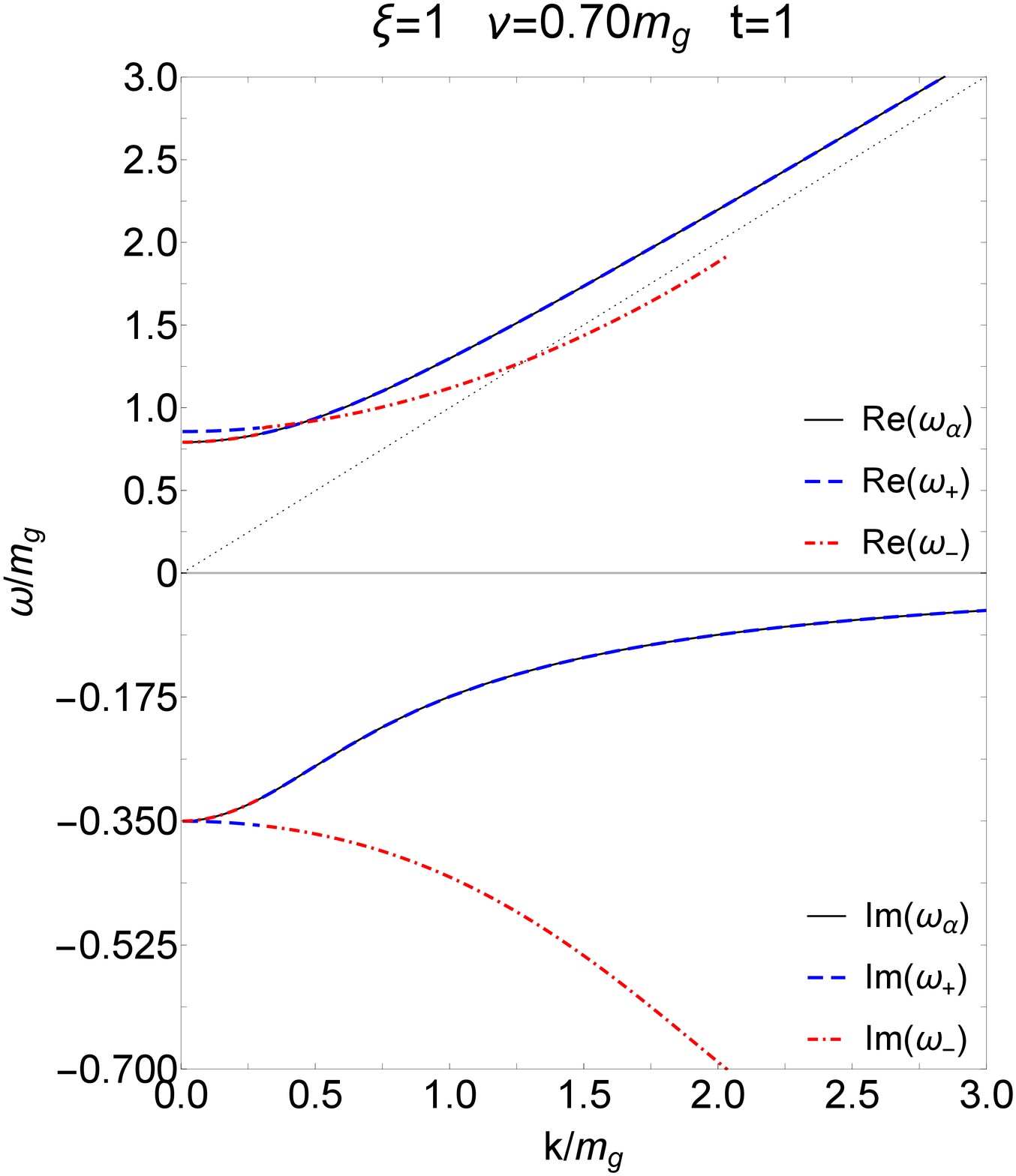}
\caption{Real and imaginary part of the dispersion relation of the stable modes with $\nu=0.7 m_g$ for different propagation angles.}
\label{sxi1v07}
\end{center}
\end{figure}

The solutions $\omega(k)$ of Eq.~(\ref{defcm}) can be obtained numerically, which determine the dispersion relations for the collective modes. The corresponding results for the stable modes are shown in Figs.~\ref{sxi1v0}-\ref{sxi1v07} where all quantities are given in units of $m_g \equiv m_D/\sqrt{3}$. As found in previous studies, the three stable modes in the collisionless limit have poles at real-valued $\omega>k$. However, with non-zero collision rate, these types of modes correspond to the complex-valued solutions $\omega(k)$ which are damped in time due to a negative imaginary part.\footnote{Following the terminology in the previous works, we still call them stable modes.} According to our results, the $\alpha$-mode behaves very similarly to the $G_+$-mode for large wave number $k$ while for small $k$, it is more like the $G_-$-mode. In particular, the $\alpha$-mode is identical to either the $G_-$-mode or the $G_+$-mode at $t=1$, thus, only two distinguishable modes exist, which were called $\alpha$-mode and $\beta$-mode in Ref.~\cite{Schenke:2006xu}.

It is interesting to consider the limit $k\rightarrow 0$ where the values of $\omega$ in the dispersion relation can be obtained analytically. We find that all the structure functions are finite in this limit and $\alpha(k\rightarrow 0)$ becomes angle-independent. Furthermore, although the other four structure functions still have a $t$-dependence, $\Omega_+^2$ and $\Omega_-^2$ as defined in Eq.~(\ref{defopm}) don't depend on $t$. Explicitly, we have
\ba
\alpha(k\rightarrow 0)&=&\Omega^2_{-}(k\rightarrow 0)=\frac{m_D^2}{4}\frac{\omega}{\xi (\omega+i \nu)}\bigg[1+(\xi-1)\frac{\arctan\sqrt{\xi}}{\sqrt{\xi}}\bigg]\, ,\nonumber \\
\Omega^2_{+}(k\rightarrow 0)&=&-\frac{m_D^2}{2}\frac{\omega}{\xi (\omega+i \nu)}\bigg[1-(\xi+1)\frac{\arctan\sqrt{\xi}}{\sqrt{\xi}}\bigg]\, .
\ea
As a result, the dispersion relations at vanishing wave number are determined by only the anisotropy parameter $\xi$ and the collision rate $\nu$. The corresponding values of $\omega_\alpha$, $\omega_{+}$, and $\omega_{-}$ in this limit are given by
\ba\label{wk0}
\omega_\alpha(k\rightarrow 0)&=&\omega_{-}(k\rightarrow 0)=\frac{m_D}{2}\sqrt{\frac{1+(\xi-1)\arctan\sqrt{\xi}/\sqrt{\xi}}{\xi}-\Big(\frac{\nu}{m_D}\Big)^2}-i \frac{\nu}{2}\, ,\nonumber \\
\omega_{+}(k\rightarrow 0)&=&\frac{m_D}{2}\sqrt{\frac{-2+2(\xi+1)\arctan\sqrt{\xi}/\sqrt{\xi}}{\xi}-\Big(\frac{\nu}{m_D}\Big)^2}-i \frac{\nu}{2}\, ,
\ea
where the imaginary part of $\omega$ equals to $-\nu/2$ independent of the anisotropy parameter $\xi$.  We note that the above complex solutions for the dispersion relations in Eq.~(\ref{defmode}) exist when the collision rate is not very large so that the square roots in the above equation are real. 
For the isotropic case, it requires $\nu/m_D<2/\sqrt{3}$. 
With increasing $\xi$, the upper bound of $\nu/m_D$ will be reduced. 

It is also worth noting that the $G_-$-mode exhibits some new features in the collisional cases. It could become space-like (${\rm Re}(\omega)<k$) at finite collision rate where a strong damping rate is also found. In addition, the $G_-$-mode terminates at certain wave number $k$ where the magnitude of its imaginary part approaches the collision rate. As mentioned in footnote~\ref{fn1}, the reason for the termination of the stable $G_-$ mode is that the stable mode solution moves from the physical Reimann sheet to a higher Reimann sheet when it passes through the cut that exists between $-1 < {\rm Re}(\omega)/k < 1$ and ${\rm Im}(\omega) = -\nu$.  This is most easy to see in the case $t=1$ where analytic forms for all structure functions can be obtained (see App.~\ref{anare} and Eqs.~(39) and (40) of Ref.~\cite{Schenke:2006xu}).  In Ref.~\cite{Schenke:2006xu} the authors understood the termination of this mode by tracking the movement of the solution from the physical Reimann sheet to higher Reimann sheets associated with the logarithm appearing in the structure functions (see Eq.~(41) of Ref.~\cite{Schenke:2006xu}).  We refer the reader to Figs.~5 and 6 of Ref.~\cite{Schenke:2006xu} and the associated discussion of this figure.  When $t \neq 1$, the cut remains in the same location, however, one must carefully analyze the discontinuities in the integrand of the master integral ${\cal I}^{[n]}$ defined in Eq.~(\ref{me}) to reach the same conclusion.  We note that the termination of this mode also occurs in the isotropic case with a finite collision rate~\cite{Carrington:2003je}.

\begin{figure}[htbp]
\begin{center}
\includegraphics[width=0.45\textwidth]{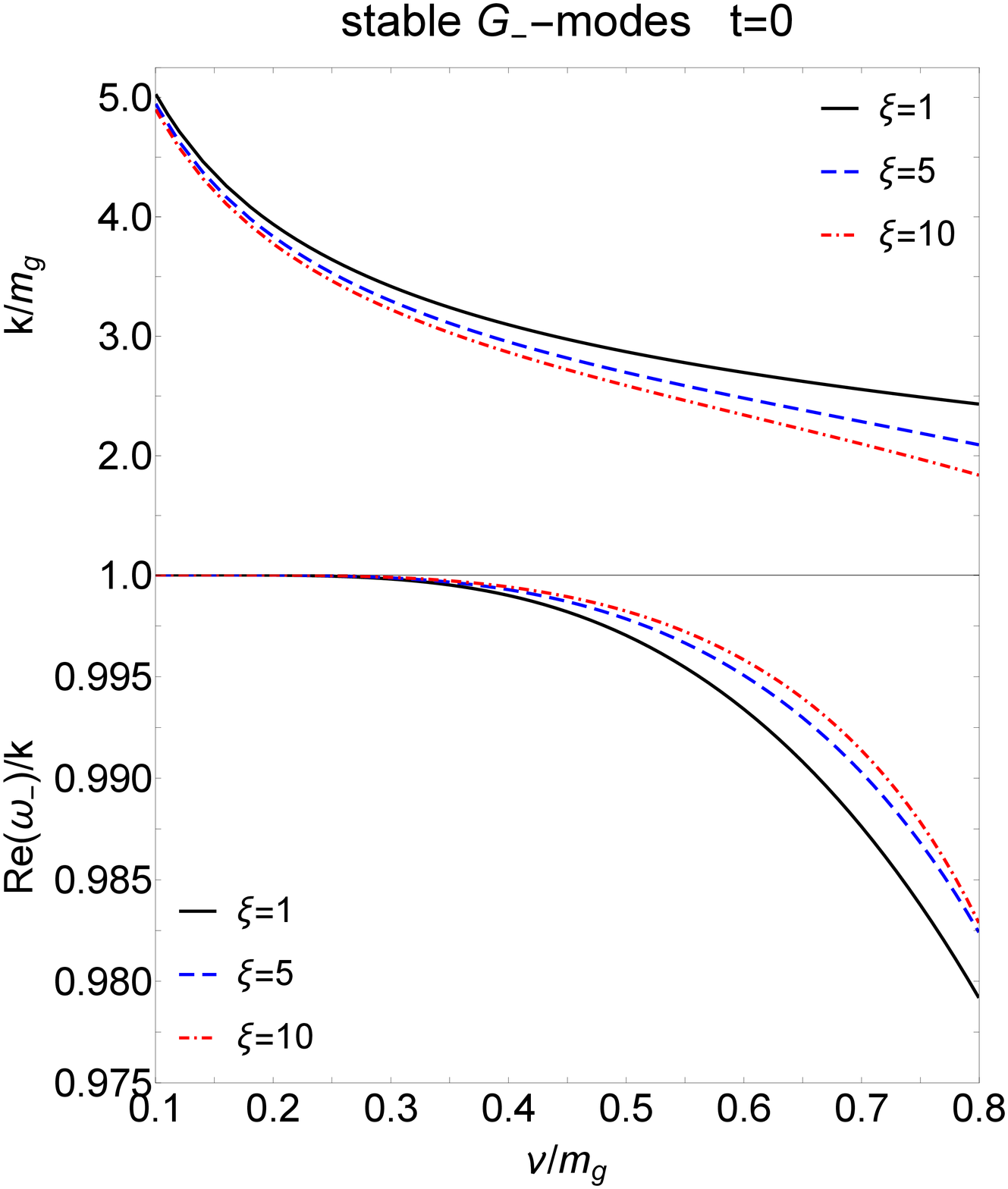} \hspace{2mm}
\includegraphics[width=0.45\textwidth]{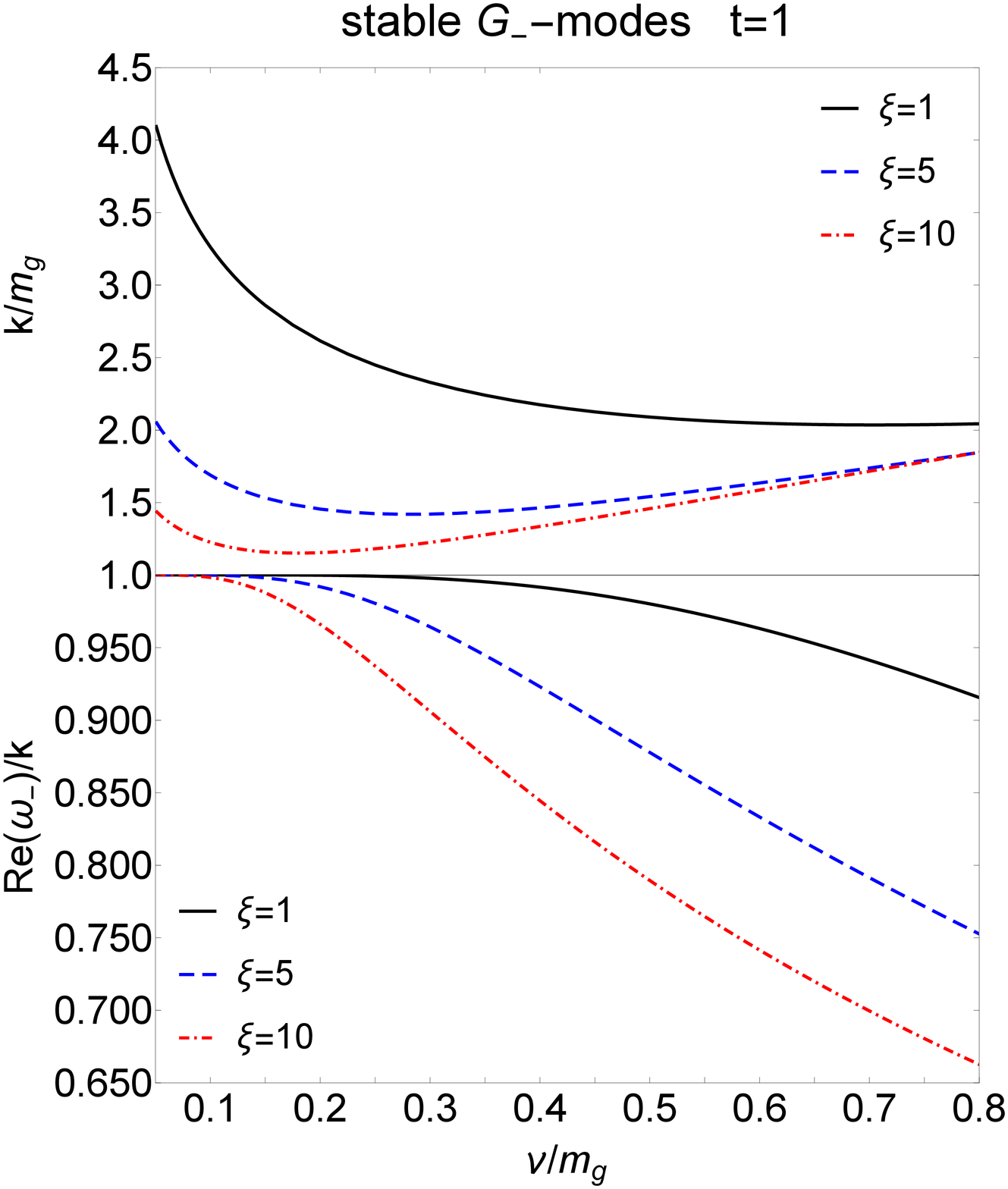}
\caption{The $\nu$-dependence of $k^\star/m_g$ and ${\rm Re}(\omega_-^\star)/k^\star$ for the stable $G_-$-modes at the termination point for $t=0$ (left) and $t=1$ (right). We present the results for small, intermediate and large anisotropies.}
\label{endpoint}
\end{center}
\end{figure}

According to Figs.~\ref{sxi1v01} and \ref{sxi1v07}, at a given anisotropy $\xi$, ${\rm Im}(\omega_-)$ drops to $-\nu$ more quickly for larger $t$. In general, determining the value of $k$ at which the $G_-$-mode terminates involves a complicated numerical evaluation. In the special case $t=0$ and $t=1$, we show in Fig.~\ref{endpoint} how the values of $k^\star/m_g$ as well as the ratio ${\rm Re}(\omega_-^\star)/k^\star$ at the termination point change with the collision rate $\nu$.  Notice that for clarity, the values of $k$ and ${\rm Re}(\omega_-)$ at the termination points are denoted by $k^\star$ and ${\rm Re}(\omega_-^\star)$, respectively. Our results show that ${\rm Re}(\omega_-^\star)/k^\star$ deviates from the light cone (which corresponds to the line ${\rm Re}(\omega_-^\star)/k^\star=1$ in this figure) more significantly when ${\bf k}$ is parallel to ${\bf n}$. For a fixed anisotropy, the deviation becomes larger with increasing $\nu$. However, as $\xi$ increases, our numerical results indicate that the deviation can become either larger or smaller depending on the propagation angle. 
Finally, we note that as a function of $\nu$, $k^\star/m_g$ can have a non-monotonic behavior depending on the degree of momentum-space anisotropy.

\begin{figure}[htbp]
\begin{center}
\includegraphics[width=0.32\textwidth]{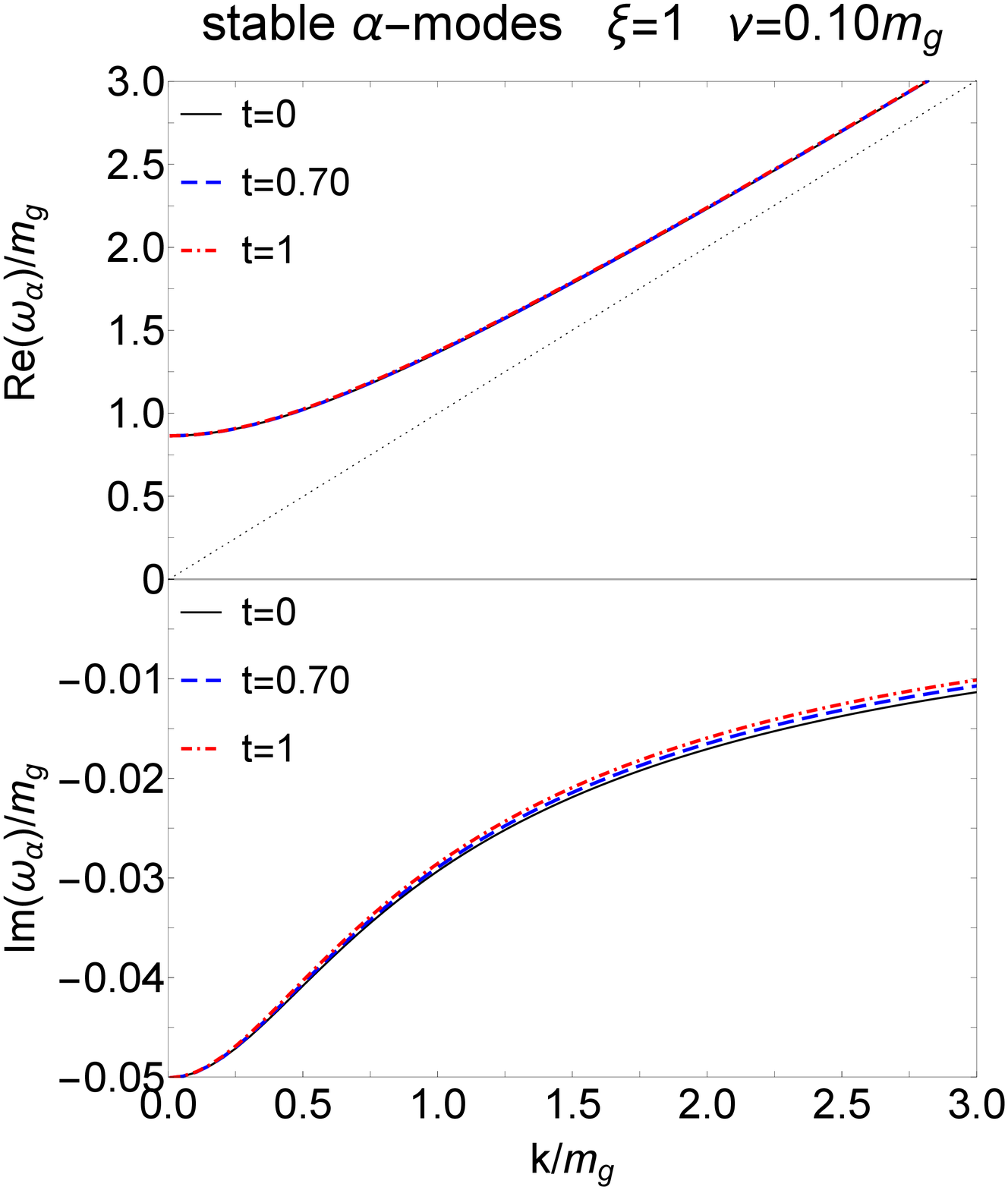}
\includegraphics[width=0.32\textwidth]{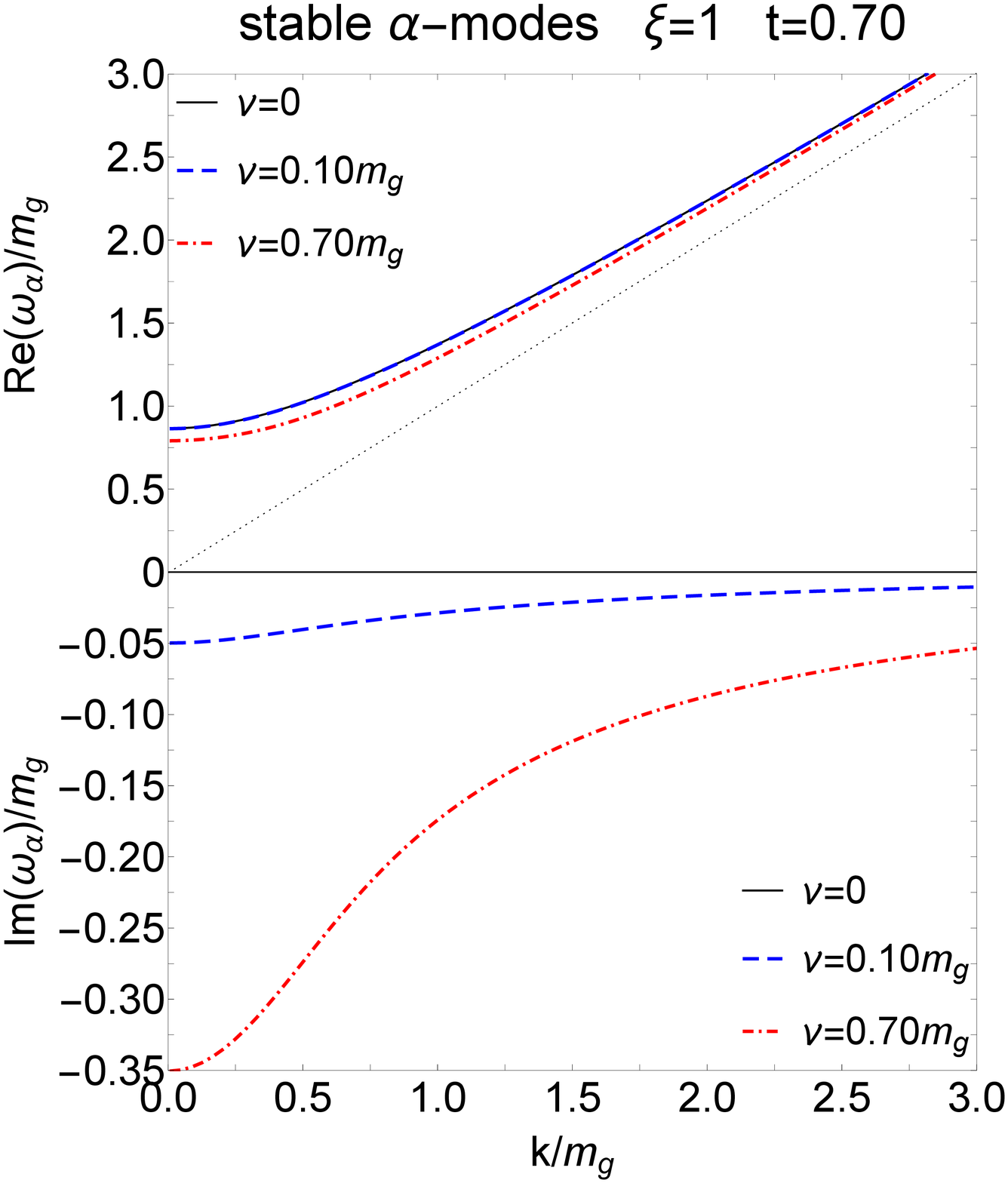}
\includegraphics[width=0.32\textwidth]{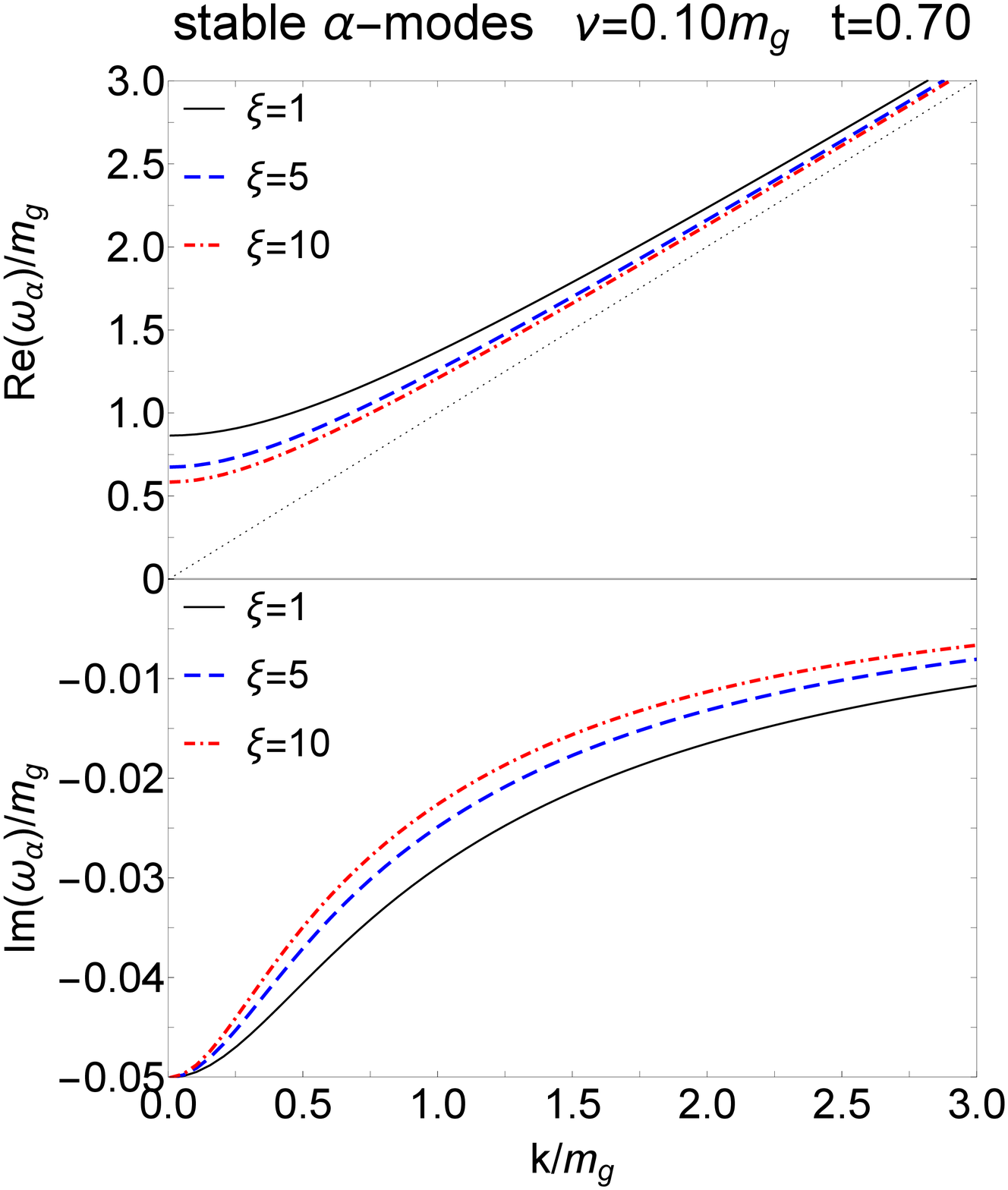}
\caption{Parameter dependence of the dispersion relations of the stable $\alpha$-modes. Left: the $t$-dependence at fixed $\xi$ and $\nu$. Middle: the $\nu$-dependence at fixed $\xi$ and $t$ where the imaginary part of $\omega$ vanishes for $\nu=0$. Right: the $\xi$-dependence at fixed $\nu$ and $t$. In these plots, $m_g=m_D/\sqrt{3}$.  }
\label{sare}
\end{center}
\end{figure}

In order to see the parameter dependence of the dispersion relations, we consider the $\alpha$-mode as an example and show the corresponding $t$-, $\nu$- and $\xi$-dependence in Fig.~\ref{sare}. As one can see from this figure, the real part of the dispersion relations is almost independent on the propagation angle, but becomes sensitive to the degree of anisotropy. On the other hand, the collision rate has a visible influence only when $\nu$ is very large. 
The dependence of the imaginary part in the dispersion relations on these parameters is more significant. Notice that similar behaviors are observed in the stable $G_+$- and $G_-$-modes. Although we only present results at a set of specific values of $t$, $\nu$, and $\xi$, we internally considered many cases which are not listed here but support the same conclusions.

\section{Dispersion relations for the unstable modes}
\label{unstable}

\begin{figure}[htbp]
\begin{center}
\includegraphics[width=0.48\textwidth]{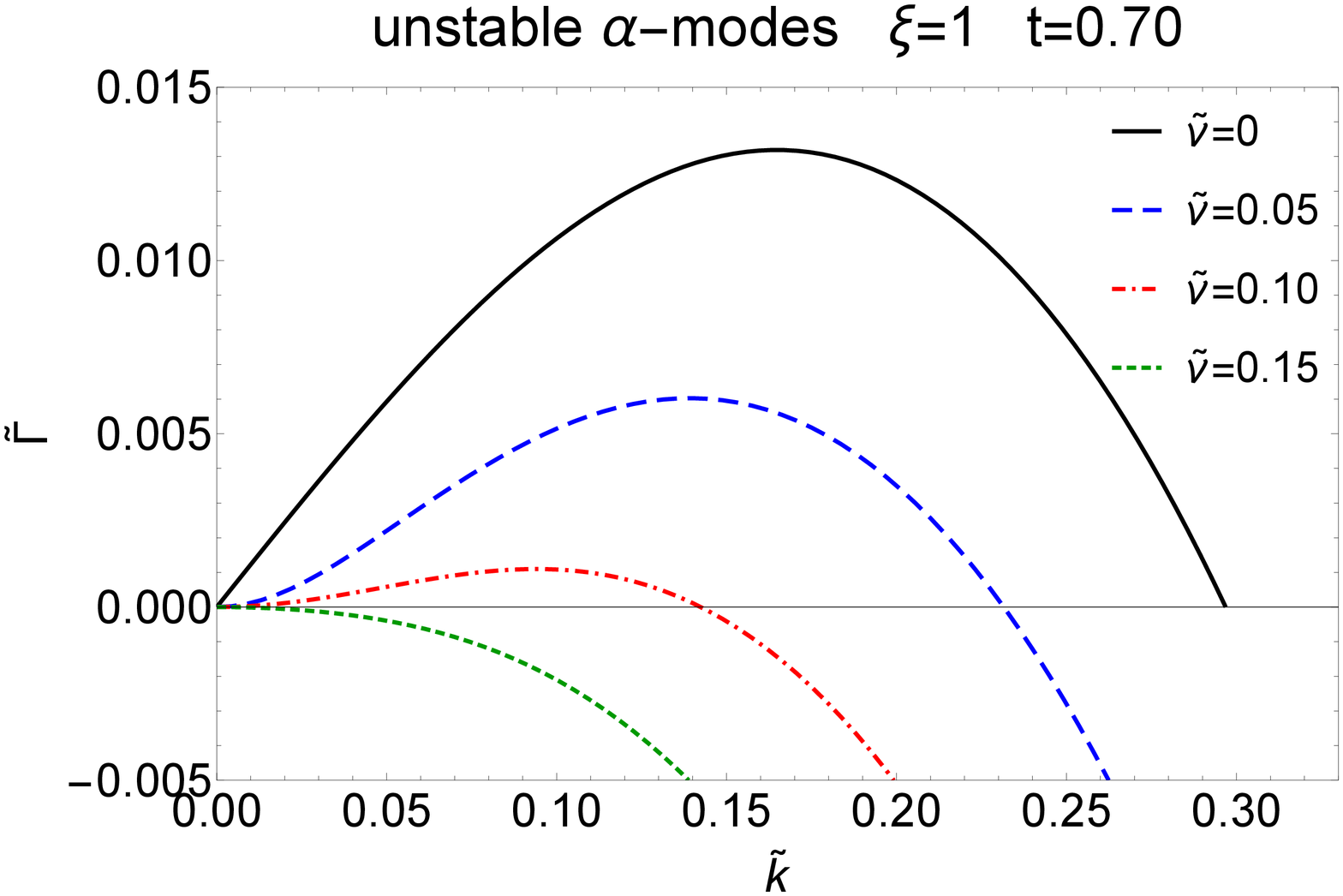}
\includegraphics[width=0.48\textwidth]{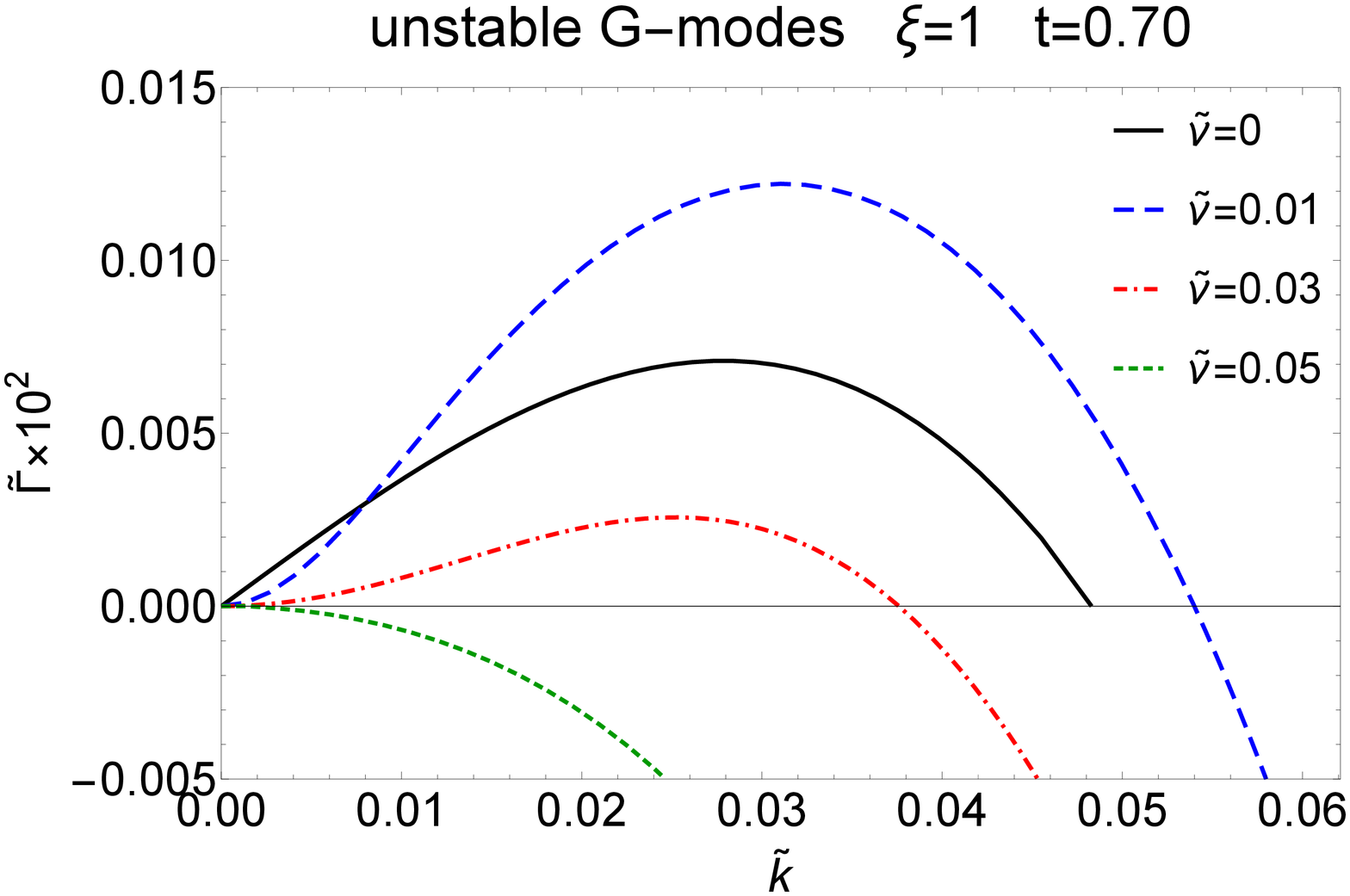}
\caption{Some examples of the dispersion relations of the unstable modes for various $\ut$ with fixed $\xi$ and $t$ where we define ${\tilde \Gamma}=\Gamma/m_D$ and ${\tilde k}=k/m_D$.}
\label{usdr}
\end{center}
\end{figure}

We can study the unstable modes by considering an imaginary-valued energy $\omega=i \Gamma$. Accordingly, the dispersion relations in Eq.~(\ref{defcm}) become
\be\label{defus}
k^2+\Gamma_\alpha^2+\alpha(i \Gamma_\alpha)=0\,,\quad\quad\Gamma_{+}^2+\Omega_{+}^2(i\Gamma_{+})=0\,,
\quad\quad
\Gamma_{-}^2+\Omega_{-}^2(i\Gamma_{-})=0\,.
\ee
Since $\Omega_{+}^2(i\Gamma_{+})$ is numerically found to be positive for $\Gamma_{+}>0$, there is only one unstable $G$-mode. Some examples can be found in Fig.~\ref{usdr}.  Notice that in the special case where $t\rightarrow 1$, one can prove that there is only one unstable mode~\cite{Schenke:2006xu} because the unstable $\alpha$-mode becomes identical to the unstable $G$-mode.

The existence of the unstable modes depends on the specific values of $\xi$, $t$ and $\ut=\nu/m_D$. As shown in Fig.~\ref{usdr}, with fixed $\xi$ and $t$, the unstable modes can only survive in certain region of $\ut$. Thus, it is interesting to find the conditions for $\km(\xi,\ut,t)\rightarrow 0$ which indicates the absence of an unstable mode. Here, $k_{max}$ corresponds to a non-zero value of the wave number $k$ at which the unstable mode terminates, i.e., $\Gamma \rightarrow 0$.  

\begin{figure}[htbp]
\begin{center}
\includegraphics[width=0.45\textwidth]{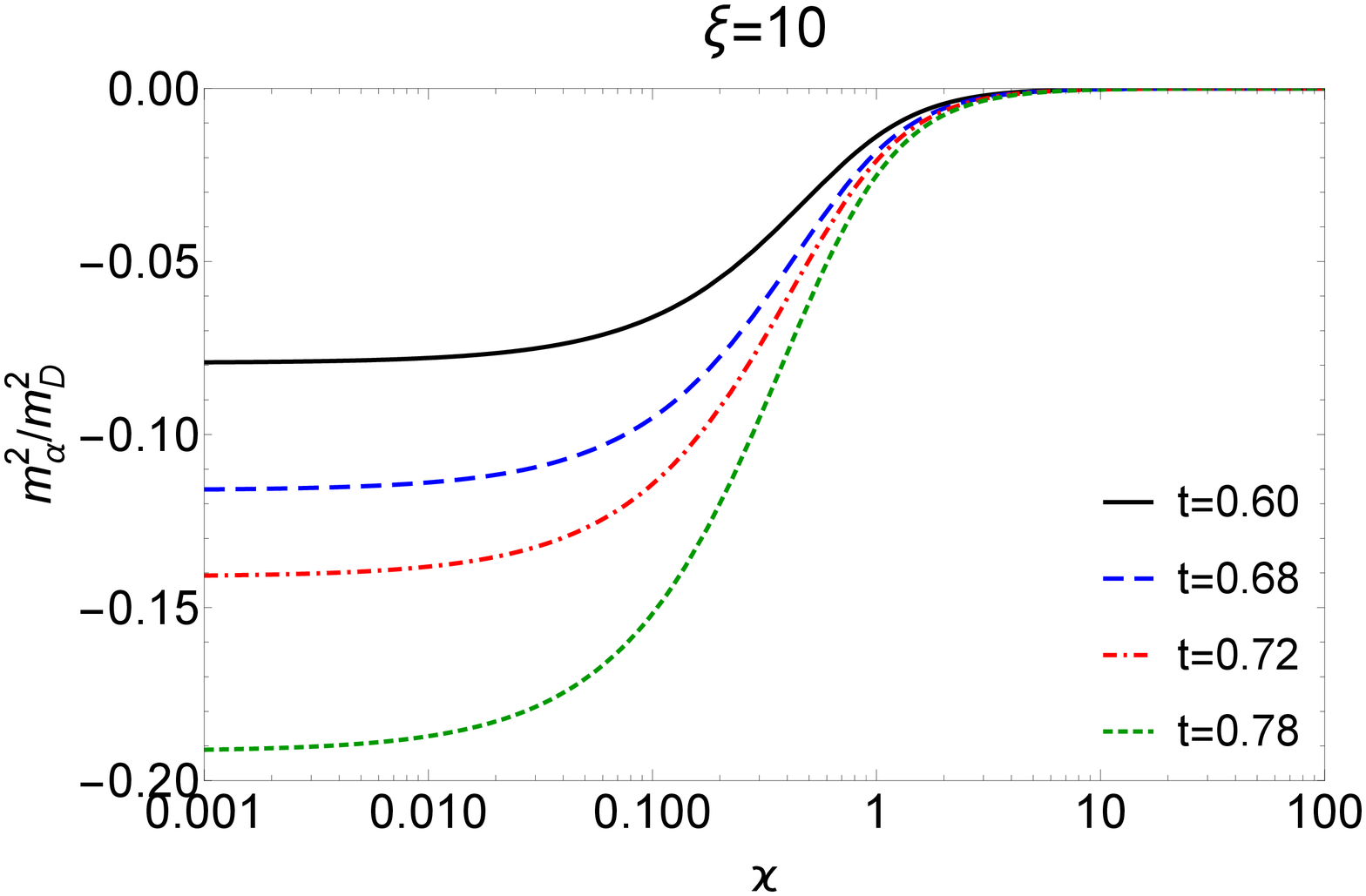}
\includegraphics[width=0.45\textwidth]{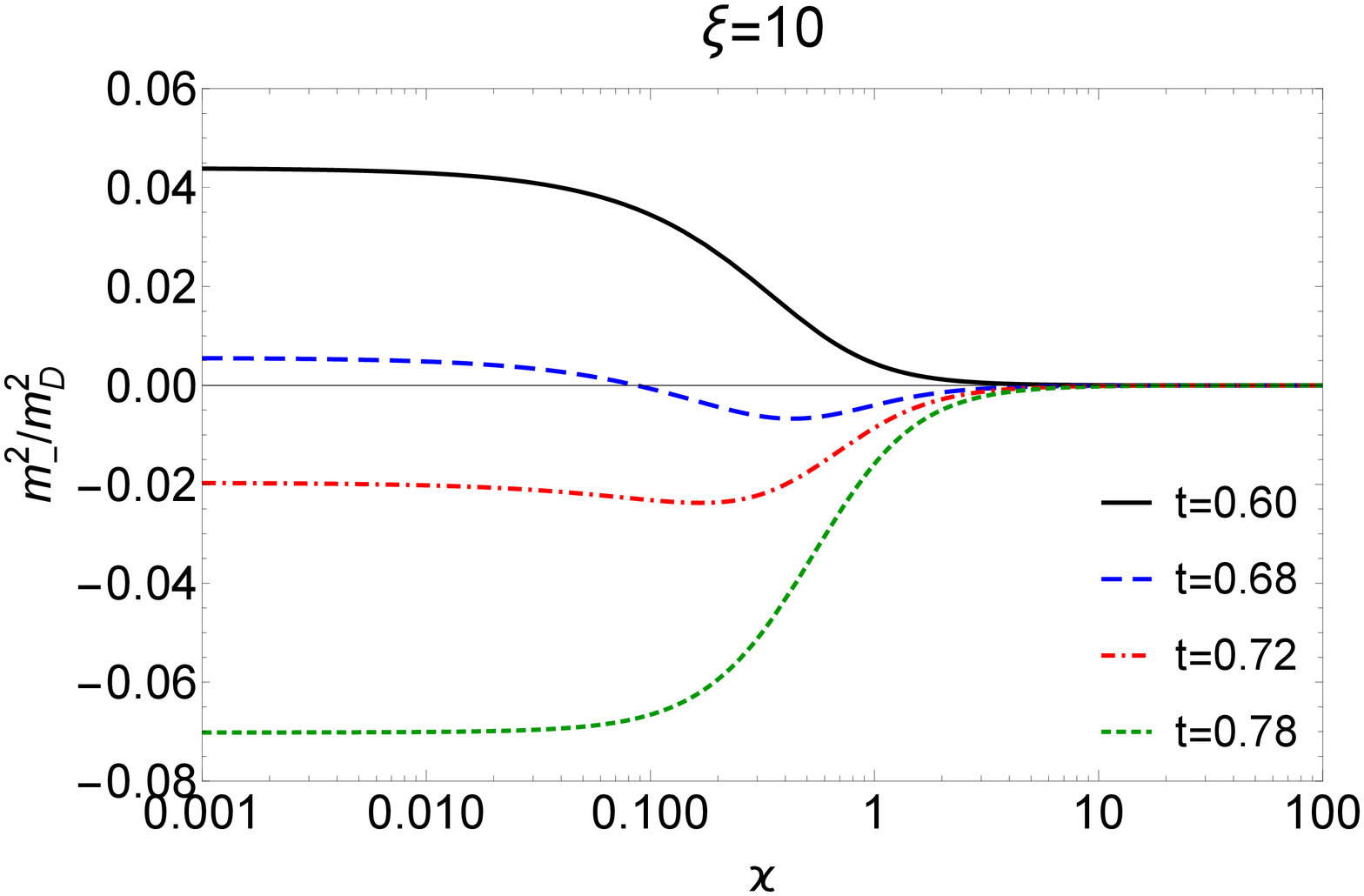}
\caption{The mass scales $m_\alpha^2$ and $m_-^2$ as a function of $x$ at fixed $\xi$ for various propagation angle $t=\cos \theta$. }
\label{maandmm}
\end{center}
\end{figure}

\begin{figure}[htbp]
\begin{center}
\includegraphics[width=0.32\textwidth]{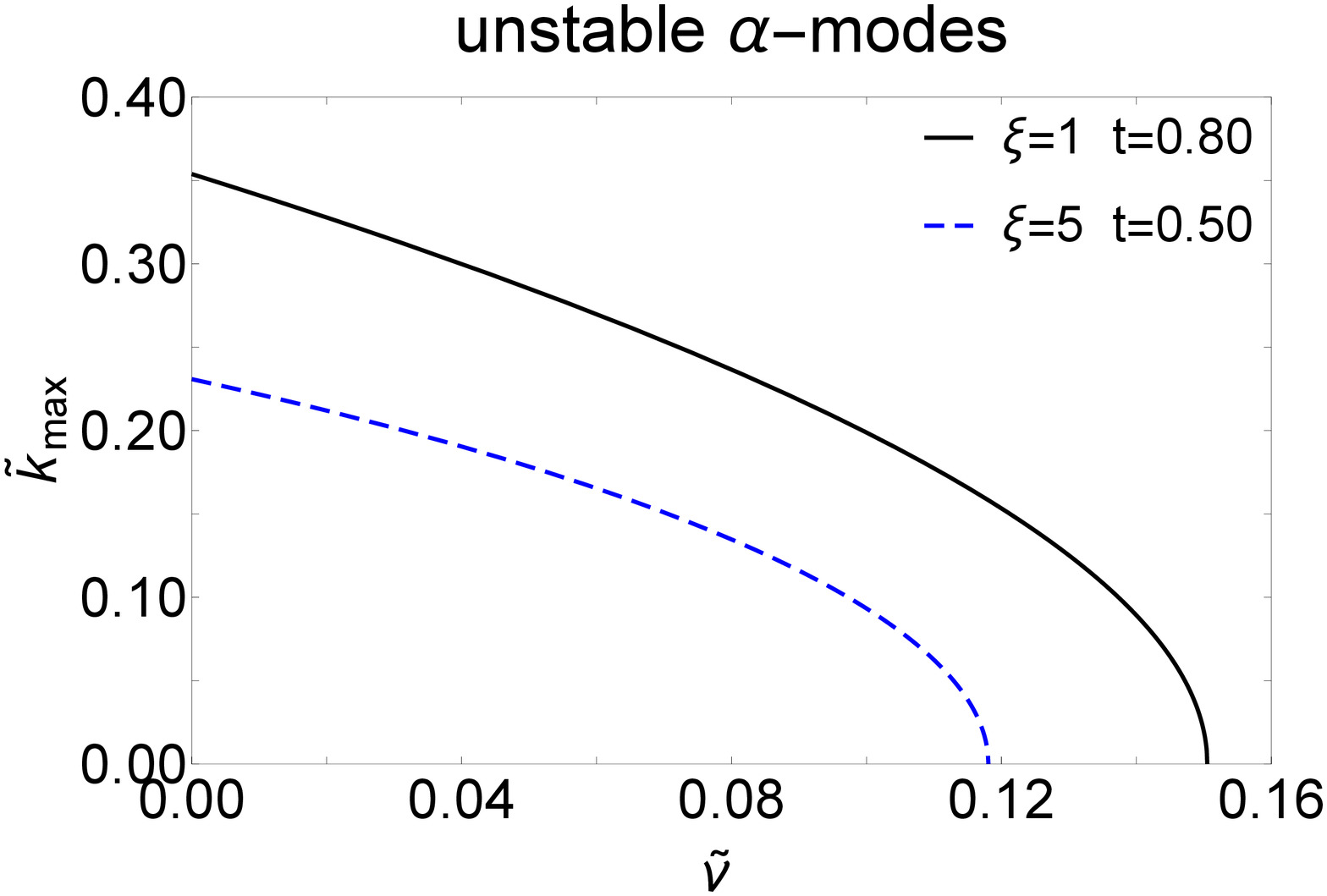}
\includegraphics[width=0.32\textwidth]{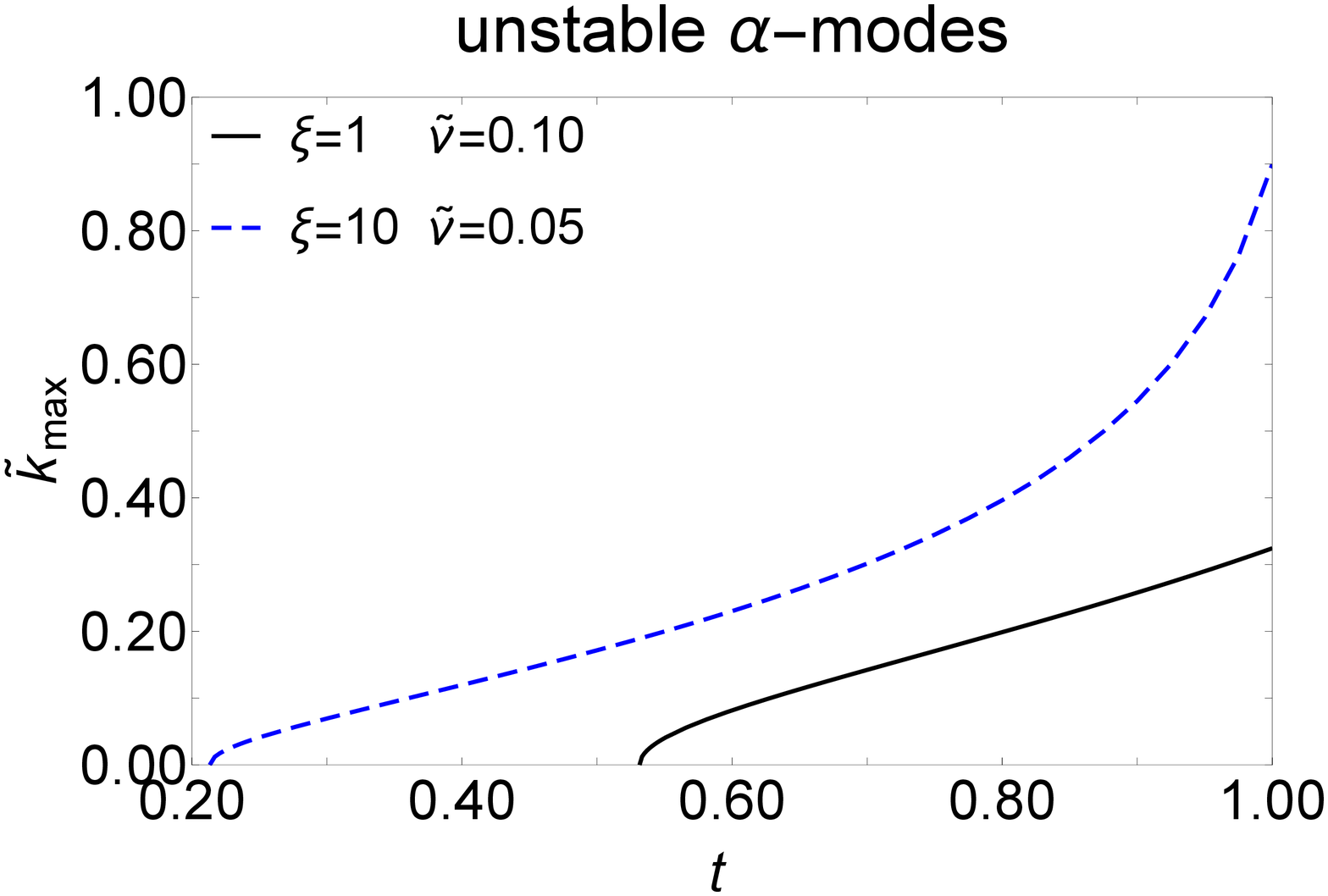}
\includegraphics[width=0.32\textwidth]{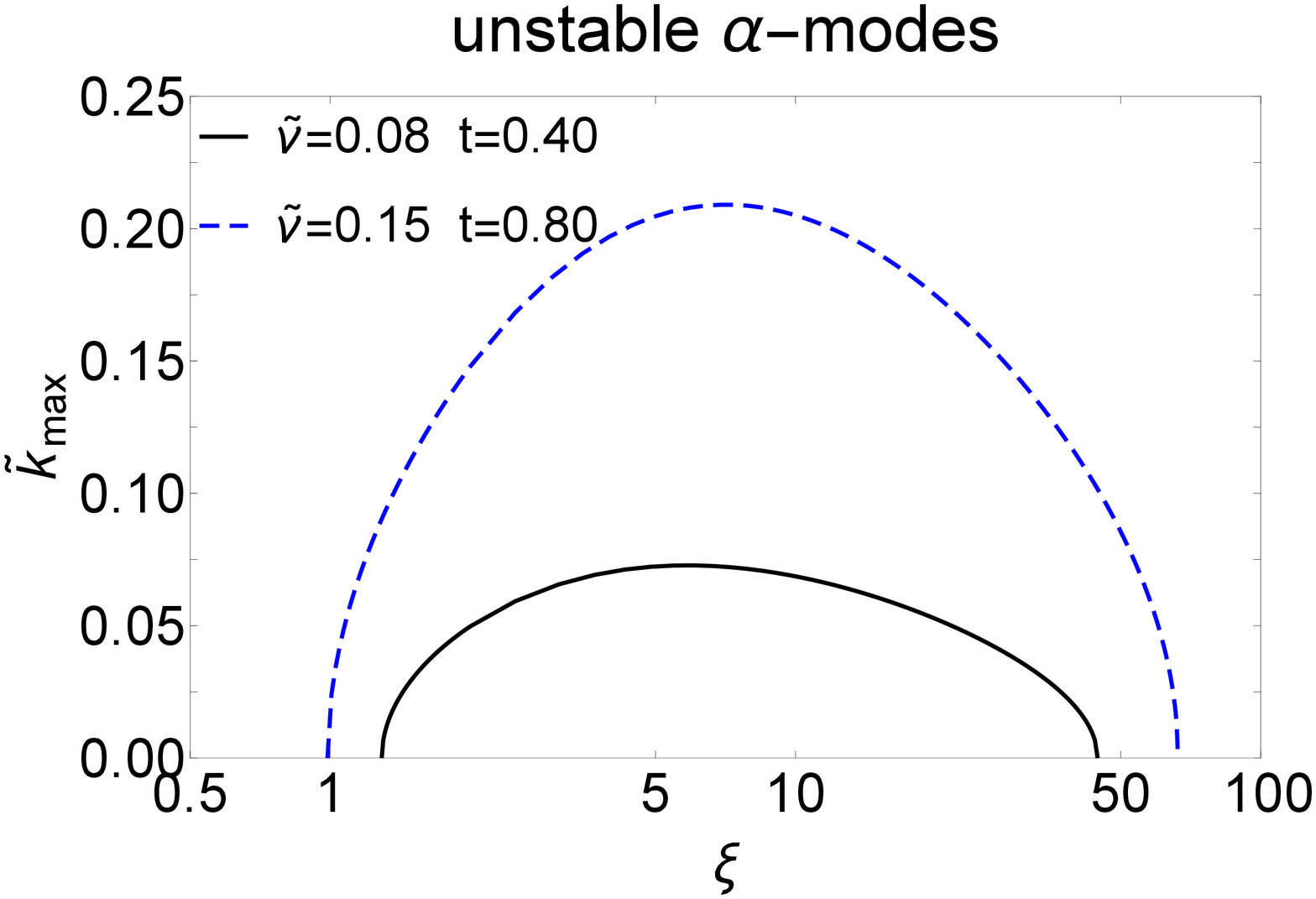}
\caption{The behavior of ${\tilde k}_{max}$ for unstable $\alpha$-modes. Left: the $\ut$-dependence of ${\tilde k}_{max}$ with fixed $\xi$ and $t$. Middle: the $t$-dependence of ${\tilde k}_{max}$ with fixed $\xi$ and $\ut$. Right: the $\xi$-dependence of ${\tilde k}_{max}$ with fixed $\ut$ and $t$.  }
\label{usalpha2}
\end{center}
\end{figure}

By setting $\Gamma=0$ in the dispersion relation, $\km(\xi,\ut,t)$ for the unstable $\alpha$-mode is determined by the following equation
\be\label{kmeq}
\km^2+m_{\alpha}^2(\xi,t,\nu/\km)=0\,.
\ee
Besides a trivial solution $\km=0$, we are actually interested in the non-zero solution $\km(\xi,\ut,t)$, which can be obtained by numerically solving the above equation. In Fig.~\ref{usalpha2}, we show how the dimensionless ${\tilde k}_{max}=\km/m_D$ depends on its variables. 

On the other hand, Eq.~(\ref{kmeq}) can be rewritten as
\be\label{utx}
\nu (x) = x \sqrt{- m_{\alpha}^2(\xi,t,x)}\, ,
\ee
with $x\equiv \nu/\km$. This equation is well defined because $m_{\alpha}^2(\xi,t,x)$ is numerically found to be negative. 
In the collisionless limit
the value of $\km$ can be simply determined by $\km (\ut \rightarrow 0)=\sqrt{- m_{\alpha}^2(\xi,t,x \rightarrow 0)}$, where the corresponding mass scales have been analytically computed in Ref.~\cite{Dumitru:2007hy}. Notice that $m_\alpha^2(\xi, t,x \rightarrow 0)$ is negative and vanishes in the small and large $\xi$ limit, therefore, the unstable $\alpha$-mode exists for any non-zero and finite anisotropy in the collisionless limit.\footnote{This conclusion is true for any propagation angle except $t=0$ where $m_{\alpha}^2(\xi,t,x)$ vanishes. In addition, the unstable $\alpha$-mode disappears as $\xi\rightarrow \infty$ because the number density of the anisotropic system drops to zero.} On the other hand, according to Fig.~\ref{usalpha2}, this is no longer true when the collision kernel is included.

In general, one can expect that the instability of an anisotropic system will be eventually eliminated by continuously increasing the collision rate. As a result, the survival of the unstable $\alpha$-mode requires $\ut < \ut(x \rightarrow \infty)$ which is the collision rate at vanishing $\km$. 
Therefore, it becomes very important to analyze the behavior of $\ut(x\rightarrow \infty)$. To do so, we Taylor expand the mass scale $m_{\alpha}^2(\xi,t,x)$ for infinitely large $x$ and in the leading-order approximation, the result reads
\be\label{lokma}
\ut (x\rightarrow \infty) = t \sqrt{f(\xi)}\, ,
\ee
where
\be\label{deffxi}
f(\xi)=\frac{-3\sqrt{\xi}+(3+\xi)\arctan\sqrt{\xi}}{4\xi^{3/2}}\, .
\ee
It can be shown that the function $f(\xi)$ is positive and vanishes as $\xi\rightarrow 0$ or $\xi\rightarrow \infty$. In addition, the maximum of $f(\xi)$ is numerically found to be $\sim 0.0562$ which is located at $\xi\approx 6.40$.

\begin{figure}[htbp]
\begin{center}
\includegraphics[width=0.32\textwidth]{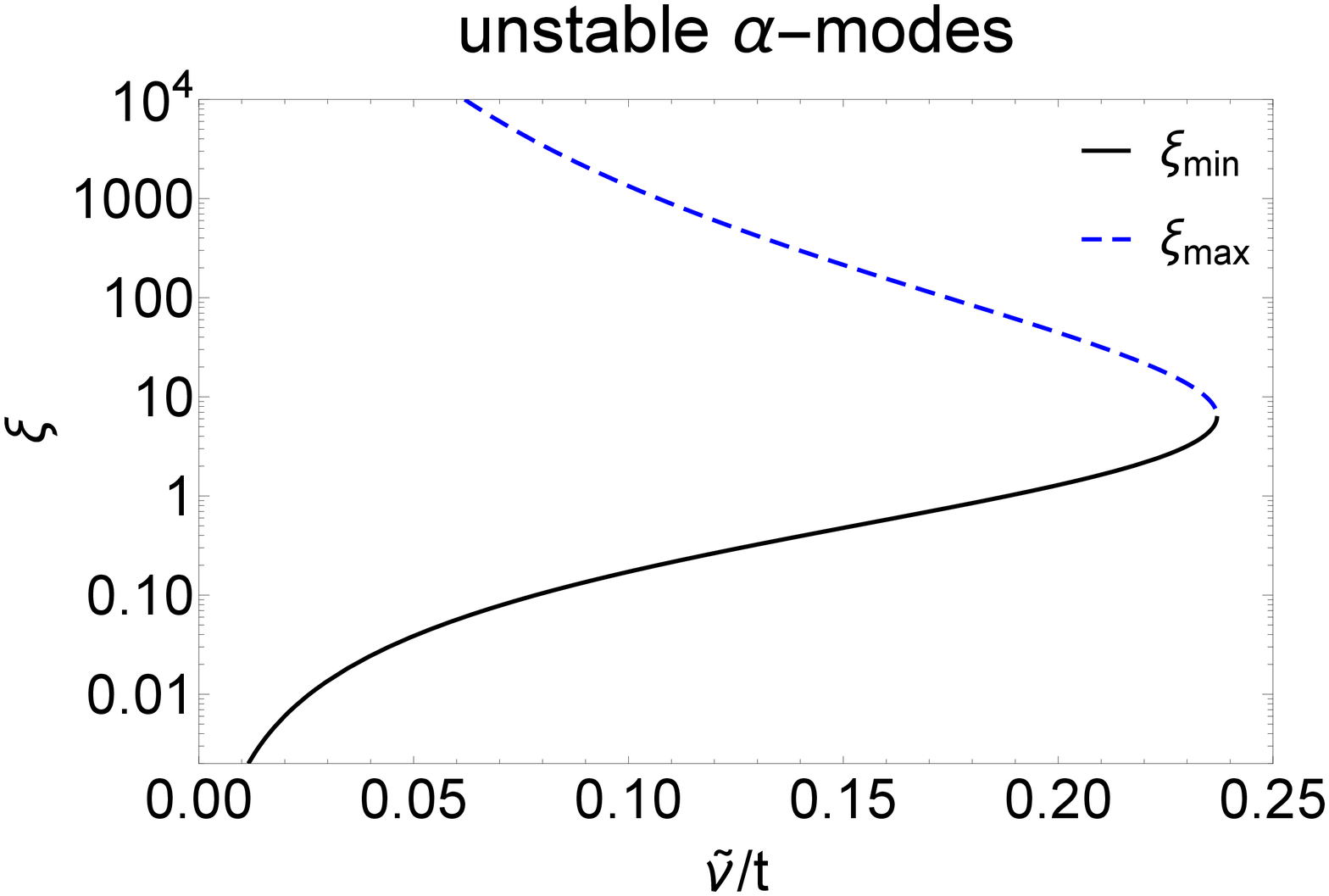}
\includegraphics[width=0.32\textwidth]{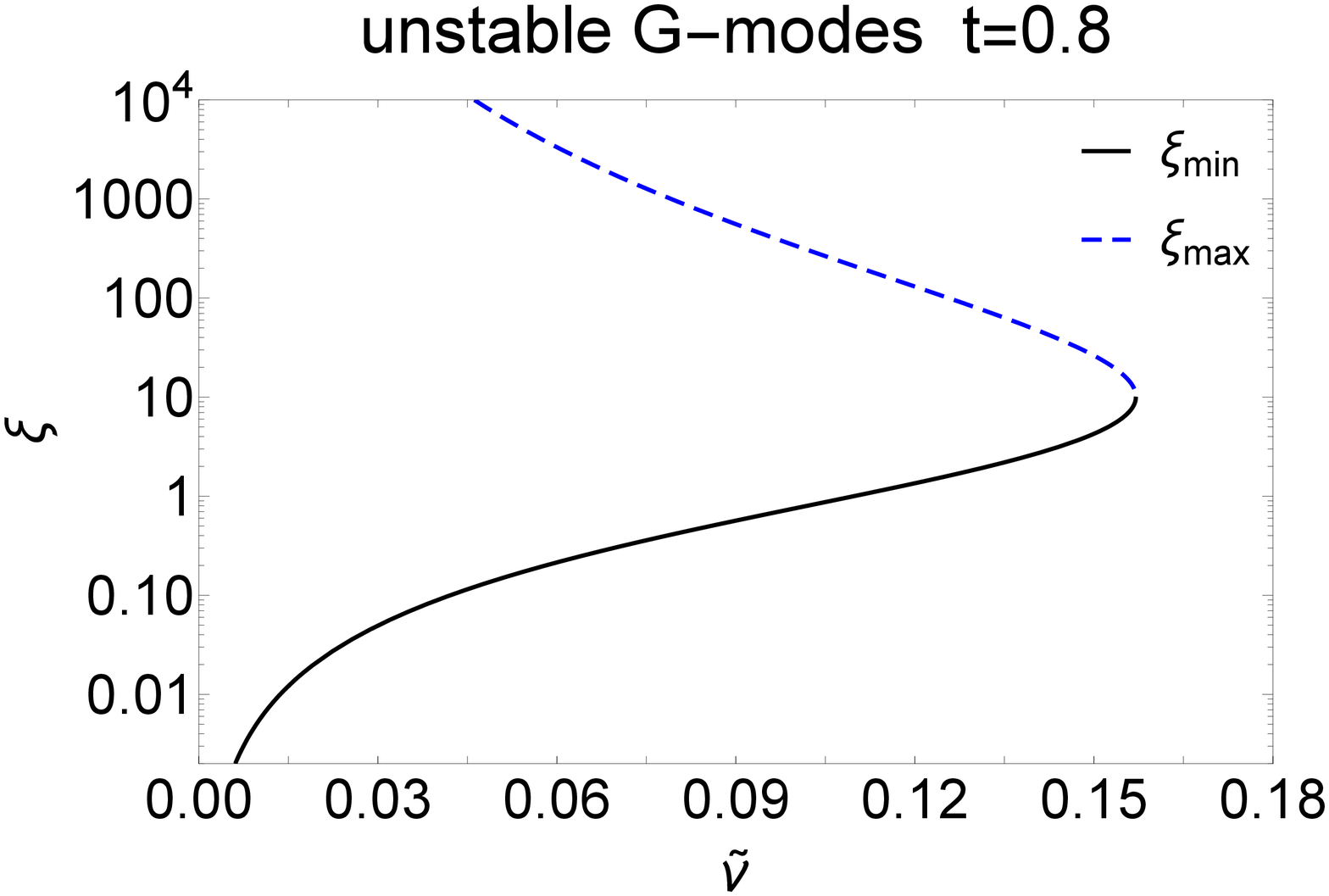}
\includegraphics[width=0.32\textwidth]{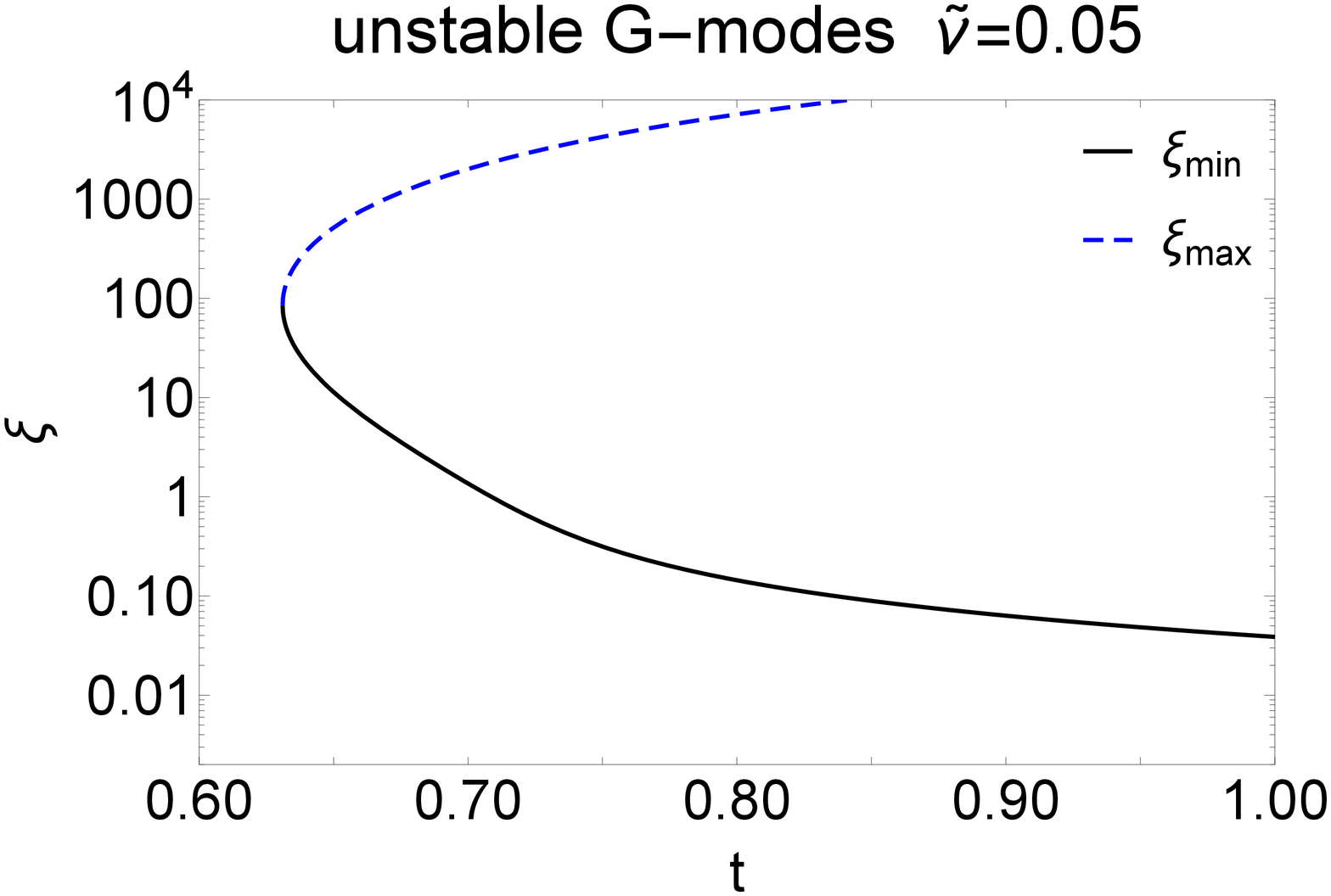}
\caption{The lower and upper limits of anisotropy parameter $\xi$ for the existence of the unstable modes. Left: the $\ut/t$-dependence of $\xi_{min}$ and $\xi_{max}$ for the unstable $\alpha$-modes. Middle: the $\ut$-dependence of $\xi_{min}$ and $\xi_{max}$ at fixed $t$ for the unstable $G$-modes. Right: the $t$-dependence of $\xi_{min}$ and $\xi_{max}$ at fixed $\ut$ for the unstable $G$-modes.}
\label{ximm}
\end{center}
\end{figure}

Given the above results, we can obtain the conditions for the existence of the 
unstable $\alpha$-mode based on Eq.~(\ref{lokma}) which are:
\begin{itemize}
\item with fixed $\xi$ and $t$, the collision rate should satisfy  $\ut <  t \sqrt{f(\xi)}$. The largest possible value for the collision rate is given by $\sqrt{f(\xi)_{max}} \approx 0.237$.
\item with fixed $\xi$ and $\ut$, the propagation angle should satisfy $t>\ut/\sqrt{f(\xi)}$. For small enough $\ut$, the propagation direction of the unstable $\alpha$-mode becomes (almost) perpendicular to the direction of anisotropy which indicates $t\rightarrow 0$.
\item with fixed $\ut$ and $t$, the anisotropy parameter should satisfy $\xi_{min}<\xi<\xi_{max}$ where $\xi_{min}$ and $\xi_{max}$ denote the two solutions of Eq.~(\ref{lokma}).
\end{itemize}
The $\ut/t$-dependence of $\xi_{min}$ and $\xi_{max}$ has to be determined numerically, which is given in Fig.~\ref{ximm}. 
In addition, the two solutions become identical at $\ut/t \approx 0.237$. It is possible to obtain analytical results for $\xi_{min}$ and $\xi_{max}$ for extremely small and large anisotropies, respectively. Explicitly, we have 
\be
\xi_{min}\approx 15(\ut/t)^{2}\,,\quad\quad \xi_{max}\approx \pi^2(\ut/t)^{-4}/64\,.
\ee
These two equations are good approximations to the exact results provided that the ratio $\ut/t$ is small. They also confirm that the unstable $\alpha$-mode exists for any non-zero and finite anisotropy in the collisionless limit $\ut\rightarrow 0$ when $t\neq 0$. 

The above discussions provide a simple way to locate the zeros of $\km(\xi,\ut,t)$ as shown in Fig.~\ref{usalpha2} which actually correspond to the critical condition for the existence of the unstable $\alpha$-mode. Thus, given a set of parameters $\xi$, $t$ and $\ut$, one can easily determine whether or not the unstable $\alpha$-mode exists.

\begin{figure}[htbp]
\begin{center}
\includegraphics[width=0.32\textwidth]{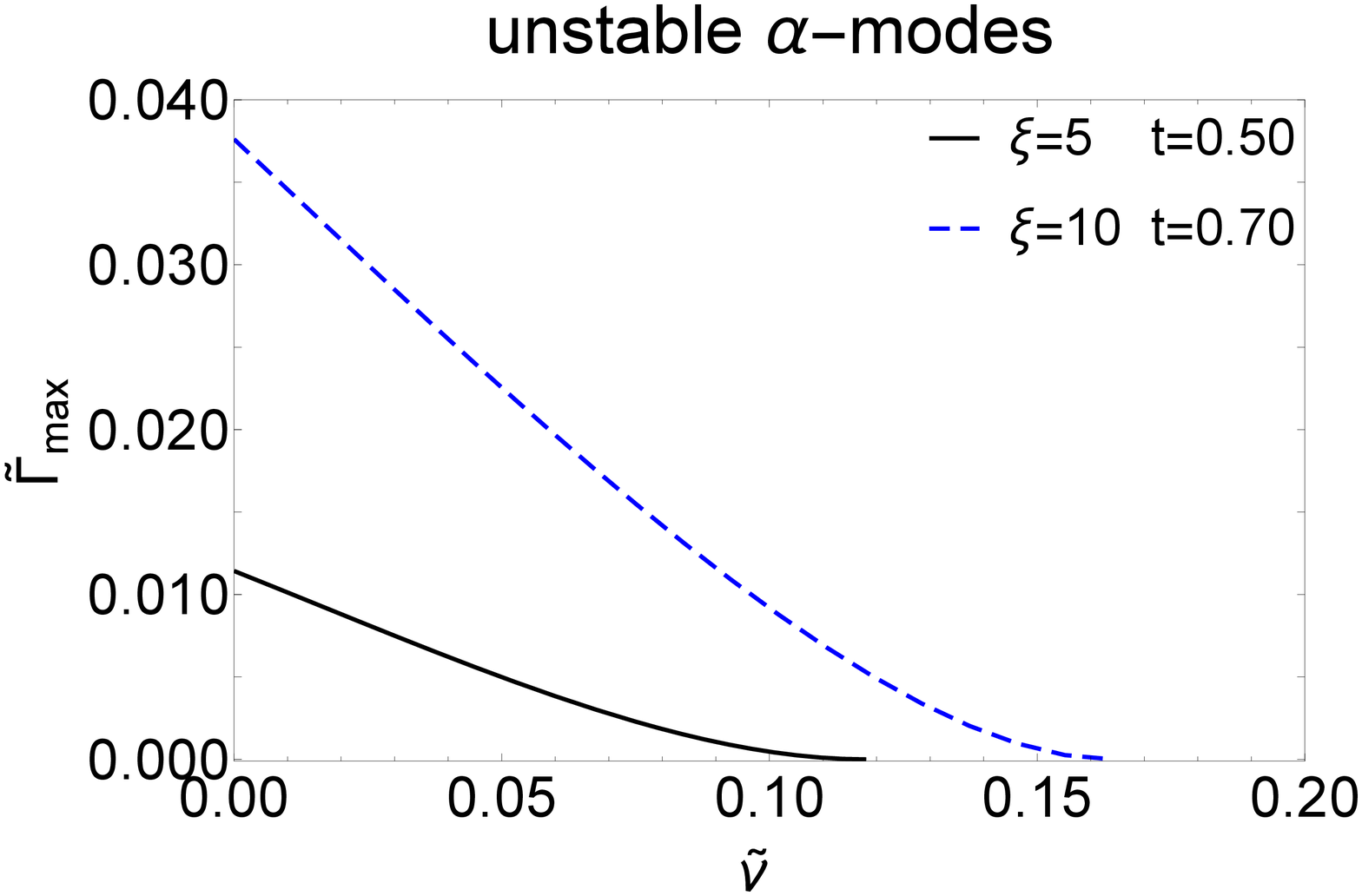}
\includegraphics[width=0.32\textwidth]{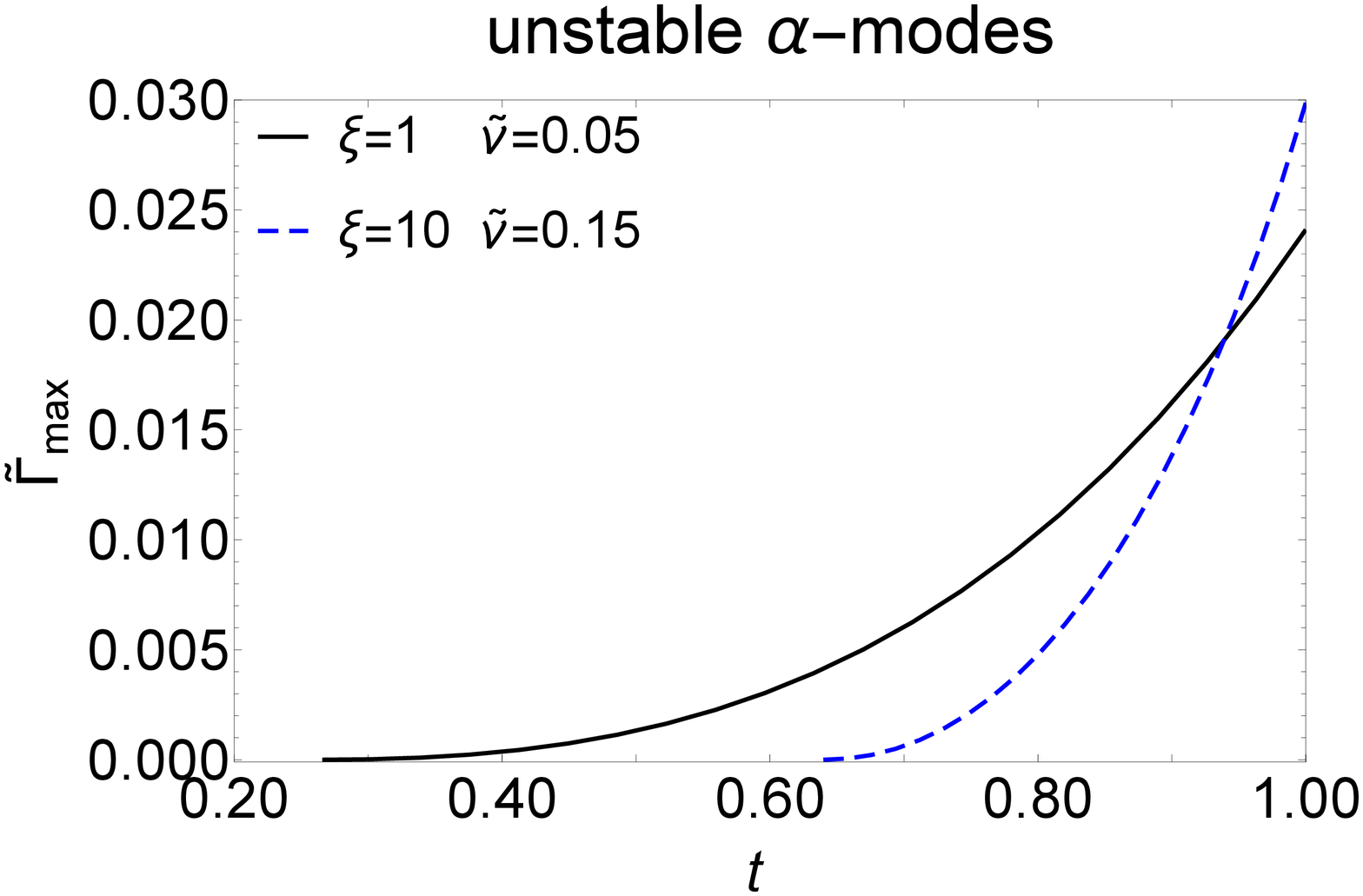}
\includegraphics[width=0.32\textwidth]{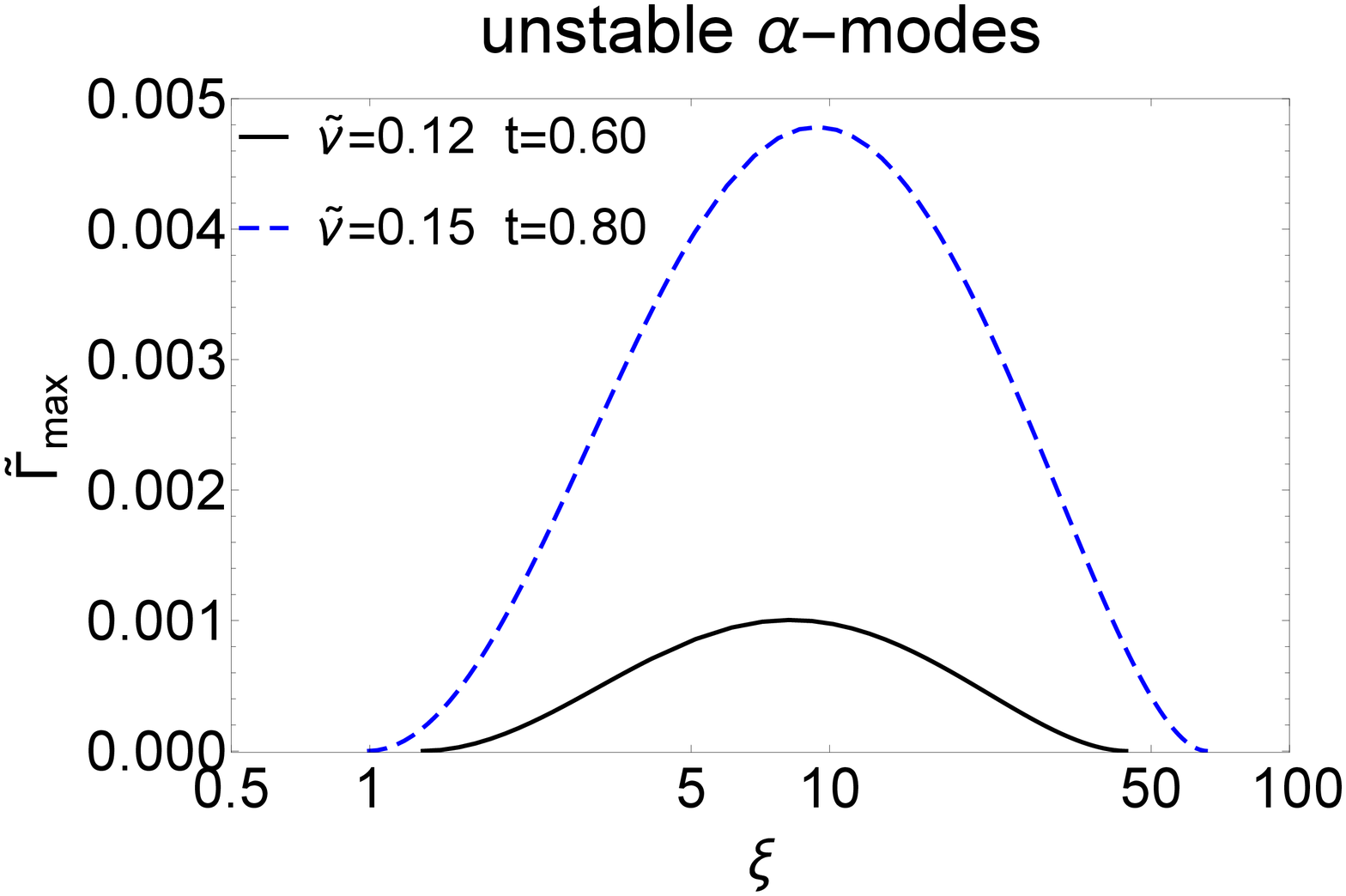}
\caption{The maximal growth rate ${\tilde \Gamma}_{max}\equiv  \Gamma_{max}/m_D$ for the unstable $\alpha$-modes as a function of $\ut$ (left), $t$ (middle) and $\xi$ (right).}
\label{agamax}
\end{center}
\end{figure}

It is also important to study the maximal growth rate $\Gamma_{max}$ in the dispersion relations of the unstable modes. As shown in Fig.~\ref{agamax}, $\Gamma_{max}$ depends on the collision rate, propagation angle, and the degree of anisotropy, which exhibits very similar behaviors when compared with the dependence on $\km$. In practice, it is of particular interest to find the maximum possible value for $\Gamma_{max}$. According to our results, at a given $\xi$ and $t$, the maximum of $\Gamma_{max}$ appears in the collisionless limit. In other words, the collision effect reduces $\Gamma_{max}$. In addition, for fixed $\xi$ and $\ut$, we find that $\Gamma_{max}$ increases with increasing $t$ and thus, confirms that the growth rate of the filamentation instability is largest when the wave vector ${\bf k}$ is parallel to the direction
of the anisotropy, i.e., $t=1$. Finally, we note that, as can be seen in the rightmost panel of Fig.~\ref{agamax}, $\Gamma_{max}$ is a non-monotonic function of the anisotropy parameter $\xi$ and the peak position needs to be determined numerically.

Next, we discuss the unstable $G$-mode which exhibits some new and complicated features as compared to the unstable $\alpha$-mode. We start with the dispersion relation in the limit $\Gamma \rightarrow 0$ which can be written as
\be\label{kmeq2}
(\km^2+m_{+}^2(\xi,t,\nu/\km))(\km^2+m_{-}^2(\xi,t,\nu/\km))=0\,.
\ee
Since the mass scale $m_{+}^2(\xi,t,\nu/\km)$ is numerically found to be always positive, one can only consider the second term in the above equation. Following the discussion of the unstable $\alpha$-mode, a similar expression for the collision rate $\nu$ can be obtained as 
\be\label{utx2}
\nu (x) = x \sqrt{- m_{-}^2(\xi,t,x)}\, , \;\;\;\;\; {\rm if}\quad m_{-}^2(\xi,t,x)\le 0 \, .
\ee
Unlike the negative mass scale $m_{\alpha}^2(\xi,t,x)$, depending on the specific values of the variables, $m_{-}^2(\xi,t,x)$ can be positive which leads to the absence of the unstable G-mode as well as an ill-defined $\ut (x)$ in Eq.~(\ref{utx2}). Therefore, the above equation applies only for $m_{-}^2(\xi,t,x)\le 0$. 

\begin{figure}[htbp]
\begin{center}
\includegraphics[width=0.32\textwidth]{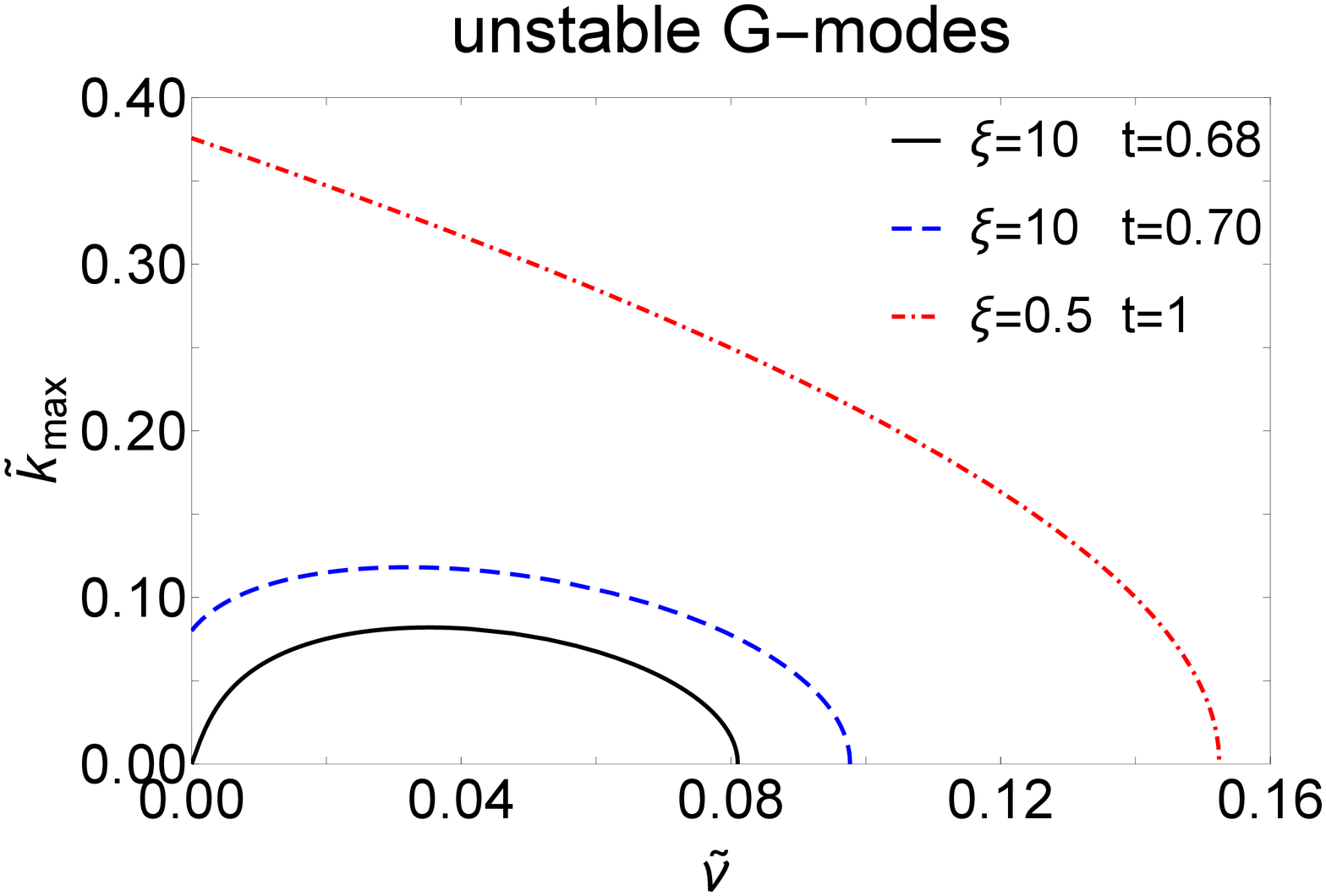}
\includegraphics[width=0.32\textwidth]{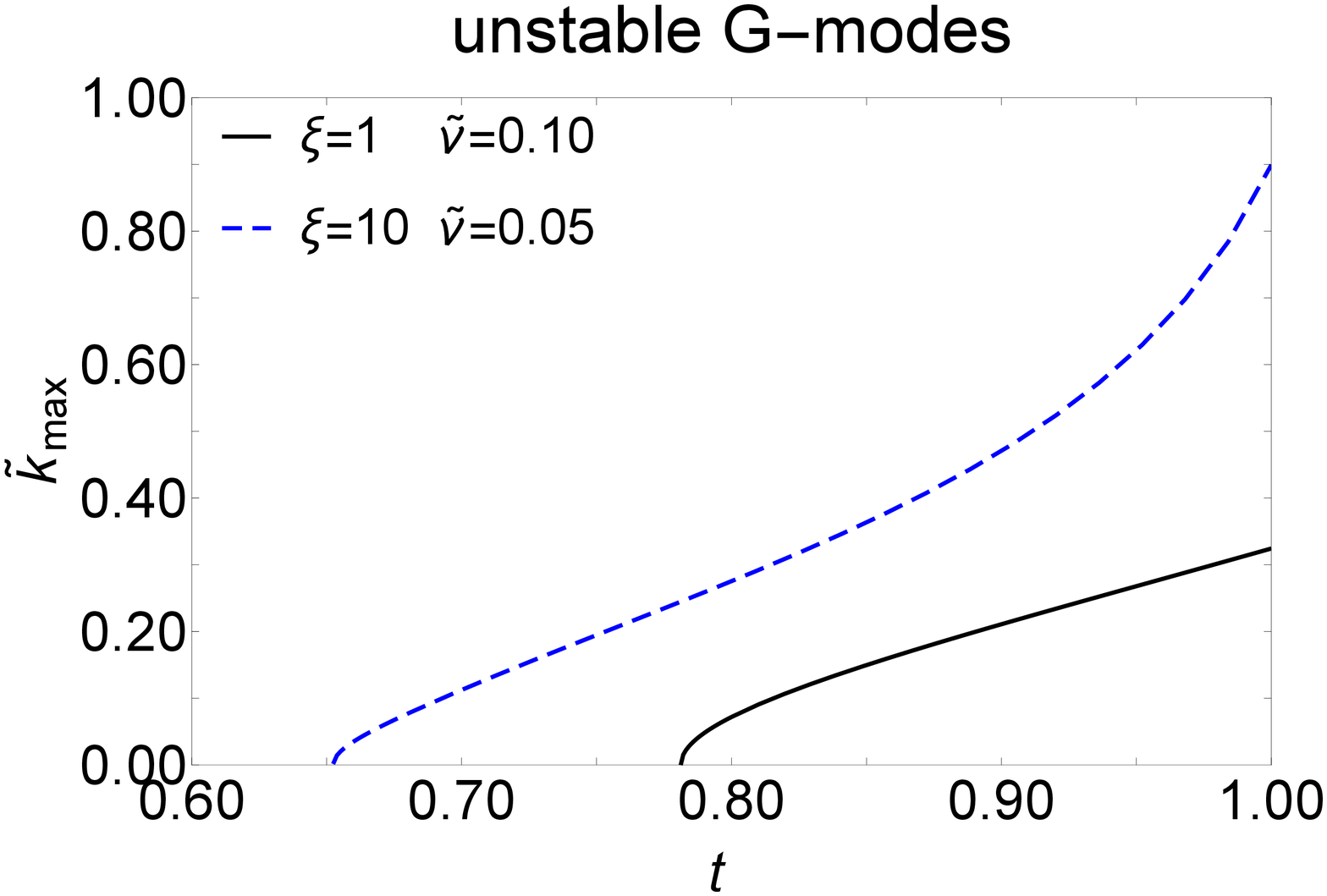}
\includegraphics[width=0.32\textwidth]{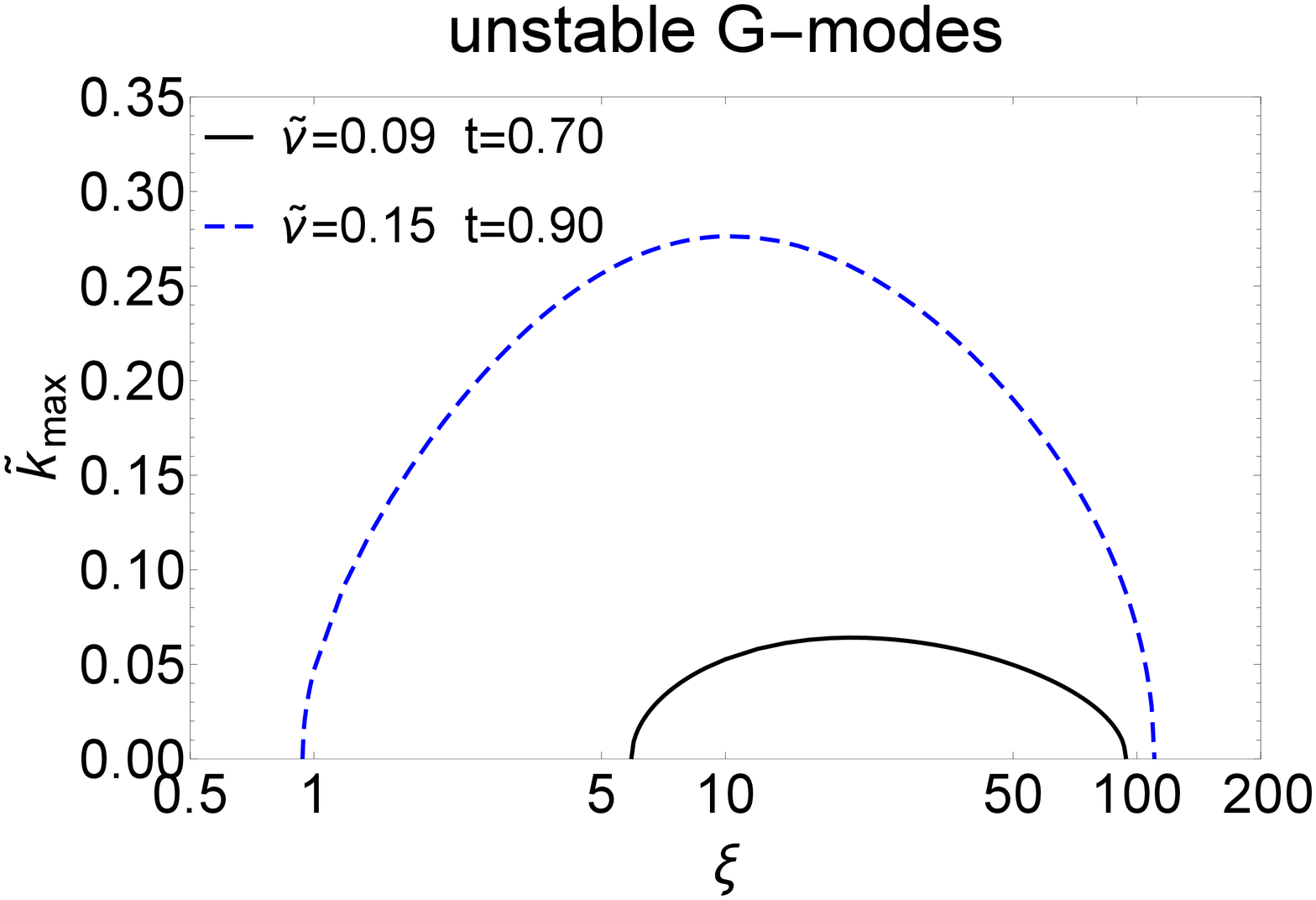}
\caption{The behavior of ${\tilde k}_{max}$ for the unstable $G$-modes. Left: the $\ut$-dependence of ${\tilde k}_{max}$ with fixed $\xi$ and $t$. Middle: the $t$-dependence of ${\tilde k}_{max}$ with fixed $\xi$ and $\ut$. Right: the $\xi$-dependence of ${\tilde k}_{max}$ with fixed $\ut$  and $t$. }
\label{usg2}
\end{center}
\end{figure}

Following a similar discussion on the unstable $\alpha$-mode, we can expand $m_{-}^2(\xi,t,x)$ for large $x$. When keeping the leading order contribution, Eq.~(\ref{utx2}) becomes
\be\label{lokm}
\ut (x\rightarrow \infty) = \sqrt{g(\xi,t)}\, ,
\ee
where
\be
g(\xi,t)=\frac{g_2(\xi)(2t^2-1)
+g_1(\xi)}{g_1(\xi)(2t^2-1)+g_2(\xi)}f(\xi)\,,
\ee
with
\be
g_1(\xi)=-3+(3+\xi)\frac{\arctan \sqrt{\xi}}{\sqrt{\xi}}\quad\quad{\rm and}\quad\quad g_2(\xi)=-1+(1+3\xi)\frac{\arctan \sqrt{\xi}}{\sqrt{\xi}}\,.
\ee
Numerically, we find that $m_{-}^2(\xi,t,x)$ is positive if it approaches to zero from above when $x \rightarrow \infty$. Consequently, only positive $g(\xi,t)$ is relevant for studying the unstable mode since $g(\xi,t)$ and $m_{-}^2(\xi,t,x\rightarrow \infty)$ have opposite signs. Based on Eq.~(\ref{lokm}), the following conclusions can be drawn.
\begin{itemize}
\item with fixed $\xi$ and $t$, the collision rate should satisfy $\ut < \sqrt{g(\xi,t)}$. Because both $g_1(\xi)$ and $g_2(\xi)$ are positive, the function $g(\xi,t)$ increases with increasing $t$ at a given $\xi$. Thus, the largest possible value for the collision rate is given by $\sqrt{f(\xi)_{max}}\approx 0.237$ which coincides with our finding for the unstable $\alpha$-mode.
\item with fixed $\xi$ and $\ut$, the propagation angle should satisfy $t>t_0(\xi,\ut)$ where $t_0$ denotes to the solution of Eq.~(\ref{lokm}) and is given by
\be
t_0(\xi,\ut)=\sqrt{\frac{g_1(\xi)-g_2(\xi)}{2 g_1(\xi)}\frac{4 \ut^2 \xi+g_1(\xi)}{4 \ut^2 \xi -g_2(\xi)}}\, .
\ee
In the above equation, the values of $\xi$ and $\ut$ cannot be arbitrary and a prerequisite $\ut<\sqrt{g(\xi,1)}$ must be met. Furthermore, at a given anisotropy $\xi$, $t_0(\xi,\ut)$ increases with increasing $\ut$ and the minimum of $t_0(\xi,\ut)$ is found to be $1/\sqrt{3}\approx 0.577$ when $\ut\ll \sqrt{g(\xi,1)}$ and $\xi$ is infinitely large.\footnote{Strictly speaking, one cannot simply take the collisionless limit where $\ut \rightarrow 0$ because in Eq.~(\ref{lokm}), $x\rightarrow \infty$ is required.} Accordingly, the propagation of an unstable $G$-mode is forbidden when $\theta$ is larger than $\arccos(1/\sqrt{3})\sim 0.3\,\pi$.
\item with fixed $\ut$ and $t$, the anisotropy parameter should satisfy $\xi_{min}<\xi<\xi_{max}$ where $\xi_{min}$ and $\xi_{max}$ denote the two solutions of Eq.~(\ref{lokm}).
\end{itemize}

As shown in Figs.~\ref{usalpha2} and \ref{usg2}, the most obvious difference between the two kinds of the unstable modes lies in the $\ut$-dependence of $\km$. In general, with fixed $\xi$ and $t$, $\km$ for the unstable $G$-mode is a non-monotonic function of $\ut$ and the peak position moves to $\ut=0$ very quickly when increasing $t$. In the limit $t\rightarrow 1$, the unstable $G$-mode becomes identical to the unstable $\alpha$-mode, and thus $\km$ decreases with increasing $\ut$. In particular, the function $\km(\ut)$ can even pass through the origin. According to Eq.~(\ref{utx2}), this happens due to the vanishing $m_-^2(\xi,t,x)$ at non-zero $x$ where both $\ut$ and $\km$ are infinitely small but their ratio $x$ is finite. 

It is worth mentioning that the peak values of $\km$ correspond to the minima of $m_-^2(\xi,t,x)$. According to Fig.~\ref{maandmm}, the minima appear at some non-zero $x$ for $m_-^2(\xi,t,x)$, nevertheless, $m_\alpha^2(\xi,t,x)$ becomes smallest as $x\rightarrow 0$. This explains the different $\ut$-dependences of $\km$ as observed in our results.

\begin{figure}[htbp]
\begin{center}
\includegraphics[width=0.55\textwidth]{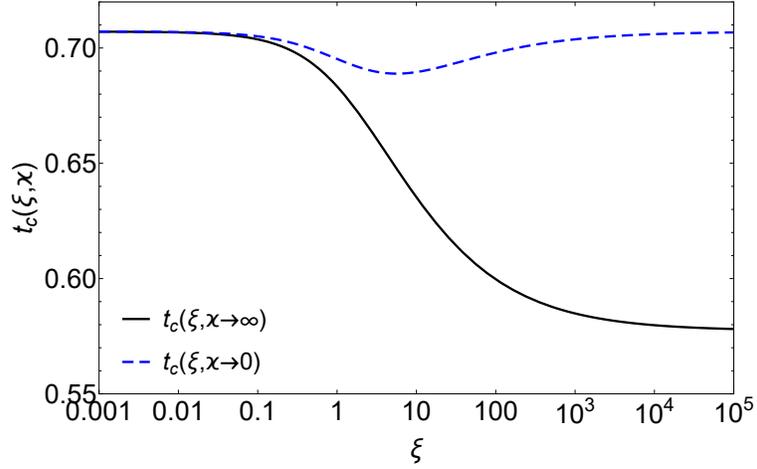}
\caption{The $\xi$-dependence of $t_c(\xi,x\rightarrow \infty)$ (solid curve) and $t_c(\xi,x\rightarrow 0)$ (dashed curve). }
\label{xitc}
\end{center}
\end{figure}

Another interesting finding is that the unstable $G$-mode survives in a wider region of $t$ in the collisional case which can be directly seen from Fig.~\ref{maandmm}. For example, given $\xi=10$ and $t=0.68$, $m_{-}^2(\xi,t,x\rightarrow 0)$ is positive and thus, no unstable mode exists in the collisionless limit where $x\rightarrow 0$. However, a negative $m_{-}^2(\xi,t,x)$ shows up at some finite $x$ which indicates the existence of the unstable modes. As a result, one can expect that for certain propagation angles, the unstable $G$-mode, which is absent in the collisionless limit, would become unstable at finite collision rate. It can be also understood by considering the sign of $m_{-}^2(\xi,t,x)$ in the small and large $x$ limits. We find that at a fixed anisotropy $\xi$, a negative mass scale exists above a critical value of $t$, which are denoted as $t_c(\xi,x\rightarrow 0)$ and $t_c(\xi,x\rightarrow \infty)$ for $m_{-}^2(\xi,t,x\rightarrow 0)$ and $m_{-}^2(\xi,t,x\rightarrow \infty)$, respectively. According to Fig.~\ref{xitc}, the critical value of $t$ in the collisionless limit is always larger than that for infinitely large $x$. Thus, for any given value of $t$ which is between the dashed and solid curves in this figure, there are unstable $G$-modes only for non-zero collision rate. Furthermore, the minimum possible value of $t$ for finding an unstable $G$-mode when $x\rightarrow 0$ is numerically found to be $\sim 0.689$ where $\xi\approx 5.64$, while it can be lowered to $1/\sqrt{3}\approx 0.577$ as $\xi\rightarrow \infty$ if the collision term is included.

On the other hand, as a function of $t$ or $\xi$, $\km$ shows a similar behavior for the two different kinds of unstable modes. However, recall that the existence of a non-zero lower bound $t_0(\xi,\ut)$ for finding an unstable $G$-mode differs from the unstable $\alpha$-mode where $t$ can take any value between $0$ and $1$ provided that $\ut/t<0.237$ can be satisfied. Given the fact that the values of $\km$ as $t\rightarrow 1$ are the same for both unstable 
$\alpha$-mode and $G$-mode, a larger increasing rate of $\km$ can be expected for the latter. In addition, unlike the unstable $\alpha$-mode, $\xi_{min}$ and $\xi_{max}$ of Eq.~(\ref{lokm}) don't simply depend on the ratio $\ut/t$. In Fig.~\ref{ximm}, both $\ut$- and $t$-dependence of these two solutions are presented. With fixed $t$, the two solutions become identical when $\ut$ equals the maximum of $\sqrt{g(\xi,t)}$. 
On the other hand, if the collision rate $\ut$ is fixed, the equality of the two appears when the propagation angle $t$ in Eq.~(\ref{lokm}) becomes smallest when varying $\xi$. 
Analytical results can be also obtained for $\xi_{min}$ and $\xi_{max}$ for extremely small and large anisotropies, respectively, which read
\be
\xi_{min}\approx \frac{15}{2 t^2-1}  \ut^2 \,,\quad\quad \xi_{max}\approx\frac{\pi^2 }{64 \ut^4}\frac{(3t^2-1)^2 }{(1+t^2)^2}\,.
\ee
These two equations are good approximations to the exact results when $\ut$ is small and $t$ is close to $1$. They also indicate that for $t>1/\sqrt{2}\approx 0.707$, the unstable $G$-mode can exist for any non-zero and finite $\xi$ in the case $\ut\rightarrow 0$.

\begin{figure}[htbp]
\begin{center}
\includegraphics[width=0.32\textwidth]{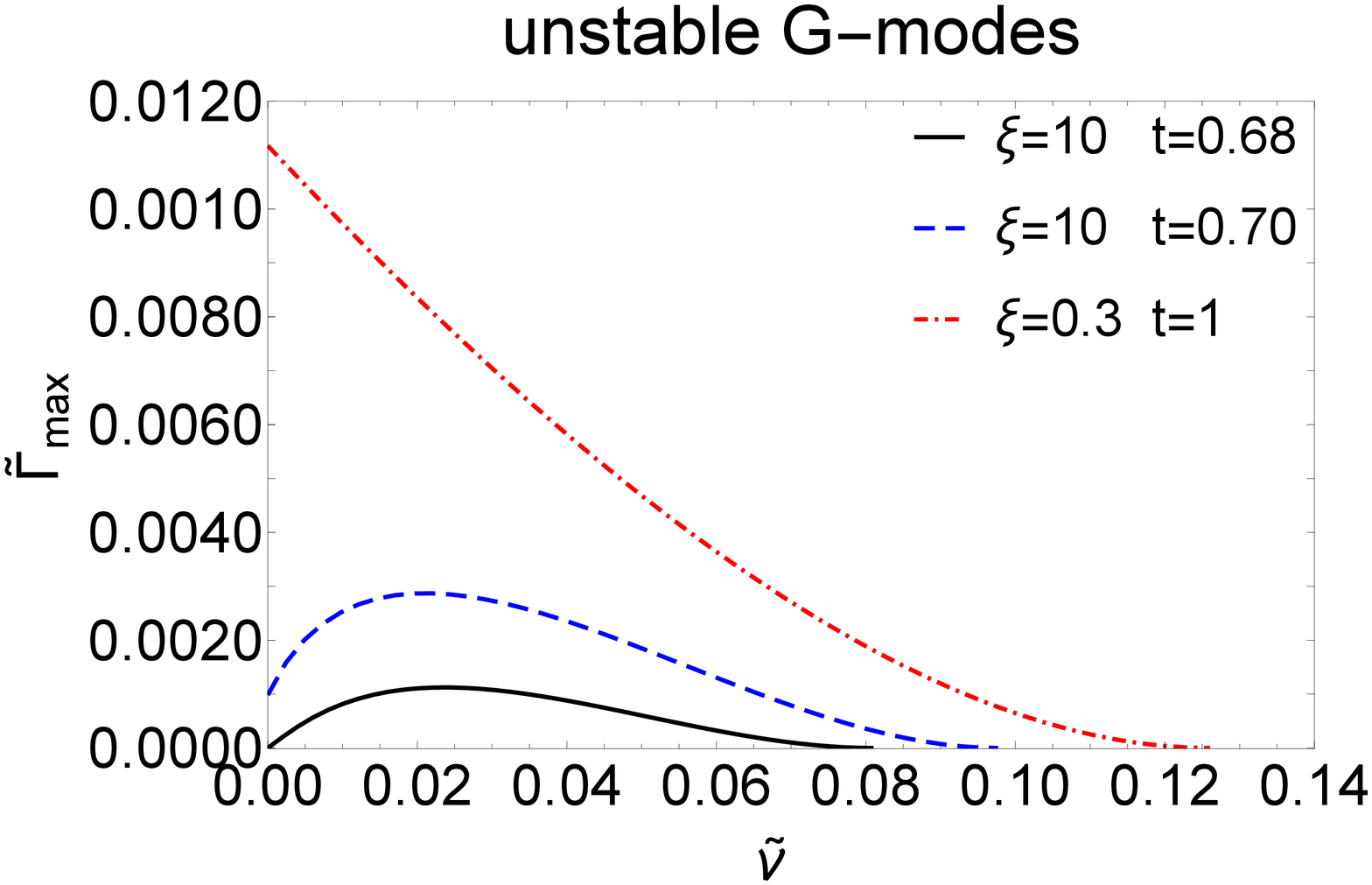}
\includegraphics[width=0.32\textwidth]{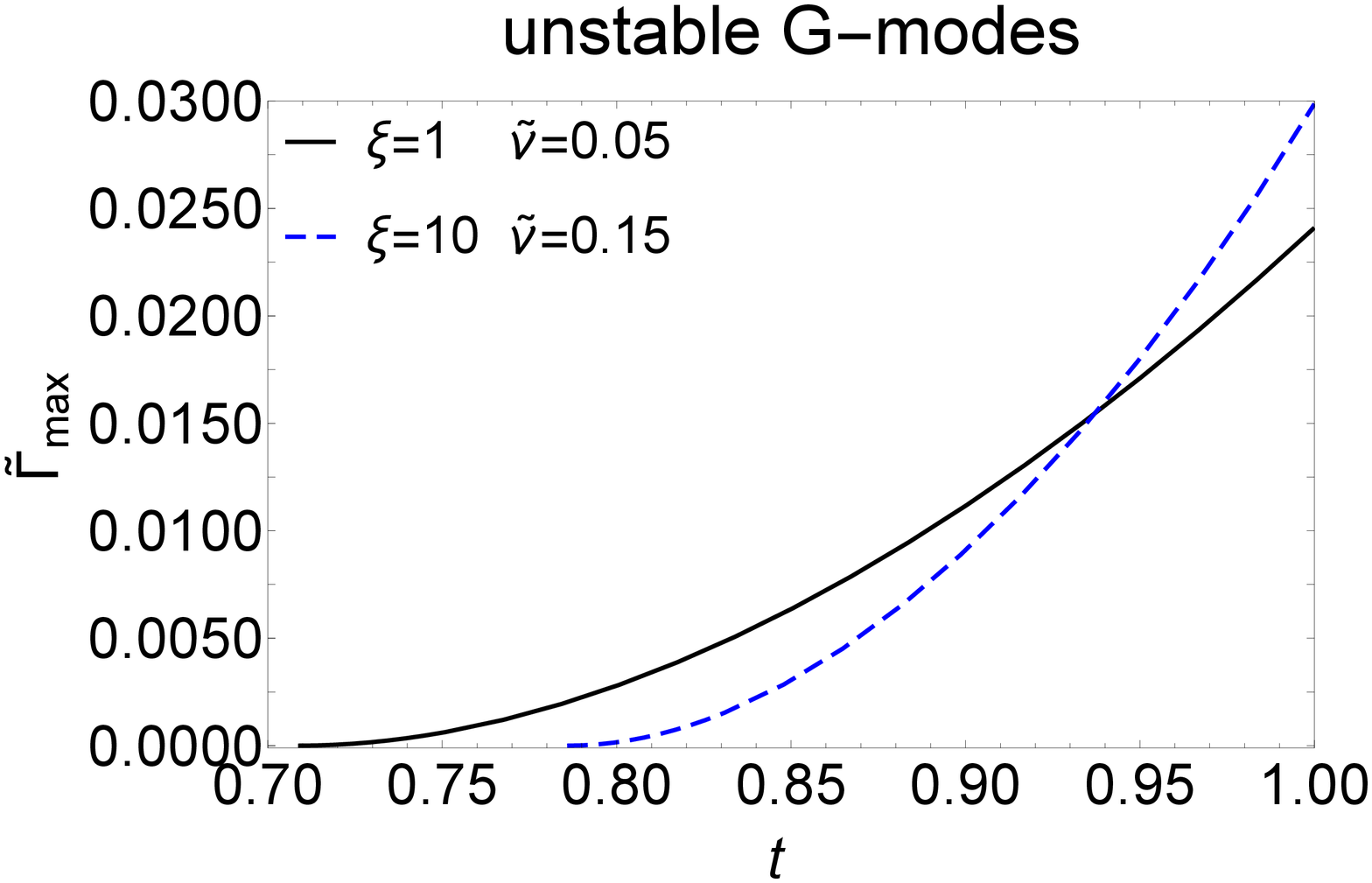}
\includegraphics[width=0.32\textwidth]{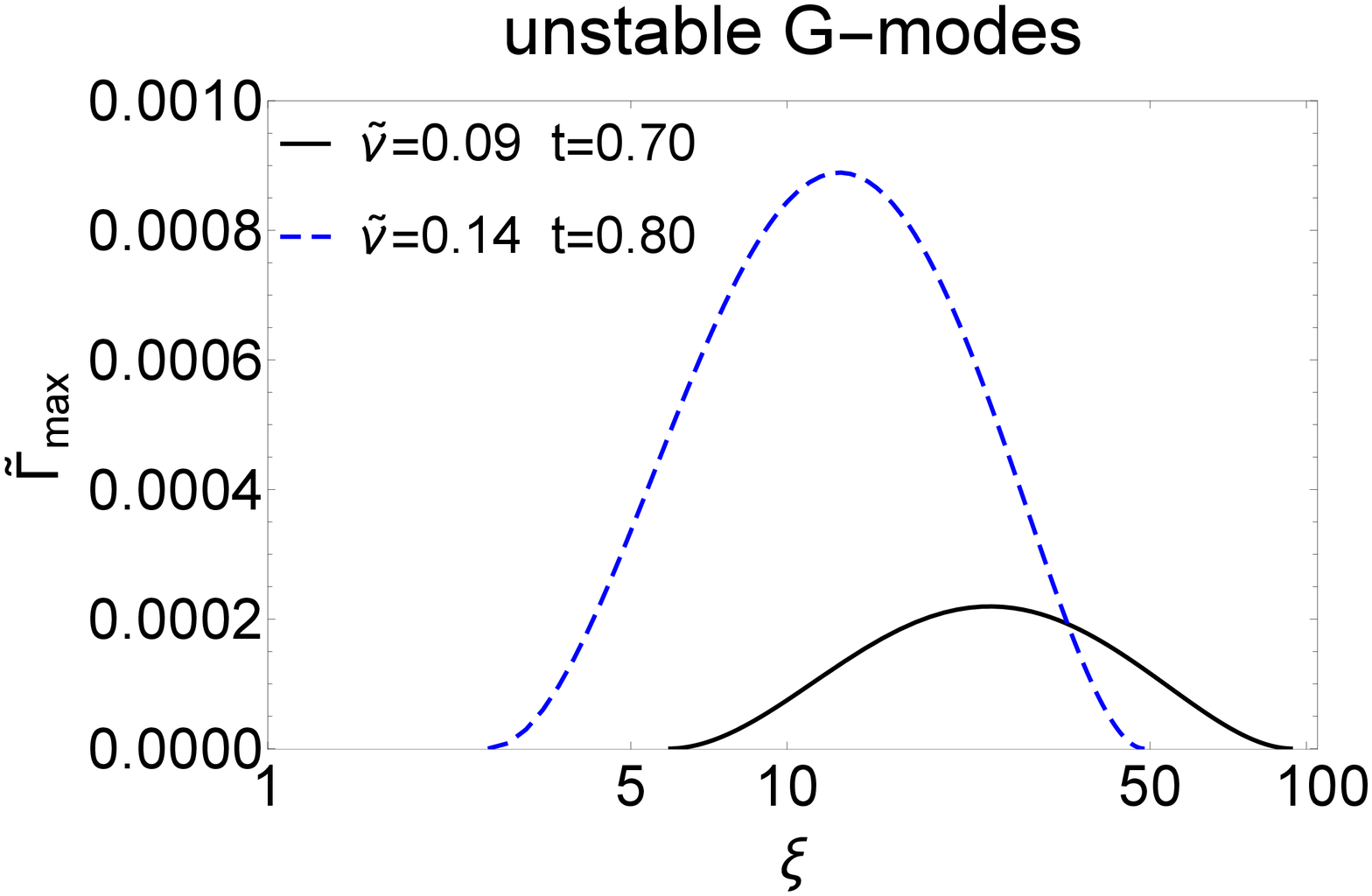}
\caption{The maximal growth rate ${\tilde \Gamma}_{max}\equiv  \Gamma_{max}/m_D$ for the unstable $G$-modes as a function of $\ut$ (left), $t$ (middle) and $\xi$ (right).}
\label{ggamax}
\end{center}
\end{figure}

In Fig.~\ref{ggamax}, we show the behavior the maximal growth rate $\Gamma_{max}$ for the unstable $G$-mode. Comparing with Fig.~\ref{agamax}, a non-monotonic dependence on $\ut$ is observed here, which is in accordance with the $\ut$-dependence of $\km$. Finally, as a function of $t$ and $\xi$, $\Gamma_{max}$ behaves similarly for the two different kinds of unstable modes. 

\section{Conclusions}\label{con}

In this paper, we have studied the collective modes of an anisotropic QCD plasma with a BGK collision kernel describing the equilibration of the plasma. This is an extension of a previous work where only a single propagation direction of the collective modes was considered. 
By solving the linearized kinetic equations of the Boltzmann-Vlasov type, we rederived the gluon self-energy. It turned out that the BGK collision kernel could break the symmetry of the gluon self-energy, namely, $\Pi^{\mu\nu}(K)\neq\Pi^{\nu\mu}(K)$ if the plasma has an anisotropic hard particle distribution in momentum space, $\xi\neq 0$. As a result, one needs a more general tensor basis to decompose the gluon self-energy for such an anisotropic system, which leads to five structure functions in the decomposition as shown in Eq.~(\ref{sede}). This corrects the previously obtained result~\cite{Schenke:2006xu} where the gluon self-energy was taken to be symmetric in the Lorentz indices. It should be noted that in Ref.~\cite{Schenke:2006xu} the wave vector ${\bf k}$ was assumed to be directed along the anisotropy direction  ${\bf n}$, i.e., $t=1$ and in this special case, the gluon self-energy obtained therein coincides with Eq.~(\ref{sede}) where only two structure functions $\alpha$ and $\beta$ are relevant. Consequently, the resulting collective modes with this specified propagation direction are indeed correct, however, for $t \neq 1$, five independent structure functions are needed.

We further determined the resummed gluon propagator, based on which we analyzed the collective modes by finding the corresponding poles. Dispersion relations for both the stable and unstable modes were presented. With the BGK collision kernel, the solutions $\omega(k)$ become complex valued for the three stable modes. It was found that ${\rm Im}(\omega)$ equals to $-\nu/2$ as $k\rightarrow 0$ and for the $G_-$-mode, it eventually drops to $-\nu$ at some finite $k$ where this mode terminates. The value of $k$ at the termination points turned out to be sensitive to the propagation angle, however, they also had a nontrivial dependence on the collision rate $\nu$ and the anisotropy parameter $\xi$.
In addition, the $k$-dependence of ${\rm Re}(\omega)$ seemed very similar as that shown in Ref.~\cite{Schenke:2006xu} which was actually consistent with our finding that ${\rm Re}(\omega)$ had a very weak dependence on the propagation angle. Here, the most notable behavior was the spacelike dispersion in the $G_-$-mode at large $k$. Our numerical results suggested that the ratio ${\rm Re}(\omega_-)/k$ at the termination point deviated from the light cone more significantly if the propagation is parallel to the anisotropy direction ${\bf n}$. It was also found that such a deviation could increase with increasing $\nu$ at a given anisotropy $\xi$.

Our results also confirmed that when propagating along the direction of anisotropy, the corresponding modes would become most unstable even in the collisional case. In other words, the largest possible value of the maximal growth rate $\Gamma_{max}$ appeared at $t=1$. On the other hand, different from what can be expected intuitively, namely that increasing the collision rate slows down the rate of growth of the unstable modes, we found that the maximal growth rate was not always a monotonically decreasing function of the collision rate. Exceptions were found in the unstable $G$-mode in the small $\nu$ region. Especially, for certain
propagation angles, the unstable G-mode which is absent in the collisionless limit can be activated by introducing a
small collision rate. We also carried out a systematic discussion of the parameter dependences of two key quantities associated with the unstable modes, namely $\km$ and $\Gamma_{max}$ which were introduced in Sec.~\ref{unstable}. The former could be used to determine whether or not the unstable modes exist, while the latter described the degree of the instability. In general, no unstable mode could survive when the collision rate $\nu>0.237 \, m_D$ and the elimination of the unstable modes occurs at both small and large anisotropies, depending on the values of $\nu$ and $t$. Furthermore, for any non-zero and finite $\xi$, the unstable $\alpha$-mode could exist for any propagation angle provided that the collision rate is small enough.  On the contrary, large angle propagation of the unstable $G$-mode was prohibited regardless of the value of $\nu$ consider herein. 

We would also like to mention that in equilibrium, with a specified temperature, other thermodynamic properties of the system such as the number density or energy density are uniquely determined. However, this is no longer true in an anisotropic system when the hard momentum scale $\lambda$ in the distribution function is specified because both the number density and energy density will change with the anisotropy parameter $\xi$. Thus, it is also possible to study the collective modes in an anisotropic QCD plasma by holding
either the number density or energy density constant. For example, as one changes the anisotropy parameter $\xi$, the energy density can be held constant which equals to that in the thermal equilibrium as long as we adjust the hard momentum scale as $\lambda\rightarrow  \lambda (f_0(\xi)/2)^{-1/4}$. Here, $f_0(\xi)$ is given by Eq.~(\ref{f0}). On the other hand, such an adjustment only leads to a simple modification on the Debye mass defined in Eq.~(\ref{md}), namely, $m_D \rightarrow m_D (f_0(\xi)/2)^{-1/4}$. As a result, it is trivial to compute the corresponding structure functions as well as the mass scales because the only change is an extra overall factor $(f_0(\xi)/2)^{-1/2}$ in Eqs.~(\ref{al})-(\ref{rh}) and (\ref{ma})-(\ref{mr}). With fixed energy density, the analysis of the collective modes doesn't involve anything new technically and the resulting collective modes have no qualitative difference as compared with those present in Secs.~\ref{stable} and \ref{unstable} where we hold the hard momentum scale fixed while varying the anisotropy parameter $\xi$. Some examples of results obtained with fixed energy density are given in App.~\ref{nor}.

As is well known, the appearance of the unstable modes is the most distinctive feature of an anisotropic QCD plasma. On the other hand, including a BGK collision kernel further complicates the gluonic collective modes. Our preliminary results indicate that there are some previously undiscovered behaviors which could emerge at large collision rate. For example, for a given anisotropy $\xi$, the complex solutions for $\omega$ as given in Eq.~(\ref{wk0}) cannot exist when the collision rate is large enough. Instead, pure imaginary solutions appear. It is certainly important to investigate these new modes and their possible physical implications. Another interesting path forward for future work is to assess the effects of the collision kernel on some other physical observables, such as the in-medium properties of the heavy quarkonia. These studies are currently in progress.

\section*{Acknowledgements}
We thank Stanis\l{}aw Mr\'{o}wczy\'{n}ski for useful discussions. The work of Y.G. is supported by the NSFC of China under Project No. 12065004 and by the Central Government 
Guidance Funds for Local Scientific and Technological 
Development, China (No. Guike ZY22096024). M.S. was supported by the U.S. Department of Energy, Office of Science, Office of Nuclear Physics Award No.~DE-SC0013470. 

\appendix

\section{Structure functions and mass scales in some limiting cases}
\label{anare}

In this appendix, we list analytical expressions for the structure functions and corresponding mass scales in some limiting cases. As already discussed in previous works, in the collisionless limit, only the four mass scales ($m_\delta^2$ becomes identical to $m_\rho^2$ in this limit) can be obtained analytically and the corresponding results can be found in Refs.~\cite{Romatschke:2003ms,Dumitru:2007hy}. With the collision kernel, when $\xi$ is small, up to linear order in $\xi$, the five structure functions read
\ba
\alpha(\xi,\oh,t,\uh)&=&\frac{\hat{\omega}}{4} \big[2z+(z^2-1)\ln\frac{z - 1}{z + 1}\big]+\frac{\xi}{24}\big[ 4t^2(3z^2-2)+2 \hat{\omega} z (3z^2-15 t^2 z^2+ 13t^2-5) \nonumber \\&+&3 (z^{2}-1  ) ( \hat{\omega}z^{2} + \hat{\omega} t^{2}-\hat{\omega} +2z t^2-5\hat{\omega} t^2 z^2){\ln\frac{z - 1}{z + 1}} \big]\, ,\nonumber\\
\beta(\xi,\oh,t,\uh)&=&- \frac{{\hat{\omega}}^{2} }{2\mathcal{W}(\hat \omega,\hat\nu )}\big(2+z {\ln\frac{z-1}{z+1}}\big)+\frac{{\xi \hat{\omega}}^{2} }{6 \mathcal{W}(\hat \omega,\hat\nu )} \big[4-6z^2+6t^2(3z^2-1)\nonumber \\
&+&3 z (1-z^2-2 t^2+3z^2 t^2) {\ln\frac{z-1}{z+1}}\big]\, ,\nonumber\\
\gamma(\xi,\oh,t,\uh)&=&
\frac{\xi}{12}(1 -t^{2}  ) ( 1 - \hat{\omega}z  )\big[4 - 6z^{2}- 3z (z^{2}-1 ){\ln\frac{z - 1}{z + 1}}\big]\, ,\nonumber\\
\delta(\xi,\oh,t,\uh)&=&\frac{\xi}{12} \hat{\omega} t \big[22z - 24z^{3}- 3 ( 1 - 5z^{2} + 4z^{4}  )\ln \frac{z-1}{z+1}\big]\, ,\nonumber\\
\rho(\xi,\oh,t,\uh)&=&\frac{\xi}{12 \mathcal{W}(\hat \omega,\hat\nu )} \hat{\omega} t \big[18 z +\hat{\omega} (4-24 z^2)- 3 ( 1-2 \hat{\omega} z -3 z^2 +4\hat{\omega} z^3 )\ln \frac{z-1}{z+1}\big]\,.
\ea
The above results, sans $\rho$, were considered first in Ref.~\cite{Kumar:2017bja} where different normalization of the distribution function was used and the gluon self-energy was decomposed using a symmetric tensor basis. 

The corresponding mass scales in the small $\xi$ limit are given by
\ba
m_{\alpha}^2(\xi,t,\hat\nu)&=&-\frac{\xi}{6}t^2\big[-1 + 3( 1 + {\hat\nu}^{2} )\mathcal{W}(0,\hat\nu )\big]\, ,\nonumber\\
m_\beta^2(\xi,t,\uh)&=&1 - \xi \big[1 - 2t^{2} + ( 1 - 3t^{2} )({\hat\nu}^{2}-\frac{1}{3\mathcal{W}(0,\hat\nu)})\big]\,,\nonumber\\
m_{\gamma}^2(\xi,t,\hat\nu)&=&\frac{\xi}{6}( 1 - t^{2} )\big[-1 + 3( 1 + {\hat\nu}^{2} )\mathcal{W}(0,\hat\nu )\big]\, ,\nonumber\\
m_\delta^2(\xi,t,\uh)&=&-\frac{\xi}{6}\frac{t}{\uh} \sqrt{1 - t^2}\big[3+4\hat\nu^2 - 3( 1 + {\hat\nu}^{2} )( 1 + 4{\hat\nu}^{2} )\mathcal{W}(0,\hat\nu)\big]\, ,\nonumber\\
m_\rho^2(\xi,t,\uh)&=&\frac{\xi}{2}\frac{t}{\uh} \sqrt{1 - t^2}\big[1+3\hat\nu^2 - \frac{1}{\mathcal{W}(0,\hat\nu)}\big]\, .
\ea

On the other hand, analytical expressions can be also obtained when we consider a special propagation direction which is parallel or perpendicular to the direction of anisotropy. When the momentum ${\bf {k}}$ is parallel to ${\bf {n}}$, i.e., $t\rightarrow 1$, we find

\ba
\alpha(\xi,\oh,\uh)&=& \frac{1}{4(1+\xi z^{2})^{2}}\Big[(\hat{\omega} z-1)(1+\xi z^{2})+(z^{2}-1)(\xi z+\hat{\omega}) \ln \frac{z-1}{z+1} \nonumber
\\&+&\big((\hat{\omega} z+1)(1+3\xi-\xi z^2+\xi^2 z^2)-4\xi(1-z^2)\big) \frac{\arctan \sqrt{\xi}}{\sqrt{\xi}}\Big]\, ,\nonumber \\
\beta(\xi,\oh,\uh)&=& \frac{\hat{\omega}^2}{2 \mathcal{W}(\hat \omega,\hat\nu )(1+\xi z^{2})^{2}}\Big[-(1+\xi z^{2})-z (1+\xi) \ln \frac{z-1}{z+1}+(1+\xi)(\xi z^2-1) \frac{\arctan \sqrt{\xi}}{\sqrt{\xi}}\Big]\, ,\nonumber \\
\delta(\xi,\oh,\uh)&=& \frac{\hat{\omega}}{4(1+\xi z^{2})^{3}}\Big[(z+4\xi z-3\xi z^3)(1+\xi z^{2})+\xi (z^{2}-1)\big(1-(4+3\xi)z^2\big) \ln \frac{z-1}{z+1} \nonumber
\\&-&\big(z-3\xi z(1+2\xi)(1-2z^2)+\xi^2z^3(2\xi-3z^2-\xi z^2)\big) \frac{\arctan \sqrt{\xi}}{\sqrt{\xi}}\Big]\,  ,\nonumber \\
\rho(\xi,\oh,\uh)&=& \frac{\hat{\omega}}{4 \mathcal{W}(\hat \omega,\hat\nu )(1+\xi z^{2})^{3}}\Big[(\hat{\omega}+4\xi z-3\hat{\omega} \xi z^2)(1+\xi z^{2})-\xi\big(1-2\hat{\omega} z-3(1+\xi)z^2+2\hat{\omega}(\xi+2)z^3+\xi z^4\big) \nonumber
\\&\times& \ln \frac{z-1}{z+1}+\big(\hat{\omega}(\xi-1)+2\xi(1+3\xi)z-6\hat{\omega}\xi(1+\xi)z^2+\xi^2(\xi+3)z^3(\hat{\omega}z-2)\big) \frac{\arctan \sqrt{\xi}}{\sqrt{\xi}}\Big]\,.
\ea
Notice that $\gamma(\xi,\oh,\uh)$ vanishes as $t\rightarrow 1$. Although $\delta(\xi,\oh,\uh)$ and $\rho(\xi,\oh,\uh)$ have a non-vanishing contribution, only $\tilde{n}^2 \delta \rho$ matters in our discussions according to Eq.~(\ref{defcm}). However, $\tilde{n}^2=0$ in this limit. These results, except $\rho(\xi,\oh,\ut)$, can be also found in Ref.~\cite{Schenke:2006xu}. Here, a typo in $\alpha(\xi,\oh,\uh)$ obtained in the previous work has been corrected. Accordingly, we can obtain the mass scales in this limit, which read
\ba
m_{\alpha}^2(\xi,\uh)&=&\frac{1}{4(1 -\xi{\hat\nu}^{2})^{2}}\Big[\xi - 1 + \xi (1 +3 {\hat\nu}^{2} )-2\xi (1 + {\hat\nu}^{2} )\mathcal{W}(0,\hat\nu ) + \big( 1 - \xi ( 1 + 3{\hat\nu}^{2} ) - {\xi}^{2} {\hat\nu}^{2} \big)\frac{\arctan\sqrt{\xi}}{\sqrt{ \xi}} \Big]\nonumber\, ,\\
m_{\beta}^2(\xi,\uh)&=&\frac{1}{2(1 -\xi{\hat\nu}^{2})^{2}\mathcal{W}(0,\hat\nu )}\Big[2(1 + \xi)\mathcal{W}(0,\hat\nu )-2\xi-(1+ \xi{\hat\nu}^{2}) + (1 + \xi)(1 + \xi{\hat\nu}^{2})\frac{\arctan\sqrt{\xi}}{\sqrt{\xi}}\Big]\, .
\ea
Here, $m_{\delta}^2(\xi,\uh)$ and $m_{\rho}^2(\xi,\uh)$ vanish as $t\rightarrow 1$ because of a factor $\sqrt{1-t}$ in their definition.

Finally, we consider $t\rightarrow 0$ where the propagation direction is perpendicular to ${\bf {n}}$. In this limit, $\delta(\xi,\oh,t,\uh)$ and $\rho(\xi,\oh,t,\uh)$ vanish and the other three structure functions take the following forms
\ba
\alpha(\xi,\oh,\uh)&=& \frac{\hat{\omega}}{2}\Big[z \frac{\arctan \sqrt{\xi}}{\sqrt{\xi}}-i\frac{1-z^2}{\sqrt{1+\xi-\xi z^2}}\arctan (\sqrt{\xi-(1+\xi)/z^2})\Big]\, ,\nonumber \\
\beta(\xi,\oh,\uh)&=& \frac{-\hat{\omega}^2}{2 \mathcal{W}(\hat \omega,\hat\nu )}\Big[\frac{1}{1 + \xi - \xi z^{2}}+\frac{\arctan \sqrt{\xi}}{\sqrt{\xi}}+i\frac{z(2+\xi-\xi z^2)}{(1 + \xi - \xi z^{2})^{\frac{3}{2}}}\arctan (\sqrt{\xi-(1+\xi)/z^2})\Big]\, ,\nonumber \\
\gamma(\xi,\oh,\uh)&=& \frac{\hat{\omega}z-1}{2}\Big[\frac{1}{1 + \xi - \xi z^{2}}-\frac{\arctan \sqrt{\xi}}{\sqrt{\xi}}-i\frac{\xi z(1-z^2)}{(1 + \xi - \xi z^{2})^{\frac{3}{2}}}\arctan (\sqrt{\xi-(1+\xi)/z^2})\Big]\, .
\ea
Then, it is easy to write down the non-vanishing mass scales in this limit, which read
\ba
m_{\beta}^2(\xi,\uh)&=&\frac{1}{2\mathcal{W}(0,\hat\nu )}\Big[\frac{1}{1 + \xi + \xi {\hat\nu}^{2}}+\frac{\arctan\sqrt{\xi}}{\sqrt{\xi}}-\frac{\uh(2 + \xi + \xi {\hat\nu}^{2})\arctan(\sqrt {\xi+(1 + \xi)/{\hat\nu}^{2}})}{(1 + \xi + \xi {\hat\nu}^{2})^{3/2}}\Big]\, ,\nonumber\\
m_{\gamma}^2(\xi,\uh)&=&\frac{1}{2}\Big[-\frac{1}{1 + \xi + \xi {\hat\nu}^{2}}+\frac{\arctan\sqrt{\xi}}{\sqrt{\xi}}- \frac{\xi\hat\nu (1+{\hat\nu}^{2})\arctan(\sqrt {\xi+(1 + \xi)/{\hat\nu}^{2}})}{(1 + \xi + \xi {\hat\nu}^{2})^{3/2}} \Big]\, .
\ea

\section{Some results obtained when fixing the energy density in the anisotropic QCD plasma}
\label{nor}

\begin{figure}[htbp]
\begin{center}
\includegraphics[width=0.32\textwidth]{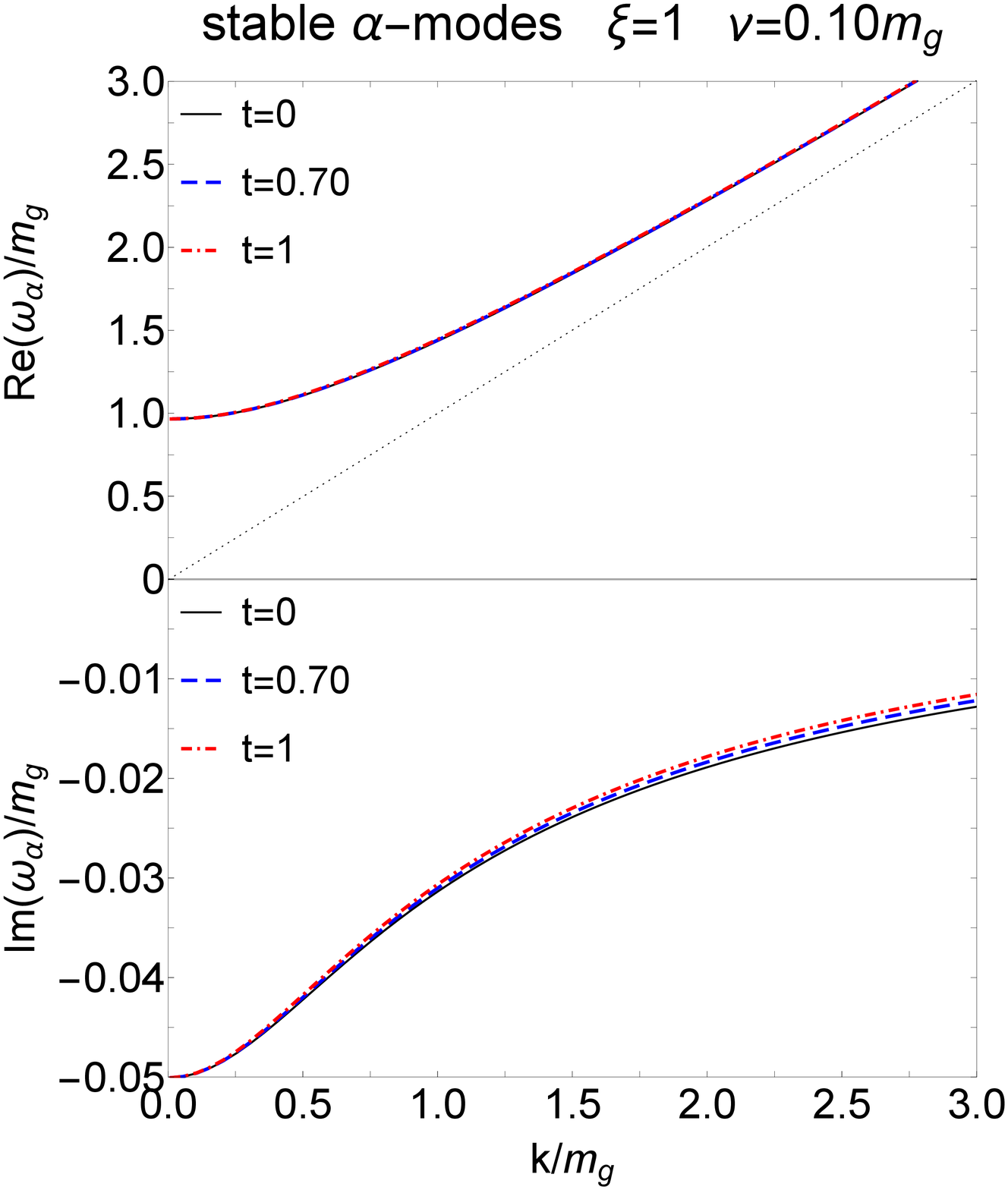}
\includegraphics[width=0.32\textwidth]{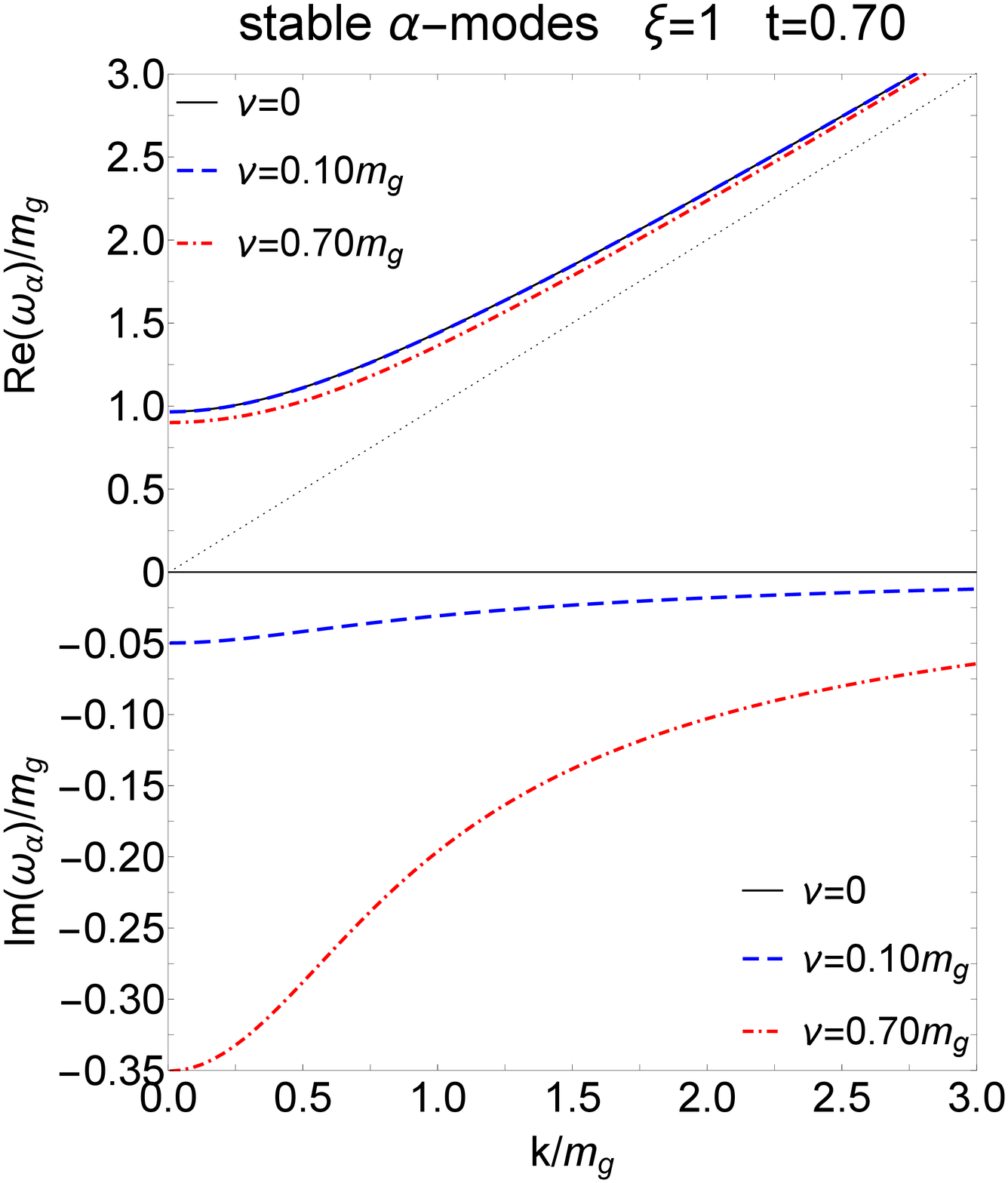}
\includegraphics[width=0.32\textwidth]{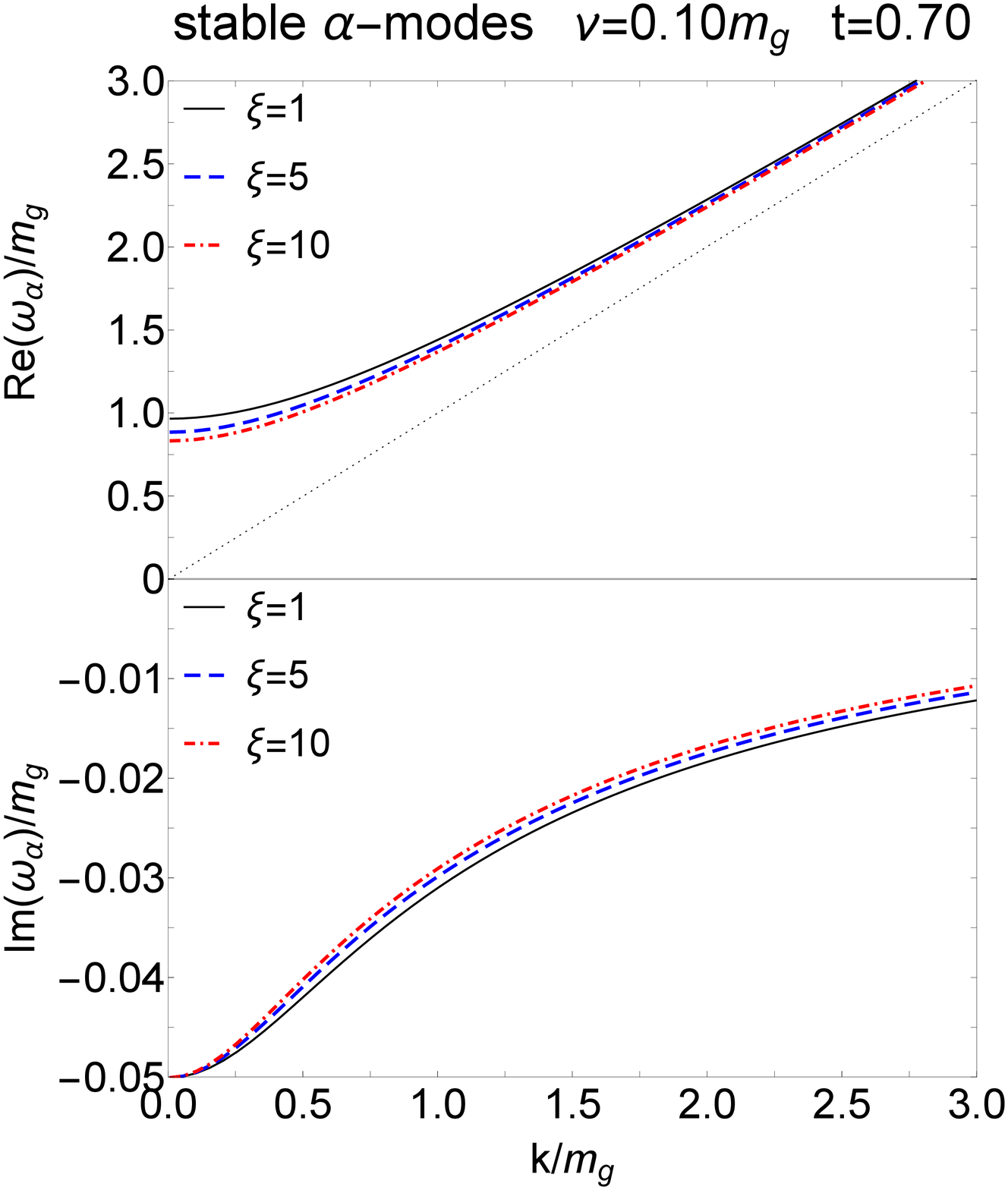}
\caption{Parameter dependence of the dispersion relations of the stable $\alpha$-modes where we fix the energy density in the anisotropic QCD plasma. Left: the $t$-dependence at fixed $\xi$ and $\nu$. Middle: the $\nu$-dependence at fixed $\xi$ and $t$. Right: the $\xi$-dependence at fixed $\nu$ and $t$. In these plots, $m_g=m_D/\sqrt{3}$.}
\label{norsare}
\end{center}
\end{figure}

\begin{figure}[htbp]
\begin{center}
\includegraphics[width=0.32\textwidth]{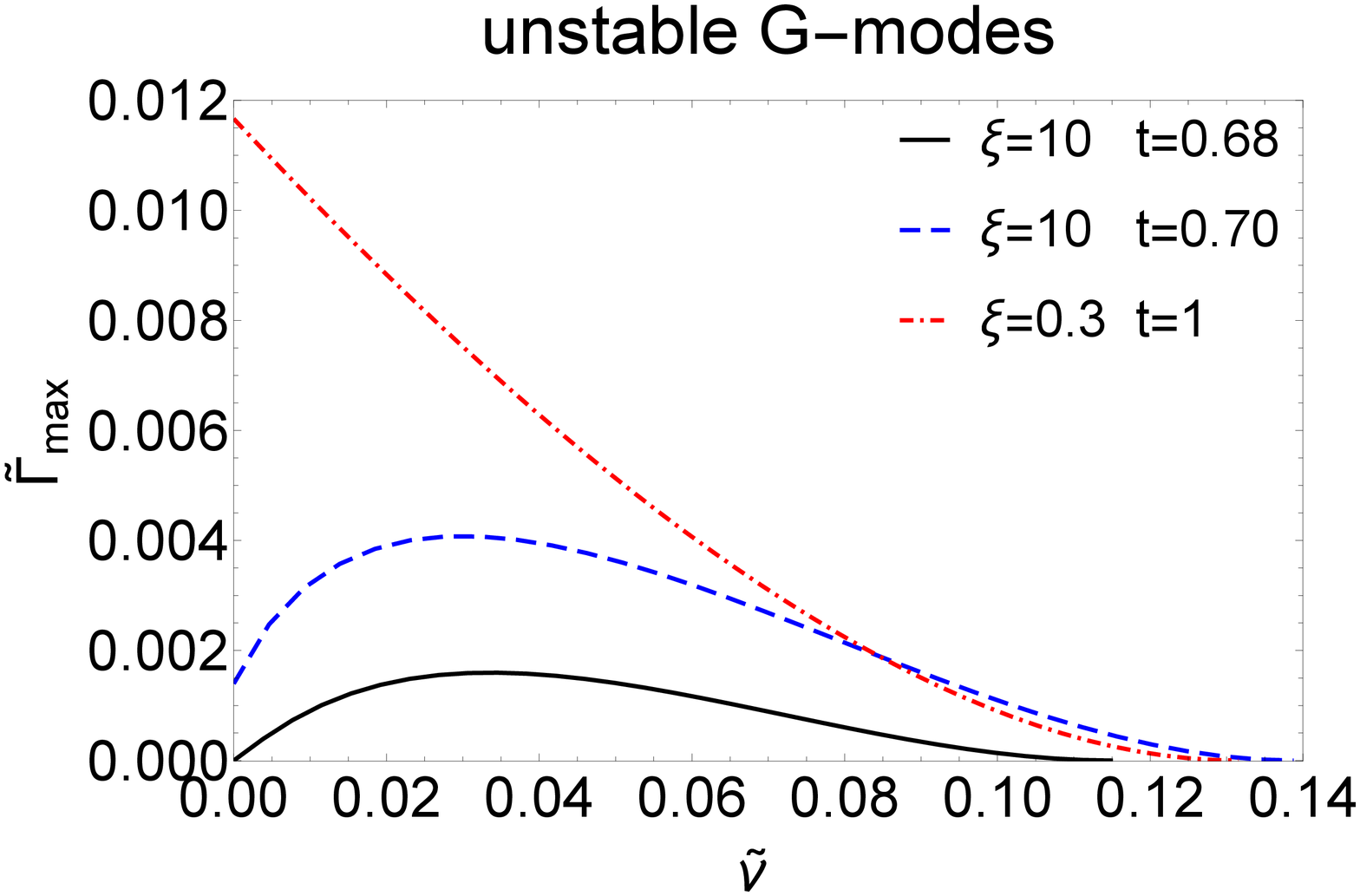}
\includegraphics[width=0.32\textwidth]{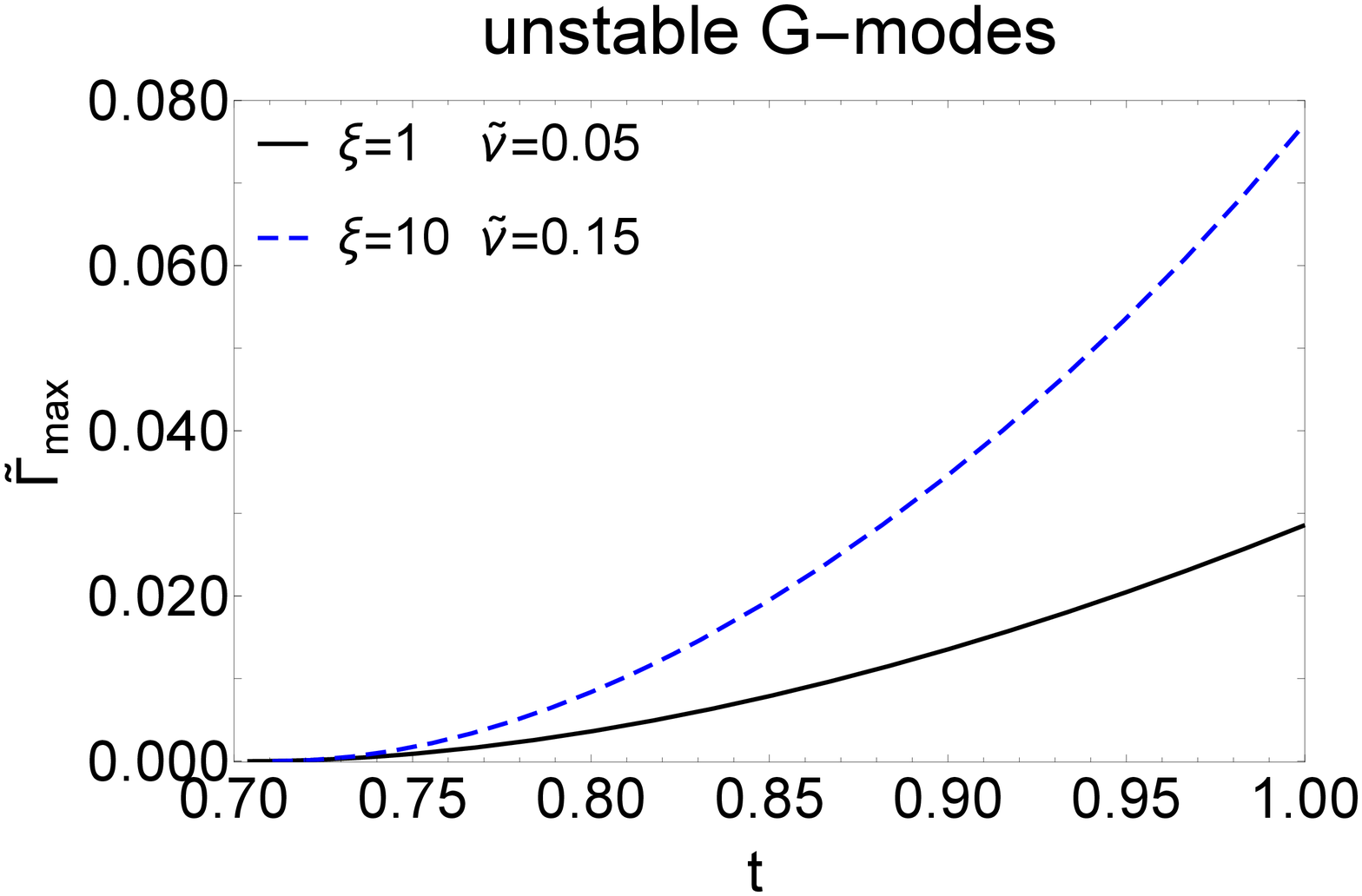}
\includegraphics[width=0.32\textwidth]{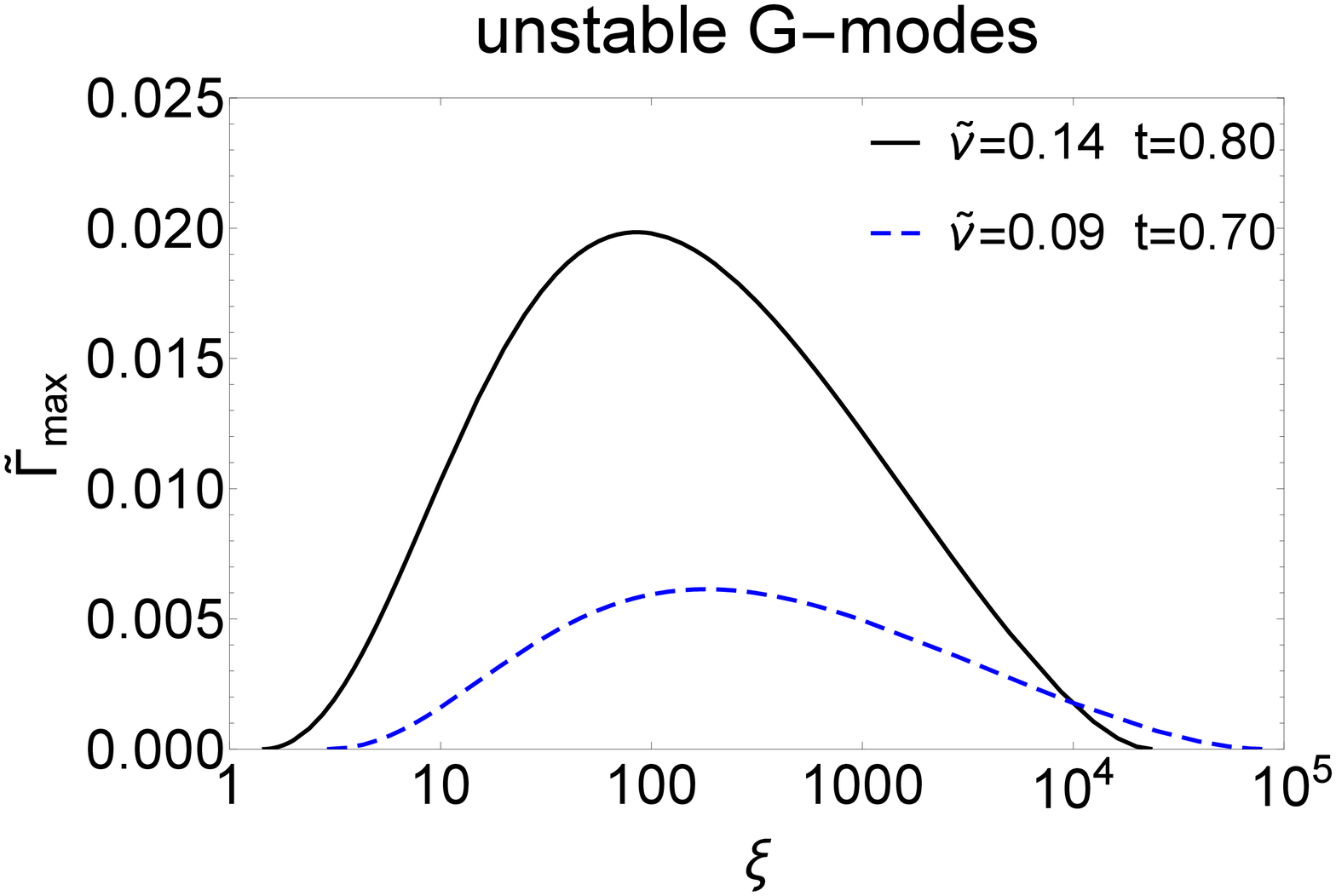}
\caption{The maximal growth rate ${\tilde \Gamma}_{max}$ for the unstable $G$-modes as a function of $\ut$ (left), $t$ (middle) and $\xi$ (right) obtained when holding the energy density fixed. }
\label{norggamax}
\end{center}
\end{figure}

In this appendix, we present some example results obtained when one requires that the energy density is held fixed as $\xi$ is varied.  Our results are shown in Figs.~\ref{norsare} and \ref{norggamax}.  As can be seen from these figures, the qualitative behavior of the stable and unstable modes is the same as in the unnormalized case presented in the main body of the text.

\bibliography{paper}

\begin{thebibliography}{24}%
\makeatletter
\providecommand \@ifxundefined [1]{%
 \@ifx{#1\undefined}
}%
\providecommand \@ifnum [1]{%
 \ifnum #1\expandafter \@firstoftwo
 \else \expandafter \@secondoftwo
 \fi
}%
\providecommand \@ifx [1]{%
 \ifx #1\expandafter \@firstoftwo
 \else \expandafter \@secondoftwo
 \fi
}%
\providecommand \natexlab [1]{#1}%
\providecommand \enquote  [1]{``#1''}%
\providecommand \bibnamefont  [1]{#1}%
\providecommand \bibfnamefont [1]{#1}%
\providecommand \citenamefont [1]{#1}%
\providecommand \href@noop [0]{\@secondoftwo}%
\providecommand \href [0]{\begingroup \@sanitize@url \@href}%
\providecommand \@href[1]{\@@startlink{#1}\@@href}%
\providecommand \@@href[1]{\endgroup#1\@@endlink}%
\providecommand \@sanitize@url [0]{\catcode `\\12\catcode `\$12\catcode
  `\&12\catcode `\#12\catcode `\^12\catcode `\_12\catcode `\%12\relax}%
\providecommand \@@startlink[1]{}%
\providecommand \@@endlink[0]{}%
\providecommand \url  [0]{\begingroup\@sanitize@url \@url }%
\providecommand \@url [1]{\endgroup\@href {#1}{\urlprefix }}%
\providecommand \urlprefix  [0]{URL }%
\providecommand \Eprint [0]{\href }%
\providecommand \doibase [0]{https://doi.org/}%
\providecommand \selectlanguage [0]{\@gobble}%
\providecommand \bibinfo  [0]{\@secondoftwo}%
\providecommand \bibfield  [0]{\@secondoftwo}%
\providecommand \translation [1]{[#1]}%
\providecommand \BibitemOpen [0]{}%
\providecommand \bibitemStop [0]{}%
\providecommand \bibitemNoStop [0]{.\EOS\space}%
\providecommand \EOS [0]{\spacefactor3000\relax}%
\providecommand \BibitemShut  [1]{\csname bibitem#1\endcsname}%
\let\auto@bib@innerbib\@empty
\bibitem [{\citenamefont {Weldon}(1982)}]{Weldon:1982aq}%
  \BibitemOpen
  \bibfield  {author} {\bibinfo {author} {\bibfnamefont {H.~A.}\ \bibnamefont
  {Weldon}},\ }\bibfield  {title} {\bibinfo {title} {{Covariant Calculations at
  Finite Temperature: The Relativistic Plasma}},\ }\href
  {https://doi.org/10.1103/PhysRevD.26.1394} {\bibfield  {journal} {\bibinfo
  {journal} {Phys. Rev. D}\ }\textbf {\bibinfo {volume} {26}},\ \bibinfo
  {pages} {1394} (\bibinfo {year} {1982})}\BibitemShut {NoStop}%
\bibitem [{\citenamefont {Braaten}\ and\ \citenamefont
  {Pisarski}(1990{\natexlab{a}})}]{Braaten:1989mz}%
  \BibitemOpen
  \bibfield  {author} {\bibinfo {author} {\bibfnamefont {E.}~\bibnamefont
  {Braaten}}\ and\ \bibinfo {author} {\bibfnamefont {R.~D.}\ \bibnamefont
  {Pisarski}},\ }\bibfield  {title} {\bibinfo {title} {{Soft Amplitudes in Hot
  Gauge Theories: A General Analysis}},\ }\href
  {https://doi.org/10.1016/0550-3213(90)90508-B} {\bibfield  {journal}
  {\bibinfo  {journal} {Nucl. Phys. B}\ }\textbf {\bibinfo {volume} {337}},\
  \bibinfo {pages} {569} (\bibinfo {year} {1990}{\natexlab{a}})}\BibitemShut
  {NoStop}%
\bibitem [{\citenamefont {Braaten}\ and\ \citenamefont
  {Pisarski}(1990{\natexlab{b}})}]{Braaten:1989kk}%
  \BibitemOpen
  \bibfield  {author} {\bibinfo {author} {\bibfnamefont {E.}~\bibnamefont
  {Braaten}}\ and\ \bibinfo {author} {\bibfnamefont {R.~D.}\ \bibnamefont
  {Pisarski}},\ }\bibfield  {title} {\bibinfo {title} {{Resummation and Gauge
  Invariance of the Gluon Damping Rate in Hot QCD}},\ }\href
  {https://doi.org/10.1103/PhysRevLett.64.1338} {\bibfield  {journal} {\bibinfo
   {journal} {Phys. Rev. Lett.}\ }\textbf {\bibinfo {volume} {64}},\ \bibinfo
  {pages} {1338} (\bibinfo {year} {1990}{\natexlab{b}})}\BibitemShut {NoStop}%
\bibitem [{\citenamefont {Frenkel}\ and\ \citenamefont
  {Taylor}(1990)}]{Frenkel:1989br}%
  \BibitemOpen
  \bibfield  {author} {\bibinfo {author} {\bibfnamefont {J.}~\bibnamefont
  {Frenkel}}\ and\ \bibinfo {author} {\bibfnamefont {J.~C.}\ \bibnamefont
  {Taylor}},\ }\bibfield  {title} {\bibinfo {title} {{High Temperature Limit of
  Thermal QCD}},\ }\href {https://doi.org/10.1016/0550-3213(90)90661-V}
  {\bibfield  {journal} {\bibinfo  {journal} {Nucl. Phys. B}\ }\textbf
  {\bibinfo {volume} {334}},\ \bibinfo {pages} {199} (\bibinfo {year}
  {1990})}\BibitemShut {NoStop}%
\bibitem [{\citenamefont {Braaten}\ and\ \citenamefont
  {Pisarski}(1992)}]{Braaten:1991gm}%
  \BibitemOpen
  \bibfield  {author} {\bibinfo {author} {\bibfnamefont {E.}~\bibnamefont
  {Braaten}}\ and\ \bibinfo {author} {\bibfnamefont {R.~D.}\ \bibnamefont
  {Pisarski}},\ }\bibfield  {title} {\bibinfo {title} {{Simple effective
  Lagrangian for hard thermal loops}},\ }\href
  {https://doi.org/10.1103/PhysRevD.45.R1827} {\bibfield  {journal} {\bibinfo
  {journal} {Phys. Rev. D}\ }\textbf {\bibinfo {volume} {45}},\ \bibinfo
  {pages} {R1827} (\bibinfo {year} {1992})}\BibitemShut {NoStop}%
\bibitem [{\citenamefont {Pisarski}(1989)}]{Pisarski:1988vb}%
  \BibitemOpen
  \bibfield  {author} {\bibinfo {author} {\bibfnamefont {R.~D.}\ \bibnamefont
  {Pisarski}},\ }\bibfield  {title} {\bibinfo {title} {{How to Compute
  Scattering Amplitudes in Hot Gauge Theories}},\ }\href
  {https://doi.org/10.1016/0378-4371(89)90525-6} {\bibfield  {journal}
  {\bibinfo  {journal} {Physica A}\ }\textbf {\bibinfo {volume} {158}},\
  \bibinfo {pages} {246} (\bibinfo {year} {1989})}\BibitemShut {NoStop}%
\bibitem [{\citenamefont {Ghiglieri}\ \emph {et~al.}(2020)\citenamefont
  {Ghiglieri}, \citenamefont {Kurkela}, \citenamefont {Strickland},\ and\
  \citenamefont {Vuorinen}}]{Ghiglieri:2020dpq}%
  \BibitemOpen
  \bibfield  {author} {\bibinfo {author} {\bibfnamefont {J.}~\bibnamefont
  {Ghiglieri}}, \bibinfo {author} {\bibfnamefont {A.}~\bibnamefont {Kurkela}},
  \bibinfo {author} {\bibfnamefont {M.}~\bibnamefont {Strickland}},\ and\
  \bibinfo {author} {\bibfnamefont {A.}~\bibnamefont {Vuorinen}},\ }\bibfield
  {title} {\bibinfo {title} {{Perturbative Thermal QCD: Formalism and
  Applications}},\ }\href {https://doi.org/10.1016/j.physrep.2020.07.004}
  {\bibfield  {journal} {\bibinfo  {journal} {Phys. Rept.}\ }\textbf {\bibinfo
  {volume} {880}},\ \bibinfo {pages} {1} (\bibinfo {year} {2020})},\ \Eprint
  {https://arxiv.org/abs/2002.10188} {arXiv:2002.10188 [hep-ph]} \BibitemShut
  {NoStop}%
\bibitem [{\citenamefont {Schenke}\ \emph {et~al.}(2006)\citenamefont
  {Schenke}, \citenamefont {Strickland}, \citenamefont {Greiner},\ and\
  \citenamefont {Thoma}}]{Schenke:2006xu}%
  \BibitemOpen
  \bibfield  {author} {\bibinfo {author} {\bibfnamefont {B.}~\bibnamefont
  {Schenke}}, \bibinfo {author} {\bibfnamefont {M.}~\bibnamefont {Strickland}},
  \bibinfo {author} {\bibfnamefont {C.}~\bibnamefont {Greiner}},\ and\ \bibinfo
  {author} {\bibfnamefont {M.~H.}\ \bibnamefont {Thoma}},\ }\bibfield  {title}
  {\bibinfo {title} {{A Model of the effect of collisions on QCD plasma
  instabilities}},\ }\href {https://doi.org/10.1103/PhysRevD.73.125004}
  {\bibfield  {journal} {\bibinfo  {journal} {Phys. Rev. D}\ }\textbf {\bibinfo
  {volume} {73}},\ \bibinfo {pages} {125004} (\bibinfo {year} {2006})},\
  \Eprint {https://arxiv.org/abs/hep-ph/0603029} {arXiv:hep-ph/0603029}
  \BibitemShut {NoStop}%
\bibitem [{\citenamefont {Arnold}\ \emph
  {et~al.}(2003{\natexlab{a}})\citenamefont {Arnold}, \citenamefont {Moore},\
  and\ \citenamefont {Yaffe}}]{Arnold:2002zm}%
  \BibitemOpen
  \bibfield  {author} {\bibinfo {author} {\bibfnamefont {P.~B.}\ \bibnamefont
  {Arnold}}, \bibinfo {author} {\bibfnamefont {G.~D.}\ \bibnamefont {Moore}},\
  and\ \bibinfo {author} {\bibfnamefont {L.~G.}\ \bibnamefont {Yaffe}},\
  }\bibfield  {title} {\bibinfo {title} {{Effective kinetic theory for high
  temperature gauge theories}},\ }\href
  {https://doi.org/10.1088/1126-6708/2003/01/030} {\bibfield  {journal}
  {\bibinfo  {journal} {JHEP}\ }\textbf {\bibinfo {volume} {01}},\ \bibinfo
  {pages} {030}},\ \Eprint {https://arxiv.org/abs/hep-ph/0209353}
  {arXiv:hep-ph/0209353} \BibitemShut {NoStop}%
\bibitem [{\citenamefont {Abraao~York}\ \emph {et~al.}(2014)\citenamefont
  {Abraao~York}, \citenamefont {Kurkela}, \citenamefont {Lu},\ and\
  \citenamefont {Moore}}]{AbraaoYork:2014hbk}%
  \BibitemOpen
  \bibfield  {author} {\bibinfo {author} {\bibfnamefont {M.~C.}\ \bibnamefont
  {Abraao~York}}, \bibinfo {author} {\bibfnamefont {A.}~\bibnamefont
  {Kurkela}}, \bibinfo {author} {\bibfnamefont {E.}~\bibnamefont {Lu}},\ and\
  \bibinfo {author} {\bibfnamefont {G.~D.}\ \bibnamefont {Moore}},\ }\bibfield
  {title} {\bibinfo {title} {{UV cascade in classical Yang-Mills theory via
  kinetic theory}},\ }\href {https://doi.org/10.1103/PhysRevD.89.074036}
  {\bibfield  {journal} {\bibinfo  {journal} {Phys. Rev. D}\ }\textbf {\bibinfo
  {volume} {89}},\ \bibinfo {pages} {074036} (\bibinfo {year} {2014})},\
  \Eprint {https://arxiv.org/abs/1401.3751} {arXiv:1401.3751 [hep-ph]}
  \BibitemShut {NoStop}%
\bibitem [{\citenamefont {Kurkela}\ and\ \citenamefont
  {Zhu}(2015)}]{Kurkela:2015qoa}%
  \BibitemOpen
  \bibfield  {author} {\bibinfo {author} {\bibfnamefont {A.}~\bibnamefont
  {Kurkela}}\ and\ \bibinfo {author} {\bibfnamefont {Y.}~\bibnamefont {Zhu}},\
  }\bibfield  {title} {\bibinfo {title} {{Isotropization and hydrodynamization
  in weakly coupled heavy-ion collisions}},\ }\href
  {https://doi.org/10.1103/PhysRevLett.115.182301} {\bibfield  {journal}
  {\bibinfo  {journal} {Phys. Rev. Lett.}\ }\textbf {\bibinfo {volume} {115}},\
  \bibinfo {pages} {182301} (\bibinfo {year} {2015})},\ \Eprint
  {https://arxiv.org/abs/1506.06647} {arXiv:1506.06647 [hep-ph]} \BibitemShut
  {NoStop}%
\bibitem [{\citenamefont {Mrowczynski}(1993)}]{Mrowczynski:1993qm}%
  \BibitemOpen
  \bibfield  {author} {\bibinfo {author} {\bibfnamefont {S.}~\bibnamefont
  {Mrowczynski}},\ }\bibfield  {title} {\bibinfo {title} {{Plasma instability
  at the initial stage of ultrarelativistic heavy ion collisions}},\ }\href
  {https://doi.org/10.1016/0370-2693(93)91330-P} {\bibfield  {journal}
  {\bibinfo  {journal} {Phys. Lett. B}\ }\textbf {\bibinfo {volume} {314}},\
  \bibinfo {pages} {118} (\bibinfo {year} {1993})}\BibitemShut {NoStop}%
\bibitem [{\citenamefont {Mrowczynski}(1997)}]{Mrowczynski:1996vh}%
  \BibitemOpen
  \bibfield  {author} {\bibinfo {author} {\bibfnamefont {S.}~\bibnamefont
  {Mrowczynski}},\ }\bibfield  {title} {\bibinfo {title} {{Color filamentation
  in ultrarelativistic heavy ion collisions}},\ }\href
  {https://doi.org/10.1016/S0370-2693(96)01621-8} {\bibfield  {journal}
  {\bibinfo  {journal} {Phys. Lett. B}\ }\textbf {\bibinfo {volume} {393}},\
  \bibinfo {pages} {26} (\bibinfo {year} {1997})},\ \Eprint
  {https://arxiv.org/abs/hep-ph/9606442} {arXiv:hep-ph/9606442} \BibitemShut
  {NoStop}%
\bibitem [{\citenamefont {Romatschke}\ and\ \citenamefont
  {Strickland}(2003)}]{Romatschke:2003ms}%
  \BibitemOpen
  \bibfield  {author} {\bibinfo {author} {\bibfnamefont {P.}~\bibnamefont
  {Romatschke}}\ and\ \bibinfo {author} {\bibfnamefont {M.}~\bibnamefont
  {Strickland}},\ }\bibfield  {title} {\bibinfo {title} {{Collective modes of
  an anisotropic quark gluon plasma}},\ }\href
  {https://doi.org/10.1103/PhysRevD.68.036004} {\bibfield  {journal} {\bibinfo
  {journal} {Phys. Rev. D}\ }\textbf {\bibinfo {volume} {68}},\ \bibinfo
  {pages} {036004} (\bibinfo {year} {2003})},\ \Eprint
  {https://arxiv.org/abs/hep-ph/0304092} {arXiv:hep-ph/0304092} \BibitemShut
  {NoStop}%
\bibitem [{\citenamefont {Arnold}\ \emph
  {et~al.}(2003{\natexlab{b}})\citenamefont {Arnold}, \citenamefont
  {Lenaghan},\ and\ \citenamefont {Moore}}]{Arnold:2003rq}%
  \BibitemOpen
  \bibfield  {author} {\bibinfo {author} {\bibfnamefont {P.~B.}\ \bibnamefont
  {Arnold}}, \bibinfo {author} {\bibfnamefont {J.}~\bibnamefont {Lenaghan}},\
  and\ \bibinfo {author} {\bibfnamefont {G.~D.}\ \bibnamefont {Moore}},\
  }\bibfield  {title} {\bibinfo {title} {{QCD plasma instabilities and bottom
  up thermalization}},\ }\href {https://doi.org/10.1088/1126-6708/2003/08/002}
  {\bibfield  {journal} {\bibinfo  {journal} {JHEP}\ }\textbf {\bibinfo
  {volume} {08}},\ \bibinfo {pages} {002}},\ \Eprint
  {https://arxiv.org/abs/hep-ph/0307325} {arXiv:hep-ph/0307325} \BibitemShut
  {NoStop}%
\bibitem [{\citenamefont {Romatschke}\ and\ \citenamefont
  {Strickland}(2004)}]{Romatschke:2004jh}%
  \BibitemOpen
  \bibfield  {author} {\bibinfo {author} {\bibfnamefont {P.}~\bibnamefont
  {Romatschke}}\ and\ \bibinfo {author} {\bibfnamefont {M.}~\bibnamefont
  {Strickland}},\ }\bibfield  {title} {\bibinfo {title} {{Collective modes of
  an anisotropic quark-gluon plasma II}},\ }\href
  {https://doi.org/10.1103/PhysRevD.70.116006} {\bibfield  {journal} {\bibinfo
  {journal} {Phys. Rev. D}\ }\textbf {\bibinfo {volume} {70}},\ \bibinfo
  {pages} {116006} (\bibinfo {year} {2004})},\ \Eprint
  {https://arxiv.org/abs/hep-ph/0406188} {arXiv:hep-ph/0406188} \BibitemShut
  {NoStop}%
\bibitem [{\citenamefont {Mrowczynski}\ \emph {et~al.}(2017)\citenamefont
  {Mrowczynski}, \citenamefont {Schenke},\ and\ \citenamefont
  {Strickland}}]{Mrowczynski:2016etf}%
  \BibitemOpen
  \bibfield  {author} {\bibinfo {author} {\bibfnamefont {S.}~\bibnamefont
  {Mrowczynski}}, \bibinfo {author} {\bibfnamefont {B.}~\bibnamefont
  {Schenke}},\ and\ \bibinfo {author} {\bibfnamefont {M.}~\bibnamefont
  {Strickland}},\ }\bibfield  {title} {\bibinfo {title} {{Color instabilities
  in the quark\textendash{}gluon plasma}},\ }\href
  {https://doi.org/10.1016/j.physrep.2017.03.003} {\bibfield  {journal}
  {\bibinfo  {journal} {Phys. Rept.}\ }\textbf {\bibinfo {volume} {682}},\
  \bibinfo {pages} {1} (\bibinfo {year} {2017})},\ \Eprint
  {https://arxiv.org/abs/1603.08946} {arXiv:1603.08946 [hep-ph]} \BibitemShut
  {NoStop}%
\bibitem [{\citenamefont {Burnier}\ \emph {et~al.}(2009)\citenamefont
  {Burnier}, \citenamefont {Laine},\ and\ \citenamefont
  {Vepsalainen}}]{Burnier:2009yu}%
  \BibitemOpen
  \bibfield  {author} {\bibinfo {author} {\bibfnamefont {Y.}~\bibnamefont
  {Burnier}}, \bibinfo {author} {\bibfnamefont {M.}~\bibnamefont {Laine}},\
  and\ \bibinfo {author} {\bibfnamefont {M.}~\bibnamefont {Vepsalainen}},\
  }\bibfield  {title} {\bibinfo {title} {{Quarkonium dissociation in the
  presence of a small momentum space anisotropy}},\ }\href
  {https://doi.org/10.1016/j.physletb.2009.05.067} {\bibfield  {journal}
  {\bibinfo  {journal} {Phys. Lett. B}\ }\textbf {\bibinfo {volume} {678}},\
  \bibinfo {pages} {86} (\bibinfo {year} {2009})},\ \Eprint
  {https://arxiv.org/abs/0903.3467} {arXiv:0903.3467 [hep-ph]} \BibitemShut
  {NoStop}%
\bibitem [{\citenamefont {Nopoush}\ \emph {et~al.}(2017)\citenamefont
  {Nopoush}, \citenamefont {Guo},\ and\ \citenamefont
  {Strickland}}]{Nopoush:2017zbu}%
  \BibitemOpen
  \bibfield  {author} {\bibinfo {author} {\bibfnamefont {M.}~\bibnamefont
  {Nopoush}}, \bibinfo {author} {\bibfnamefont {Y.}~\bibnamefont {Guo}},\ and\
  \bibinfo {author} {\bibfnamefont {M.}~\bibnamefont {Strickland}},\ }\bibfield
   {title} {\bibinfo {title} {{The static hard-loop gluon propagator to all
  orders in anisotropy}},\ }\href {https://doi.org/10.1007/JHEP09(2017)063}
  {\bibfield  {journal} {\bibinfo  {journal} {JHEP}\ }\textbf {\bibinfo
  {volume} {09}},\ \bibinfo {pages} {063}},\ \Eprint
  {https://arxiv.org/abs/1706.08091} {arXiv:1706.08091 [hep-ph]} \BibitemShut
  {NoStop}%
\bibitem [{\citenamefont {Dumitru}\ \emph {et~al.}(2008)\citenamefont
  {Dumitru}, \citenamefont {Guo},\ and\ \citenamefont
  {Strickland}}]{Dumitru:2007hy}%
  \BibitemOpen
  \bibfield  {author} {\bibinfo {author} {\bibfnamefont {A.}~\bibnamefont
  {Dumitru}}, \bibinfo {author} {\bibfnamefont {Y.}~\bibnamefont {Guo}},\ and\
  \bibinfo {author} {\bibfnamefont {M.}~\bibnamefont {Strickland}},\ }\bibfield
   {title} {\bibinfo {title} {{The Heavy-quark potential in an anisotropic
  (viscous) plasma}},\ }\href {https://doi.org/10.1016/j.physletb.2008.02.048}
  {\bibfield  {journal} {\bibinfo  {journal} {Phys. Lett. B}\ }\textbf
  {\bibinfo {volume} {662}},\ \bibinfo {pages} {37} (\bibinfo {year} {2008})},\
  \Eprint {https://arxiv.org/abs/0711.4722} {arXiv:0711.4722 [hep-ph]}
  \BibitemShut {NoStop}%
\bibitem [{\citenamefont {Dumitru}\ \emph {et~al.}(2009)\citenamefont
  {Dumitru}, \citenamefont {Guo},\ and\ \citenamefont
  {Strickland}}]{Dumitru:2009fy}%
  \BibitemOpen
  \bibfield  {author} {\bibinfo {author} {\bibfnamefont {A.}~\bibnamefont
  {Dumitru}}, \bibinfo {author} {\bibfnamefont {Y.}~\bibnamefont {Guo}},\ and\
  \bibinfo {author} {\bibfnamefont {M.}~\bibnamefont {Strickland}},\ }\bibfield
   {title} {\bibinfo {title} {{The Imaginary part of the static gluon
  propagator in an anisotropic (viscous) QCD plasma}},\ }\href
  {https://doi.org/10.1103/PhysRevD.79.114003} {\bibfield  {journal} {\bibinfo
  {journal} {Phys. Rev. D}\ }\textbf {\bibinfo {volume} {79}},\ \bibinfo
  {pages} {114003} (\bibinfo {year} {2009})},\ \Eprint
  {https://arxiv.org/abs/0903.4703} {arXiv:0903.4703 [hep-ph]} \BibitemShut
  {NoStop}%
\bibitem [{\citenamefont {Carrington}\ \emph {et~al.}(2004)\citenamefont
  {Carrington}, \citenamefont {Fugleberg}, \citenamefont {Pickering},\ and\
  \citenamefont {Thoma}}]{Carrington:2003je}%
  \BibitemOpen
  \bibfield  {author} {\bibinfo {author} {\bibfnamefont {M.~E.}\ \bibnamefont
  {Carrington}}, \bibinfo {author} {\bibfnamefont {T.}~\bibnamefont
  {Fugleberg}}, \bibinfo {author} {\bibfnamefont {D.}~\bibnamefont
  {Pickering}},\ and\ \bibinfo {author} {\bibfnamefont {M.~H.}\ \bibnamefont
  {Thoma}},\ }\bibfield  {title} {\bibinfo {title} {{Dielectric functions and
  dispersion relations of ultrarelativistic plasmas with collisions}},\ }\href
  {https://doi.org/10.1139/p04-035} {\bibfield  {journal} {\bibinfo  {journal}
  {Can. J. Phys.}\ }\textbf {\bibinfo {volume} {82}},\ \bibinfo {pages} {671}
  (\bibinfo {year} {2004})},\ \Eprint {https://arxiv.org/abs/hep-ph/0312103}
  {arXiv:hep-ph/0312103} \BibitemShut {NoStop}%
\bibitem [{\citenamefont {Kumar}\ \emph {et~al.}(2018)\citenamefont {Kumar},
  \citenamefont {Jamal}, \citenamefont {Chandra},\ and\ \citenamefont
  {Bhatt}}]{Kumar:2017bja}%
  \BibitemOpen
  \bibfield  {author} {\bibinfo {author} {\bibfnamefont {A.}~\bibnamefont
  {Kumar}}, \bibinfo {author} {\bibfnamefont {M.~Y.}\ \bibnamefont {Jamal}},
  \bibinfo {author} {\bibfnamefont {V.}~\bibnamefont {Chandra}},\ and\ \bibinfo
  {author} {\bibfnamefont {J.~R.}\ \bibnamefont {Bhatt}},\ }\bibfield  {title}
  {\bibinfo {title} {{Collective excitations of a hot anisotropic QCD medium
  with the Bhatnagar-Gross-Krook collisional kernel within an effective
  description}},\ }\href {https://doi.org/10.1103/PhysRevD.97.034007}
  {\bibfield  {journal} {\bibinfo  {journal} {Phys. Rev. D}\ }\textbf {\bibinfo
  {volume} {97}},\ \bibinfo {pages} {034007} (\bibinfo {year} {2018})},\
  \Eprint {https://arxiv.org/abs/1709.01032} {arXiv:1709.01032 [nucl-th]}
  \BibitemShut {NoStop}%
\bibitem [{\citenamefont {Wang}\ and\ \citenamefont
  {Shovkovy}(2021)}]{Wang:2021ebh}%
  \BibitemOpen
  \bibfield  {author} {\bibinfo {author} {\bibfnamefont {X.}~\bibnamefont
  {Wang}}\ and\ \bibinfo {author} {\bibfnamefont {I.}~\bibnamefont
  {Shovkovy}},\ }\bibfield  {title} {\bibinfo {title} {{Photon polarization
  tensor in a magnetized plasma: Absorptive part}},\ }\href
  {https://doi.org/10.1103/PhysRevD.104.056017} {\bibfield  {journal} {\bibinfo
   {journal} {Phys. Rev. D}\ }\textbf {\bibinfo {volume} {104}},\ \bibinfo
  {pages} {056017} (\bibinfo {year} {2021})},\ \Eprint
  {https://arxiv.org/abs/2103.01967} {arXiv:2103.01967 [nucl-th]} \BibitemShut
  {NoStop}%
\end{thebibliography}%

\end{document}